\documentclass[pra,aps,eqsecnum,showpacs,amsmath,10pt]{revtex4-1}
\usepackage{esint}
\usepackage{enumitem}
\usepackage{subfigure}
\usepackage{xcolor}
\usepackage{bbm}
\usepackage{amsfonts}
\usepackage{float}
\newcommand{\bra}[1]{\langle{#1}|}
\newcommand{\ket}[1]{|{#1}\rangle}
\newcommand{\braket}[2]{\langle{#1}|{#2}\rangle}
\usepackage{mathrsfs}
\usepackage{url}
\usepackage{hyperref}
\usepackage{graphicx}
\usepackage{array}
\usepackage[english]{babel}
\begin{document}
\title{Geometric Phase as the Key to Interference in Phase Space : Integral Representations for States and Matrix Elements}
\author{Mayukh N. Khan }
\email{hereismayukh@gmail.com}
\affiliation{Institute of Mathematical Sciences,
CIT Campus, Chennai 600113, India}
\author{S. Chaturvedi}
\email{subhash@iiserb.ac.in}
\affiliation{Indian Institute of Science Education and Research, Bhopal, Bhopal Bypass Road, Bhauri, Bhopal 462066, India}
\author{N. Mukunda}
\email{nmukunda@gmail.com}
\affiliation{INSA Distinguished Professor, Indian Academy of Sciences, C. V. Raman Avenue, Sadashivnagar Post, Bengaluru 560080, India}
\author{R. Simon}
\email{simon@imsc.res.in}
\affiliation{
SERB Distinguished Fellow, Institute of Mathematical Sciences,
CIT Campus, Chennai 600113, India}
\begin{abstract}
We apply geometric phase ideas to coherent states to shed light on interference phenomenon
in the phase space description of continuous variable Cartesian quantum systems. In contrast to Young's
interference characterized by path lengths, phase space interference turns out to be determined by areas.
The motivating idea is Pancharatnam's concept of ``being in phase" for Hilbert space vectors.   
Applied to the overcomplete family of coherent states, we are led to preferred 
one-dimensional integral representations for various states of physical significance, 
such as the position, momentum, Fock states and the squeezed vacuum. These 
are special in the sense of being ``in-phase superpositions".
Area considerations emerge naturally within a fully
quantum mechanical context. 
Interestingly, the Q-function is maximized along the line of such superpositions. 
We also get a fresh perspective on the Bohr-Sommerfeld quantization 
condition. Finally, we use our exact integral representations to obtain asymptotic expansions for 
state overlaps and matrix elements, leading to phase space area considerations similar to the ones noted earlier in the 
seminal works of Schleich, Wheeler and collaborators, but now from the perspective of geometric phase.\end{abstract}
\maketitle
\section{Introduction}{\label{Intro}}
It is a well recognized and long appreciated fact that the physical basis of quantum mechanics is so profound that 
it permits being formulated in a variety of mathematical forms, each emphasizing particular aspects of the subject. 
Thus, the original Heisenberg and Schr\"odinger discoveries of matrix and wave mechanics respectively highlight the noncommutativity 
of dynamical variables, and the superposition principle for pure states of quantum systems. In later years there appeared first the 
phase space approach to quantum mechanics pioneered by Wigner\,\cite{Wigner}, and then the elucidation by Dirac\,\cite{Dirac} of the 
role of the Lagrangian in quantum mechanics. The latter resulted in the path integral formulation due to Feynman\,\cite{Feynman1,*Feynman2}, 
and the operator action principle due to Schwinger\,\cite{Schwinger}.

Parallel to the above, with the passage of time and the continuing attempts at refining and reformulating the interpretation of the 
mathematical structure of quantum mechanics, every now and then some suprising new features and implications are discovered. 
Some examples which may be cited are superselection rules\,\cite{Superselection1,*Superselection2,*Superselection3}, systems of coherent states\,
\cite{CoherentKlauder},~\cite{perelomov1972coherent,*CoherentPerelomov}, 
the Zeno effect\,\cite{Zeno}, and geometric phases\,\cite{Berry,Wilczek_1989,*Bohm,*NiuRmpBerryPhase}. Geometric phases
have found applications across various fields in atomic physics, optics, and condensed matter. 
In condensed matter, there has been a deluge of activity in the context of the quantum Hall effect
~\cite{Girvin1990,*macdonald1994introduction,*girvin1999quantum,*stern2008anyons,*tong2016lectures},
topological insulators~\cite{TopologicalInsulatorHasan,*RMPZhangQi,*bernevig2013topological,*franz2013contemporary}, and Weyl semi-metals~\cite{rao2016weyl,*Burkov,*ReviewArmitageVishwanath}.
There are of course many more examples, a prominent 
one of very great current interest being entanglement in states of composite quantum systems\,
\cite{Einstein_1935,*Amicoreview,*Horodeckireview}.

Our aim in the present work is to link two of the themes recalled above -- phase space methods on the one hand and geometric phase 
ideas on the other. The original definitive formulation of quantum mechanics was at the level of wave functions or probability 
amplitudes on configuration space, based on vectors in Hilbert spaces. As mentioned earlier the Wigner phase space approach came 
slightly later. Here one deals directly with density matrices rather than wave functions. It then became clear that this approach 
is the counterpart of the earlier Weyl rule\,\cite{Weyl_rule1,*Weyl_rule2} of association of a quantum mechanical operator with each classical 
dynamical variable, for systems based on the usual Cartesian coordinate and momentum variables. A comprehensive formulation of this 
entire approach was achieved by Moyal\,\cite{Moyal} about a decade and a half later. Phase space methods for quantum systems with 
finite dimensional Hilbert spaces not possessing the more familiar kinds of classical analogues is another activity of  
considerable interest\,\cite{finiteWigner1,*finiteWigner2}.

The geometric phase concept is of relatively more recent origin, though not long after its discovery \cite{Berry} in 1983, some important 
precursors were recognized. The original work by Berry \,\cite{Berry} was in the context of adiabatic cyclic quantum evolution 
governed by the Schr\"odinger equation -- at the end of such evolution, the wave function picks up a phase of a geometric nature 
as it returns to its original form.
Barry Simon clarified the mathematical structure underlying Berry's work in a paper which, interestingly, appeared in print~\cite{BSimonBerryPhase}
ahead of Berry's own work.
In later work, the restrictions of adiabaticity\,\cite{adiabaticity} and cyclic evolution\,
\cite{cyclicevolution} were both removed, until finally a fully kinematic formulation\,\cite{kinematicformulation1,*kinematicformulation2} not even 
requiring the Schr\"odinger equation was developed. Among others, two significant precursors to this development are the work of
Pancharatnam in 1956 in the arena of classical polarization optics\,\cite{Pancharatnam}, and that of Bargmann in 1964 
\cite{Bargmanninvariant} connected 
with the Wigner theorem on symmetry transformations in quantum mechanics\,\cite{Wigner-Theorem},\cite{NewProof1,*NewProof2}.

In attempting to combine geometric phase ideas with phase space methods, we are motivated to some degree by 
the enormous success of phase space methods in quantum optics, especially in the theoretical and experimental studies of nonclassical 
features of radiation states. 


Some general remarks about using phase space methods in quantum mechanics are appropriate at this stage. 
These methods can be at two distinct levels: the density matrix, and  state vectors or wave functions. In the former we have the 
three well known examples of the Sudarshan-Glauber diagonal coherent state $\phi$  representation, the Wigner distribution $W(q,p)$, 
and the Husimi $Q$ function. The first and third both depend heavily on properties of coherent states, while the second 
has a simpler behaviour under the unitary action of linear canonical transformations. Further, $\phi$ is in 
general a quite singular distribution, thus limiting its practical uses; while $W(q,p)$ though real, often fails to be pointwise non-negative. 
The $Q$ function on the other hand has all the mathematical properties of a probability distribution on phase space. However, it 
cannot be interpreted as a probability in a physically significant way, since its values over different regions of phase space do not 
refer to distinct mutually exclusive experimental alternatives. 
Also every probability distribution in phase space is not a valid $Q$ function of some state.
At the wave function level, we have the Bargmann analytic function description of 
states\,\cite{bargmann1961hilbert}, again based on coherent states.

\subsection{Precedents}
The formalism developed here offers a fresh perspective on a topic that has broadly come to be known as ``interference in phase space".
To put our work in the proper context, we briefly recapitulate here the existing literature on this subject,  highlighting the 
ideas underlying the various approaches that have been proposed. Interference in phase space was first discussed by Wheeler~\cite{Wheeler_1985}
in the context of the Frank-Condon effect, drawing parallels to squeezed state physics. 
Schleich et al., wrote a series of papers using ideas of interference in phase space outlined below
to explain oscillations 
in photon number distribution as a signature of non-classical features in squeezed coherent modes of radiation
\cite{Schleich_1987,*Schleich_1988foundations},~\cite{Schleich_1988},\cite{Dowling_1991},~\cite{Schleich_2001}. 
In recent years, it has become possible to probe such phenomenon experimentally~\cite{xue2017controlling,MahmetValbruchSqueezingexperimental}.
Schleich et al. also pointed out the power of 
their technique in other situations, such as determining asymptotic expressions for 
matrix elements of the displacement operator in the Fock basis. Later, Dutta et al.~\cite{Dutta1993} uncovered further features in the photon number 
distribution of squeezed states, a giant oscillatory envelope over rapid oscillations. 
Even this was explained by Ref~\cite{dainty1994} using interference methods.
Since then, similar methods have been used in a variety of other 
situations, such as interpreting squeezing resulting from the superposition of coherent states \cite{Janzky1},\cite{Schleich_1991,*Buzek1991,*Janszky_1993} 
and photon distribution in squeezed Fock states\cite{KnightSqueezednumberstates}. 

The approach in \cite{Schleich_1987,*Schleich_1988foundations},\cite{Schleich_1988},\cite{Dowling_1991} 
hinges on the semiclassical WKB approximation. The authors of these works develop the picture of a state as a Bohr-Sommerfeld band with an area of $2\pi$. 
For example the Fock state $\ket{m}$ is represented by the band caught between the two circles of radius $\sqrt{2m}$ and $\sqrt{2m+2}$ respectively. 
They motivate this approach by appealing to the Wigner function associated with the state $\ket{m}$, whose last ripple lies at a distance 
of $\sqrt{2(m+1/2)}$. Similarly, a coherent state corresponds to a circle of radius $\sqrt{2}$ and a squeezed state to a cigar. 
The inner product of two states is taken to be essentially determined by the area of overlap of the associated bands.
In the cases where there is more than one area of overlap, one observes interesting interference phenomenon.
The amplitude of interference is determined by the areas of overlaps of the bands, and the relative phase is related to the  
area caught between the center lines of the corresponding bands. In particular, the inner product between two states $\ket{\chi}$ and $\ket{\psi}$ with
two regions of intersection is given by
\begin{equation}
	\bra{\chi}\psi\rangle\approx 2\sqrt{A_{\chi\psi}}\,\cos(\theta_{\chi\psi})
	\label{eqn:overlapSchleich}
\end{equation}
where $A_{\chi\psi}$ corresponds to the area of overlap of the bands and $\theta_{\chi\psi}$ 
is half of the the area caught between the center lines of the bands corresponding to the two states 
(apart from extra factors of $\pi/4$ arising from the turning points). To make the correspondence
between Eqn.~\eqref{eqn:overlapSchleich} and the WKB approximation exact (in the case of photon oscillations of a highly squeezed coherent state),
the area of overlap $A_{\chi\psi}$ had to be weighed by the Wigner distribution of the squeezed state.~\cite{Schleich_1988}
This picture of understanding oscillations in the inner product is intuitively very appealing and results from a judicious combination of 
ideas taken from  semiclassical description and two source interference.

The work of Schleich et al. was closely followed by that of Milburn \cite{Milburn_1989} and then later Mundarain et al.~\cite{Mundarain1} 
who interpreted overlap of states using the $Q$ distribution.
They noted that we can rewrite the inner product $\braket{\chi}{\psi}$ as 
(the notation used is standard, for clarification refer to definitions in the main text)
\begin{align}
 \braket{\chi}{\psi}&=\int\frac{d^2 z}{\pi}\braket{\chi}{z}\braket{z}{\psi}\nonumber\\
		    &=\int d^2z\sqrt{Q_\chi(z,z^\ast)\,Q_\psi(z,z^\ast)}\,\,e^{i\left[{\psi}(z)-{\chi}(z)\right]}\nonumber\\
 \text{ where }\braket{z}{\psi}& =\sqrt{\pi\, Q_\psi(z,z^\ast)}\,e^{i\,\psi(z)}
 \label{eqn:Qfunctionoverlap}
\end{align}
Then, the regions in the phase space that contribute to the overlap are determined by looking at places where the product of the two $Q$ distributions
is different from zero. This is similar in spirit to restricting the integral to the regions where the bands of the two states intersect.
Again, interference results when there is more than one such region.
Mundarain et al \cite{Mundarain1} use this idea to look at the photon distribution for displaced Fock states and squeezed 
displaced states.
They determine the inner product by calculating
$Q_\chi(z,z^\ast)$ and  $Q_\psi(z,z^\ast)$ and evaluate the interfering points by determining where the maximum contours of the $Q$ functions intersect.
The phases $\chi(z),\psi(z)$ are determined
by computing the inner products $\braket{z}{\chi}$ and $\braket{z}{\psi}$ respectively and these phases, in turn, control the interference pattern.
To calculate the amplitude of interference they go back to the Bohr Sommerfeld picture of bands with areas of $2\pi$ in phase space and calculate the 
areas of intersection. 
\subsection{Structure of present paper and summary}
In contrast to the works cited above, our approach, though inspired by them, is entirely based on the mathematical properties of coherent states and  
the notion of geometric phase in quantum mechanics.. It exploits the flexibility in the expansion of state vectors in terms 
of coherent states afforded by the overcompleteness of the latter to arrive at certain privileged expansions using ideas 
rooted in the idea of geometric phase. Area considerations emerge here naturally through the Pancharatnam phases and Bargmann 
invariants associated with coherent states without directly invoking any semiclassical considerations.

The material of this paper and the key ideas are arranged as follows. 

Section~\ref{sec:coherentgeometricreview}
reviews the main properties of coherent states, and the kinematic 
approach to the geometric phase, before bringing them together. This is done to set up basic notations, and  for the sake of 
completeness. The role of the Pancharatnam concept of the relative phase between two nonorthogonal Hilbert space vectors, and the 
consequences for superpositions of coherent states, are brought out in some detail. Of special importance is how area considerations
arise when we analyze in-phase continuous superpositions of coherent states along a curve in phase space. 

Next, with the help of the simplest example of superposition of two coherent states, we link our notion of in-phase 
superposition with the resulting interference pattern in the Husimi-Kano $Q$ 
distribution, and with the ideas on ``interference in phase space" as developed in \cite{Milburn_1989}. We also see how squeezing can arise
from such in-phase superpositions. 

In section~\ref{sec:h4generators},
we develop one-dimensional 
phase space integral representations for position, momentum eigenstates and Fock states in terms of coherent states such that they are 
in-phase in the Pancharatnam sense. These states are the eigenstates of the elements of the harmonic oscillator algebra $h_4$ (spanned by the Hermitian generators
$\{\mathbbm{1},\hat{p},\hat{q},\hat{a}^\dagger \hat{a}\}$). The generators of $h_4$ turn
out to have relatively simple actions in the phase plane, and the in-phase integral representations of their eigenstates are along their corresponding orbits.
We also discuss  how  our ideas lead to a fresh perspective on the Bohr-Sommerfeld quantization rule.
Finally, we see that the maximum of the $Q$ distribution of the eigenstates of position, momentum and number operators lie
along their corresponding orbits. 

Section~\ref{sec:applications} gives illustrative applications of the results of section~\ref{sec:h4generators} by calculating some familiar 
overlap integrals, as well as selected matrix elements of the displacement, oscillator time evolution, and squeezing operators. 
These examples aim to bring out clearly the 
simplicity and ease with which the computations can be carried out in our phase space approach.
Further, these examples lead quite naturally to the asymptotic expansions of Hermite and generalized Laguerre polynomials and show how area considerations 
determine the behaviour of oscillations in these cases. As an aside, we compare our approximations with ``standard" results and WKB analysis 
using root mean square criterion. It seems that as far as estimating values of the special functions are concerned, 
the WKB analysis produces marginally better results (albeit with errors of the same order). However, our results match those
obtained in the literature~\cite{dominici2007asymptotic} on estimation of zeros of Hermite polynomials. A
more detailed analysis is beyond the scope of this paper, and we do not investigate it further.

We note that the in-phase integral representations of the position, momentum and Fock states along
the orbits of the corresponding generators appear very naturally. 
This is not a coincidence. Coherent states of a single mode of radiation can be 
understood using Perelomov's~\cite{perelomov1972coherent} general formalism as arising from the Heisenberg-Weyl algebra spanned by $\{\mathbbm{1},\hat{q},
\hat{p}\}$,
or its complexified equivalent $\{\mathbbm{1},\hat{a},\hat{a}^\dagger\}$, with the vacuum state $\ket{0}$ as the fiducial state. 
The coherent state system does not change when we add the number operator $\hat{a}^{\dagger}\hat{a}$ to this algebra to get the aforementioned harmonic
oscillator algebra $h_4$. Only the isotropy subgroup of the fiducial state gets enlarged from $U(1)$ to $U(1)\times U(1)$ with the addition of
$\hat{a}^{\dagger}\hat{a}$ to the algebra.
\cite{CoherentPerelomov}. 
This underlying structure leads to the simple actions of the generators of $h_4$ on coherent states.

Section~\ref{sec:conclusions} contains concluding comments.
We hope that the insights gained from in-phase superpositions will prove relevant to efforts in quantum state engineering and 
squeezing using superpositions of coherent states which has now become an experimentally feasible and rapidly burgeoning field 
\cite{QuantumEngineeringNobel,*QuantumEngineering,*QuantumEngineeringconf,*QuantumEngineeringrev1,*QuantumEngineeringrev2}.
\section{Review of Coherent states, Geometric phases, their interrelations}\label{sec:coherentgeometricreview}
In this section, in order to make this paper reasonably self--contained, we recall 
briefly some of the essential properties of harmonic oscillator coherent states and of (the kinematic approach to) geometric phases. 
This also helps us to set up notations suited to our later applications.
\subsection{ Harmonic oscillator coherent states}
We deal with operators and states associated with a single canonical pair of Cartesian quantum mechanical variables. 
The basic Heisenberg commutation relations for Hermitian position and momentum operators, and for their usual complex combinations are :
{\allowdisplaybreaks\begin{align}\label{Fundamental Commutation}
 [\hat{q},\hat{p}] &= i; \nonumber\\
\hat{a}	,\hat{a}^\dagger &= \frac{1}{\sqrt{2}}(\hat{q} \pm i \hat{p})~:~\quad[\hat{a},\hat{a}^{\dagger}]=1.
\end{align}}
The unitary phase space displacement operators $D$ can also be given using either complex or real parameters:

{\allowdisplaybreaks\begin{align}\label{Setting notation}
z = \frac{1}{\sqrt{2}}\left(q + i p\right),&\quad -\infty < q,p < \infty~:~\nonumber\\
 D(z) &= D(q,p) = e^{z\hat{a}^\dagger - z^{*}\hat{a}} = e^{-\frac{1}{2}{|z|}^2}e^{z\hat{a}^\dagger }e^{- z^{*}\hat{a}}\nonumber\\
&= e^{i(p\hat{q}-q\hat{p})}\nonumber\\
&= e^{\frac{i}{2}qp}e^{-i q \hat{p}}e^{i p \hat{q}}\nonumber\\
&= e^{-\frac{i}{2}qp}e^{i p \hat{q}}e^{-i q \hat{p}}.
\end{align}}
where $q,p,z$ are c-numbers.
Their basic algebraic properties are
{\allowdisplaybreaks\begin{align}\label{Composition Law}
 D(z)^{-1} &= D(z)^{\dagger} = D(-z);\nonumber\\
D(z')D(z) &= e^{i \text{Im} (z^{*}z')} D(z+z'),\nonumber\\
\text{i.e.},\,\, D(q',p')D(q,p) &= e^{\frac{i}{2}\left( q p' -p q'\right)}\,D\left(q+q',p+p'\right).
\end{align}}
Under conjugation by these, the operators $\hat{q},\hat{p},\hat{a},{\hat{a}}^\dagger$ experience c-number shifts:
\begin{equation}
 {D(z)}^{-1}\{\hat{q},\hat{p},\hat{a},{\hat{a}}^\dagger\}D(z) = \{\hat{q}+q,\hat{p}+p,\hat{a}+z,{\hat{a}}^\dagger+z^{*}\}.
\end{equation}
The phase factor appearing in the composition law in Eqn.~\eqref{Composition Law} has a {\em geometrical meaning}, being 
the area of a triangle in the phase plane, with vertices $(0,0),(q,p),(q',p')$. We have 
\begin{equation}\label{area_triangle}
 A(0,0;q,p;q',p') = \text{Im}(z^* z') = \frac{1}{2}(qp'-pq'),
\end{equation}
where $A$ denotes area, the basic element being $dqdp$. (This differs from $d^{2}z = d\text{Re} z d\text{Im} z$ in the complex form by a 
factor of 2). The area in Eqn.~\eqref{area_triangle} is counted positive if the sequence of vertices indicated is anticlockwise, 
and negative otherwise; it of course vanishes if $(q,p),(q',p')$ lie on a straight line through the origin.

Coherent states $\ket{z}$ are the result of the displacement operators acting on the oscillator ground state, the latter being 
the unique normalized state annihilated by $\hat{a}$:
{\allowdisplaybreaks\begin{align}
 \hat{a}\ket{0}=&0, \,\,\braket{0}{0} =1~:~\nonumber\\
\ket{z} =& D(z)\ket{0}=  e^{-\frac{1}{2}{|z|}^2}\sum_{n=0}^{\infty}\frac{z^n}{\sqrt{n!}}\ket{n},\nonumber\\
\ket{n} =& \frac{{{\hat{a}}^{\dagger\,n}}}{\sqrt{n!}}  \ket{0},\,\, \braket{n}{n'}=\delta_{nn'};\nonumber\\
\hat{a}\ket{z}=z\ket{z};\quad&{\hat{a}^\dagger} \hat{a}\ket{n} = n\ket{n}.
\end{align}}
The Fock states $\ket{n}$ are the orthonormal eigenstates of the harmonic oscillator Hamiltonian, with of course $\ket{z=0}=\ket{n=0}$. 
We will frequently denote the states $\ket{z}$ by $\ket{q,p}$, the connection between the labels being as in  
Eqn.~\eqref{Setting notation}. Rewriting Eqn.~\eqref{Composition Law}, we have under displacements
{\allowdisplaybreaks\begin{align}\label{Displacement}
 D(z')\ket{z} &= e^{i \text{Im}(z^{*}z')}\ket{z+z'},\nonumber\\
i.e.,D(q',p')\ket{q,p} &= e^{\frac{i}{2}(qp'-pq')}\ket{q+q',p+p'}.
\end{align}}
We have the expectation values, uncertainties, and inner products:
{\allowdisplaybreaks
\begin{subequations}
\begin{align}\label{Coherent_state_relations}
 \bra{q,p}\hat{q}\,\,\text{or}\,\,\hat{p}\ket{q,p} &= q\, \text{or}\, p,\nonumber\\
(\Delta q)^2 &= \bra{q,p}(\hat{q}-q)^2\ket{q,p} = 1/2,\nonumber\\
(\Delta p)^2 &= \bra{q,p}(\hat{p}-p)^2\ket{q,p} = 1/2,\nonumber\\
\Delta(q,p) &= \bra{q,p}\frac{1}{2}[(\hat{q}-q)(\hat{p}-p)+(\hat{p}-p)(\hat{q}-q)]\ket{q,p} =0;\\
\braket{z'}{z} &= \braket{q',p'}{q,p}\nonumber\\
&= \exp[{-\frac{1}{2}{|z'-z|}^2 + i \text{Im}(z'^*z)}]\nonumber\\
&= \exp[{-\frac{1}{4}(q'-q)^2-\frac{1}{4}(p'-p)^2-\frac{i}{2}(qp'-pq')}].\label{eqn:coherentinnerproduct}
\end{align}
\end{subequations}
}
To distinguish these coherent states from the continuum (delta function normalized) position and momentum eigenstates, 
we use the following notations for the latter:
{\allowdisplaybreaks\begin{align}
 \hat{q}\ket{q,\text{pos}} &= q\ket{q,\text{pos}},\,\, \braket{q',\text{pos}}{q,\text{pos}} = \delta(q-q');\nonumber\\
\hat{p}\ket{p,\text{mom}} &= p\ket{p,\text{mom}},\,\, \braket{p',\text{mom}}{p,\text{mom}} = \delta(p-p');\nonumber\\
\braket{q,\text{pos}}{p,\text{mom}}&= \frac{1}{\sqrt{2\pi}}e^{i qp},-\infty  <q,p,\, q',p'<\infty
\end{align}}
Under displacements, using equation \eqref{Setting notation}, we have the actions:
\begin{align}
 D(q',p')\ket{q,\text{pos}}&= e^{i p'(q+q'/2)}\ket{q+q',\text{pos}},\nonumber\\
D(q',p')\ket{p,\text{mom}}&= e^{-i q'(p+p'/2)}\ket{p+p',\text{mom}}.
\end{align}
The position space and momentum space wave functions of the coherent states are
\begin{align}\label{eqn:wavefncoherent}
 \braket{q',\text{pos}}{q,p} &= \frac{1}{\pi^{1/4}}\exp[-\frac{1}{2}{(q'-q)}^2 + i p(q'-q/2)],\nonumber\\
\braket{p',\text{mom}}{q,p} &= \frac{1}{\pi^{1/4}}\exp[-\frac{1}{2}{(p'-p)}^2 - i q(p'-p/2)].
\end{align}

In this resum\'{e} of properties of coherent states, we turn finally to their overcompleteness. This is intimately related to 
the existence of the Bargmann entire analytic function representation of the fundamental commutation relation 
(\ref{Fundamental Commutation}). The more familiar representations use Schr\"odinger wavefunctions $\psi(q) = 
\braket{q,\text{pos}}{\psi}$ and $\hat{p}=-i\frac{d}{dq}$, or momentum space wavefunctions $\tilde{\psi}(p)=
\braket{p,\text{mom}}{\psi}$ and $\hat{q}=i\frac{d}{dp}$, to describe a general state vector $\ket{\psi}$. 
Both these are square integrable over the real line, but may otherwise be chosen arbitrarily. The Bargmann representation 
associates with $\ket{\psi}$ an entire analytic function $f_{\psi}(z)$ of a certain class given by
\begin{align}
 \ket{\psi}\rightarrow f_{\psi}(z) &= e^{\frac{1}{2}{|z|}^2}\braket{z^*}{\psi}\nonumber\\
&=\sum_{n=0}^{\infty}\frac{z^n}{\sqrt{n!}}\braket{n}{\psi}.
\end{align}
In this representation we have
\begin{align}
 (\hat{a}^\dagger f)(z) &= zf(z), \,\,(\hat{a}f)(z) = \frac{d}{dz}f(z),\nonumber\\
\braket{\psi'}{\psi} &= \int\frac{d^2 z}{\pi}e^{-{|z|}^2}f_{\psi^{'}}(z)^{*}f_{\psi}(z).
\end{align}
This representation makes it possible to appreciate the following facts:

\noindent
(a) There is a resolution of the identity,
\begin{equation}
 \int\frac{d^2 z}{\pi}\ket{z}\bra{z}=\int\int\frac{dqdp}{2\pi}\ket{q,p}\bra{q,p} = 1,
\end{equation}
as a result of which any $\ket{\psi}$ can certainly be expanded in the form
\begin{equation}
 \ket{\psi}= \int\frac{d^2 z}{\pi} e^{-\frac{1}{2} {|z|}^2}f_{\psi}(z^{*})\ket{z}.
\end{equation}
Here the `expansion coefficient' $f_{\psi}(z^{*})$ is very special in that it is entire analytic\,(in $z^{*}$).
\\ \noindent
(b) If one gives up this property, the expansion becomes highly non--unique, since 
\begin{align}
	\int d^2z\,\,g(|z|)\,z^n\ket{z} &=0;\quad n=1,2,3,\cdots
\end{align}
for a suitable function $g(|z|)$ for which the integral exists.
In this sense the set of all coherent states is not an independent set; rather they form an \emph{overcomplete} set.

 \noindent
(c) This overcompleteness is a reflection of the fact that an entire function $f(z)$ is completely determined, in principle, if 
one knows its values at all points in any so-called `characteristic set' $S$ in the complex plane
\cite{bargmann1961hilbert},\cite{bargmann1971completeness}. A subset $S\subset \mathcal{C}$ 
is characteristic if $f(z)=0$ for all $z\in S$ implies $f(z)=0$ identically. Examples are : any two dimensional region with non 
vanishing area; any sufficiently smooth line of finite or infinite length; any infinite sequence $\{z_n \}$ with a finite 
limit point. 

 \noindent
(d) All these possibilities mean that we can have `expansions' of a general $\ket{\psi}$ in terms of double integrals over suitable 
portions of $\mathcal{C}$; one dimensional line integrals over one-parameter families of coherent states 
$\{\ket{z(s)},s_1\leq s\leq s_2\}$; and even discrete approximations to $\ket{\psi}$ to any desired accuracy using 
states $\ket{z_n}$ for $z_n$ in some discrete infinite characteristic set.

\noindent
(e) Examples of one-dimensional line integral representations are, for instance,
\begin{equation}
 \ket{\psi} = \int_{-\infty}^{\infty}dq\, u(q) \ket{q,0} = \int_{-\infty}^{\infty} dp\, v(p)\ket{0,p},
\end{equation}
which have been used in\,\cite{MukundaSudarshan}.  In each such case one has to 
examine carefully the nature and properties of the `functions' $u(q), v(p)$ needed to recover a given $\ket{\psi}$, for all 
possible choices of $\ket{\psi}$. These and other similar expansions, also along circles in $\mathcal{C}$,  will be exploited in 
section~\ref{sec:h4generators}, guided by insights from geometric phase theory.
\subsection{Geometric Phase Theory}
We recapitulate the essentials using the kinematic formulation\,\cite{kinematicformulation1,*kinematicformulation2}. Let $\mathcal{H}$ be the Hilbert space 
pertaining to some quantum system. Let $\mathcal{B}$ be the unit sphere in $\mathcal{H}$:
\begin{equation}
 \mathcal{B} = \{\ket{\psi}\,\in \,\mathcal{H}\,|\,\braket{\psi}{\psi}=1\}\subset{\mathcal{H}}.
\end{equation}
The space $\mathcal{R}$ of unit rays is obtained from $\mathcal{B}$ via the canonical projection $\pi$:
\begin{align}
 \mathcal{R} = \{\hat{\rho}(\psi)&= \ket{\psi}\bra{\psi};\,\,\,\,\ket{\psi}\in\mathcal{B}\};\nonumber\\
\pi :\,\,\mathcal{B}\rightarrow\mathcal{R}\,\,:\,&\ket{\psi}\in\mathcal{B}\rightarrow\hat{\rho}(\psi)\in \mathcal{R}.
\end{align}
Neither $\mathcal{B}$ nor $\mathcal{R}$ is a vector space. Given any parametrised (piecewise) once differentiable curve 
$\mathscr{C}\subset\mathcal{B}$, by projection we obtain its  image $C\subset \mathcal{R}$:
\begin{align}
 \mathscr{C} = \{\ket{\psi{(s)}}\in\mathcal{B}\,\,|\,\, s_1\leq s\leq{s_2}\}\subset{\mathcal B} \rightarrow\nonumber\\
C = \pi(\mathscr{C}) = \{\hat{\rho}(s) = \ket{\psi(s)}\bra{\psi(s)} \in \mathcal{R}\,\,|\,\, s_1\leq s\leq s_2 \} \subset{\mathcal{R}}.
\end{align}
Then, a geometric phase is associated with $C$: it is the difference between the total - or Pancharatnam, and dynamical phases, 
each of which depends on $\mathscr{C}$ :
\begin{align}\label{definitions_phase}
 \phi_{\text{geom}}(C) &= \phi_{P}(\mathscr{C})- \phi_{\text{dyn}}(\mathscr{C}),\nonumber\\
\phi_{P}(\mathscr{C})&= \text{arg}(\psi(s_1),\psi(s_2)),\nonumber\\
\phi_{\text{dyn}}(\mathscr{C}) &= \text{Im}\int_{s_1}^{s_2}ds (\psi(s),\frac{d\psi(s)}{ds})\nonumber\\
&=-i \int_{s_1}^{s_2}ds(\psi(s),\frac{d\psi(s)}{ds}).
\end{align}
(For flexibility, we write vectors as $\ket{\psi}$ or $\psi$, inner products as $\bra{\,}\,\rangle$ or $(\,,\,)$ as convenient). 
This geometric phase is invariant under smooth local phase changes\,\,--\,\,$U(1)$ gauge transformations\,\,--\,\,of the form
\begin{equation}\label{gauge}
 \mathscr{C}\rightarrow\mathscr{C'}:\,\,\psi'(s) = e^{i\alpha(s)}\psi(s),
\end{equation}
as well as under monotonic reparametrizations:
\begin{align}
 s\rightarrow s'=f(s),&\,\,\frac{df(s)}{ds} > 0, \nonumber\\
\psi'(s') = \psi(s).
\end{align}
Even though the individual terms $\phi_{P}(\mathscr{C}),\, \phi_{\text{dyn}}(\mathscr{C})$ do change under (\ref{gauge}), their 
difference $\phi_{\text{geom}}(C)$ is invariant, which is why it is shown as dependent on the ray space image $C\subset \mathcal{R}$. 

We note that when the curve $\mathscr{C}$ is generated by the Hamiltonian, then the dynamical phase coincides with the familiar phase accumulated
during time evolution, thus justifying its moniker. For details, see Eqn.~\eqref{GeomPhaseHamil}.

Now we briefly recall Pancharatnam's \cite{Pancharatnam} work in the context of geometric phase. In his pioneering investigations in classical 
polarization optics, Pancharatnam introduced the concept of two complex two--component transverse electric vectors being 
``in phase'' if their superposition leads to maximum possible intensity. In case they are not in phase, he gave a quantitative measure 
of their relative phase. Pancharatnam came to these conclusions while studying interference of coherent beams of polarization.
Each state of polarization can be represented by a point on the Poincar\'e sphere. 
He also noted the non-transitivity of the in-phase relation, if there are three states of light $a,b,c$ such that $a$ and $b$ are in phase and 
so is $b$ and $c$, then $a$ and $c$ are, in general, not in phase, their phase difference is quantified geometrically by $\delta=\frac{1}{2}\Omega_{abc}$,
where $\Omega_{abc}$ is the solid angle subtended by the geodesic triangle on the Poincar\'e sphere. One is immediately drawn to recognising 
a formal similarity between Pancharatnam's result and the Berry phase \cite{Berry} arising in the context of a quantum mechanical description 
of a spin $1/2$ particle in an adiabatically rotating magnetic field. However, it should be borne in mind that no notion of adiabaticity 
was involved in Pancharatnam's work and that it preceded Berry's by almost twenty-eight years. 

Pancharatnam's idea can be generalized to the Hilbert space of any dimension in the context of quantum mechanics, and 
it then leads to the following definitions. Two mutually non-orthogonal vectors $\psi_1,\psi_2\,\in\,\mathcal{H}$ will be said to 
be ``in phase" if $(\psi_1,\psi_2)$ is real positive; if they are not, their relative phase is the argument of their inner 
product :
\begin{align}
\text{Phase of} \,\psi_2\, \text{relative to}\,\, \psi_1 &= \text{arg}(\psi_1,\psi_2),\nonumber\\
 \psi_1,\psi_2 \,\,\text{are in phase}& \Leftrightarrow (\psi_1,\psi_2)\,\, \text{real} >0.
\end{align}
(This explains the notation for the first term $\phi_P(\mathscr{C})$in Eqn.~(\ref{definitions_phase}) defining the geometric 
phase: $\phi_{P}(\mathscr{C})$ is the Pancharatnam phase of $\psi(s_2)$ relative to $\psi(s_1)$ )

The geometrical meaning and connection to maximum constructive interference can be brought out very simply. Assume that
$\psi_1$ and $\psi_2$ are normalized and non-orthogonal, and consider their superposition
\begin{equation}
 \ket{\psi} = \ket{\psi_1} + e^{i\theta}\ket{\psi_2},
\end{equation}
incorporating a relative phase. We can always unambiguously decompose $\ket{\psi_2}$ into parts parallel and orthogonal to 
$\ket{\psi_1}$:
{
\begin{align}
 \ket{\psi_2} &= re^{i\phi}\ket{\psi_1}+\sqrt{1-r^2}\ket{\psi_{1}^{\bot}},\nonumber\\
\braket{\psi_1}{\psi_1^{\bot}} &=0,\,\,0<r\leq 1.
\end{align}

Then
\begin{align}
 \phi_P(\psi_1,\psi_2) &= \phi,\nonumber\\
\ket{\psi} &= (1+ r\,e^{i(\theta+\phi)})\ket{\psi_1} + \sqrt{1-r^2}e^{i\theta}\ket{\psi_{1}^{\bot}},\nonumber\\
{||\psi||}^2 &= 2(1+r\cos(\theta+\phi)).
\end{align}}
It is clear that $||\psi||$ is a maximum, corresponding to maximum constructive interference, when $\theta = -\phi$, i.e., for 
\begin{align}\label{constructive_Interference}
 \ket{\psi} &= \ket{\psi_1} + e^{-i\phi_{P}(\psi_1,\psi_2)}\ket{\psi_2},\nonumber\\
{||\psi||}^2 &= 2(1+r),
\end{align}
when the phase $\theta$ is chosen to cancel $\phi$ and the two terms in the superposition are ``in phase" in the Pancharatnam sense. 
In short, ``in-phase superposition" leads to maximum constructive interference of vectors in Hilbert space.

As noted earlier, Pancharatnam also recognized that the ``in phase" property is not transitive: if $\psi_1,\psi_2$ and similarly $\psi_2,\psi_3$ are 
in phase pairs, then in general $\psi_1$ and $\psi_3$ will be out of phase and that he had also computed this final relative phase when the $\psi$'s are 
two component complex vectors. In the general quantum mechanical context, the signature and measure of this nontransitivity are 
shown very simply by the phase of the so--called ``three vertex Bargmann invariant"\,\cite{Bargmanninvariant}, an expression defined 
by
\begin{align}
 \text{arg} \Delta_{3}(\psi_1,\psi_2,\psi_3) &= \text{arg}(\psi_1,\psi_2)(\psi_2,\psi_3)(\psi_3,\psi_1)\nonumber\\
&= \text{arg}\,\, \text{Tr}\left(\hat{\rho}(\psi_1)\hat{\rho}(\psi_2)\hat{\rho}(\psi_3)\right),
\end{align}
for any three pairwise non-orthogonal vectors. 

This is easily understood by taking three arbitrary non-orthogonal vectors $\psi_1,\psi_2,\psi_3$. 
Adjacent states can be brought in phase with one another
by adding appropriate $U(1)$ phases.
\begin{align}
 \psi'_1=\psi_1,\psi'_2=e^{-i\phi_P(\psi_1,\psi_2)}\psi_2,\psi'_3=e^{-i\phi_P(\psi_1,\psi_2)}e^{-i\phi_P(\psi_2,\psi_3)}\psi_3
\end{align}
However, now 
\begin{align}
\phi_P(\psi'_1,\psi'_3)&=-\phi_P(\psi_1,\psi_2)-\phi_P(\psi_2,\psi_3)+\phi_P(\psi_1,\psi_3)\\
&=-\text{arg}(\psi_1,\psi_2)(\psi_2,\psi_3)(\psi_3,\psi_1)\nonumber\\
&= -\text{arg}\,\, \text{Tr}\left(\hat{\rho}(\psi_1)\hat{\rho}(\psi_2)\hat{\rho}(\psi_3)\right)
\end{align}
Further, the invariance of this result to local $U(1)$ gauge transformations is obvious, 
since it is expressed in terms of elements $\hat{\rho}\,\in \mathcal{R}$, the space of units rays.

The key fact is that the product of inner products is in general complex, 
so even if the pairs $\psi_1,\psi_2$ and $\psi_2,\psi_3$ are rendered in phase by adjoining phase factors to $\psi_2$ and $\psi_3$, 
the `gauge invariant' phase of $\Delta_3(\psi_1,\psi_2,\psi_3)$ will remain, and cannot be transformed away:
\begin{equation}\label{eqn:inphaseBargmann}
 \phi_P(\psi_1,\psi_2) = \phi_P(\psi_2,\psi_3)=0\,\Rightarrow\,\phi_P(\psi_1,\psi_3)= -\text{arg}\Delta_{3}(\psi_1,\psi_2,\psi_3).
\end{equation}

Let us next consider a sequence of normalized vectors $\psi_1,\psi_2,\cdots,\psi_{n},\cdots$ where each successive pair of vectors 
are in phase:
\begin{equation}
 \phi_{P}(\psi_{j},\psi_{j+1}) = \text{arg}(\psi_{j},\psi_{j+1})=0,\,\, j = 1,2,\cdots
\end{equation}
We can say we have a \emph{locally in phase sequence}, but non-transitivity means that vectors two or more steps apart may be 
out of phase. Nevertheless, given these vectors \emph{in the stated sequence} we can attach special significance to the sum
\begin{equation}\label{discrete}
 \ket{\psi}= \ket{\psi_1}+\ket{\psi_2} +\ket{\psi_3} + \cdots +\ket{\psi_n}+\cdots
\end{equation}
and say that $\ket{\psi}$ arises by \emph{locally in-phase superposition} of the sequence of vectors
$\{\psi_{n}\}$. This is a generalization of Eqn.~\eqref{constructive_Interference}, and can be further generalized to a continuous 
family of vectors as well.

With these ideas in mind, we return to the geometric phase $\phi_{\text{geom}}(C)$ in Eqn.~\eqref{definitions_phase}. Starting  
with the curve $\mathscr{C}=\{\psi(s)\}\subset\mathcal{B}$, we can always switch to a gauge transformed curve 
$\mathscr{C'}=\{\psi'(s)\}$ such that the vectors along the latter are `locally in phase':
\begin{align}
 \psi'(s) &= e^{i\alpha(s)}\psi(s),\,\,\alpha(s) = -\int_{s_1}^{s}ds'\text{Im} (\psi(s'),\dot{\psi}(s'));\nonumber\\
(\psi'(s),\dot{\psi'}(s)) &=0,\quad  (\psi'(s),\psi'(s+ds)) = 1+O(\delta s)^2,\nonumber\\
\phi_{\text{dyn}}(\mathscr{C'})&=0,\nonumber\\
\phi_{\text{geom}}(C) &= \phi_{P}(\psi'(s_1),\psi'(s_2)).
\end{align}
Thus, the dynamical phase disappears along $\mathscr{C'}$. Further, the Pancharatnam phase between infinitesimally separated
points along $\mathscr{C'}$ is also zero.
In the terminology of principal fibre bundles, $\mathscr{C'}$ is a horizontal lift of $C$, and so may be written as 
$\mathscr{C_{\text{hor}}}$. It is completely determined by its starting point in $\mathcal{B}$ and the ray space image $C$. 
Then the vector $\psi'(s)$ along $\mathscr{C_{\text{hor}}}$ experiences (locally) `in phase transport', and the geometric 
phase $\phi_{\text{geom}}(C)$ is entirely the accumulated measure of the non-transitivity of such transport in going 
from $\psi'(s_1) = \psi(s_1)$ to $\psi'(s_2)$. 

One can also say that the result of the superposition of these vectors,
\begin{align}
 \ket{\psi(\mathscr{C_{\text{hor}}})} = 
 &\underset{{\text{along } }~\mathscr{C}_{\text {hor}} }{\int_{s_1}^{s_2}} ds \ket{\psi'(s)}
\end{align}
an integral of vectors along a horizontal curve, has special mathematical (and physical) significance as being the result of  
maximum (local)  constructive interference. This is the continuous version of equation (\ref{discrete}).

An instructive example of the passage   $\mathscr{C}\rightarrow\mathscr{C_{\text{hor}}}$ is in the case of 
Schr\"odinger evolution. If $\ket{\psi(t)}$ is any solution of
\begin{align}
 \frac{d\psi(t)}{dt} = -i\hat{H}(t)\psi(t),
\end{align}
where $\hat{H}(t)$ is a given Hamiltonian operator, we can define
\begin{align}
	\mathscr{C} &= {\{\psi(t),\quad t_1\leq t \leq t_2}\}.
\end{align}
Then we easily find
{\allowdisplaybreaks\begin{align}
 \phi_{P}(\mathscr{C})&= \text{arg}(\psi(t_1),\psi(t_2)),\nonumber\\
\phi_{\text{dyn}}(\mathscr{C}))&= -\int_{t_1}^{t_2}dt\, E(t),\,\, E(t) =(\psi(t),\,\hat{H}(t)\psi(t));\nonumber\\
\mathscr{C_{\text{hor}}} &= \{\psi'(t)= \exp(i\int_{t_1}^{t}dt'E(t'))\,\,\psi(t)\},\nonumber\\
\phi_{\text{dyn}}(\mathscr{C_{\text{hor}}}) &=0,\nonumber\\
\phi_{\text{geom}}(C)&=\phi_{P}(\psi'(t_1),\psi'(t_2)) = \text{arg}(\psi(t_1),\psi'(t_2)).
\label{GeomPhaseHamil}
\end{align}}
\subsection{Geometric phases for coherent states}
\begin{figure}[htbp]
\centering
\includegraphics[width=0.25\textwidth]{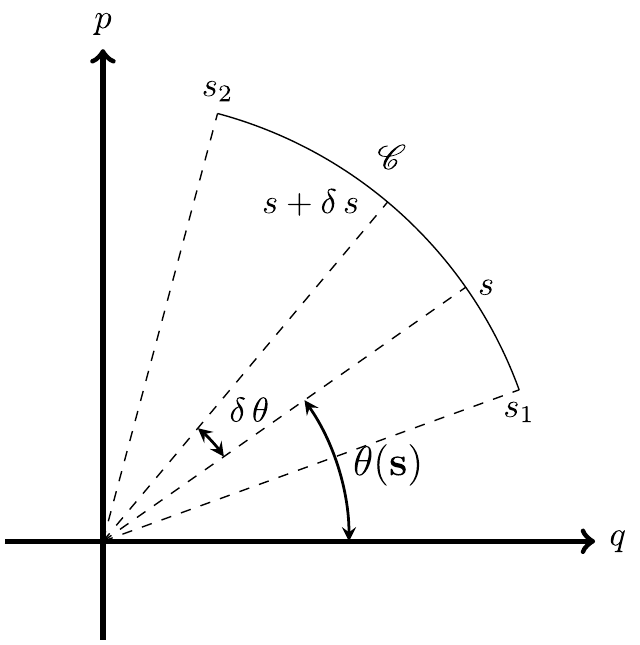}
\caption{Curve $\mathscr{C}$ in phase plane}
\label{fig:Curve_phase}
\end{figure}

In concluding this section, we apply geometric phase ideas to the case of (single mode) coherent states. 
Although, geometric phase in coherent states have been considered~\cite{chaturvedi1987berry,*yong1990berry,*field2004geometric},
our emphasis in this section is somewhat different.
To some extent, we now 
go beyond mere recall of known material.
In dealing with Hilbert space curves $\mathscr{C}$ within the manifold of coherent states,
\begin{equation}\label{manifold_coherent}
 \mathscr{C} = \{\ket{z(s)}=\ket{q(s),p(s)}\,\,|\,\,s_{1}\leq s\leq{s_2}\},
\end{equation}
we can when useful picture $\mathscr{C}$ as lying in the phase plane :
\begin{equation}
 \mathscr{C} =\{z(s)=\frac{1}{\sqrt{2}}(q(s)+i p(s))\,\,|\,\,s_1\leq s \leq s_2\}.
\end{equation}
Referring to Figure \ref{fig:Curve_phase}, and using radial and polar variables, we have for a small increment $\delta s$ in parameter value:
\begin{align}
 \phi_{P}(\ket{z(s)},\ket{z(s+\delta s)})&= \text{arg}\braket{q(s),p(s)}{q(s+\delta{s}),p(s+\delta{s})}\nonumber\\
&= \frac{1}{2}{r(s)}^2\frac{d\theta(s)}{ds}\delta s.
\end{align}

The curve $\mathscr{C}$  in Eqn. (\ref{manifold_coherent}) is in general not horizontal (in the Hilbert space sense), 
the gauge--transformed horizontal curve being:
\begin{equation}
 \mathscr{C_{\text{hor}}} = \{\ket{\psi(s)} = \exp(-\frac{i}{2}\int_{s_1}^{s}ds'\,\,r(s')^2\frac{d\theta{(s')}}{ds'})\,\,
 \ket{q(s),p(s)}\}.
 \label{eqn:coherentstatehorizontal}
\end{equation}
This is an important equation, we pause to note that the phase $-\frac{1}{2}\int_{s_1}^{s}ds'\,\,r(s')^2\frac{d\theta{(s')}}{ds'}$
is negative of the area swept along the curve from $s_1$ to $s$. In section~\ref{sec:h4generators}, we will see specific examples
of this equation in the context of integral representations of position, momentum and Fock states. This will lead in section~\ref{sec:applications} to the 
the result that the oscillations in the inner product of two states is determined by half the area caught between the lines of
integral representations denoting the states.

Therefore the geometric phase for $C= \pi(\mathscr{C})$ is:
\begin{align}
 \phi_{\text{geom}}(C) = \phi_{P}(\mathscr{C_{\text{hor}}}) &= \text{arg}\braket{\psi(s_1)}{\psi(s_2)}\nonumber\\
&=\frac{1}{2}(q(s_1)p(s_2)-p(s_1)q(s_2)) - \frac{1}{2}\int_{s_1}^{s_2}d\theta(s)~ r(s)^2 .
\end{align}
The second term is the negative of {\em{the area bounded by $\mathscr{C}$ and the initial and final radial lines}}. In case $\mathscr{C}$ is closed, we have:
\begin{align}\label{closed_curve}
 \partial \mathscr{C} = 0,\,&\, z(s_2)=z(s_1):\,\,&\nonumber\\
\phi_{\text{geom}}(C) &= -(\text{area enclosed by}\,\, \mathscr{C}\,\, \text{in phase plane}).
\end{align}

Now, we are ready to point out the connections between Pancharatnam phases and Bargmann invariants in the specific case of coherent 
states. From Eqn.~\eqref{eqn:coherentinnerproduct}, remembering that two coherent states are never orthogonal, we have:
\begin{equation}\label{phase_area}
 \phi_{P}(\ket{q_1,p_1},\ket{q_2,p_2})=\frac{1}{2}(q_1 p_2 - q_2 p_1) = A(0,0;q_1,p_1;q_2,p_2),
\end{equation}
the area of the relevant triangle. 

For the Bargmann invariant among three coherent states we have a similar 
geometric result:
\begin{align}\label{phase_argument}
 \text{arg}\Delta_3(\ket{z_1},\ket{z_2},\ket{z_3})&= \text{arg}\braket{z_1}{z_2}\braket{z_2}{z_3}\braket{z_3}{z_1}\nonumber\\
&= \text{Im}(z_{1}^{*}z_{2}+ z_{2}^{*}z_{3}+ z_{3}^{*}z_{1})\nonumber\\
&=\text{Im}(z_{2}^{*}-z_{1}^{*})(z_3-z_1)\nonumber\\
&=A(0,0;q_2-q_1,p_2-p_1;q_3-q_1,p_3-p_1)\nonumber\\
&=A(q_1,p_1;q_2,p_2;q_3,p_3).
\end{align}
Thus, this is exactly the area of the triangle in the phase plane, with appropriate sign. Comparing this with Eqn. (\ref{closed_curve}) 
we see that the Bargmann invariant phase  is the negative of the geometric phase for a triangle viewed as a closed curve in the phase plane. 

The 
connection between Eqn. (\ref{phase_area}) and Eqn.~(\ref{phase_argument}) should
not come as a surprise: this is because by convention the phases 
of the coherent states $\ket{z}$ are so chosen that every $\ket{z}$ is in phase with the fiducial state $\ket{0}$; 
\begin{align}
	\phi_P\left(\ket{z_1},\ket{z_2}\right)&=-\mbox{arg }\Delta_3\left(\ket{z_1},\ket{0},\ket{z_2}\right)\nonumber\\
	&=\mbox{arg }\Delta_3\left(\ket{0},\ket{z_1},\ket{z_2}\right)~\mbox{using Eqn.}~\eqref{eqn:inphaseBargmann}\nonumber\\
	&=A(0,0;q_1,p_1;q_2,p_2)\label{eqn:inphaseBargmanncoherent}
\end{align}
\subsection{A preparatory exercise}\label{subsec:preparatory}
\begin{figure}
\centering
\subfigure[b][\scriptsize{$\ket{\psi}=\frac{1}{N_1}(\ket{-q_0,0}+\ket{q_0,0})$\newline
In phase superposition of two coherent states along the x -axis.\newline $q_0$=0.4
}]{ \includegraphics[width=0.48\textwidth]{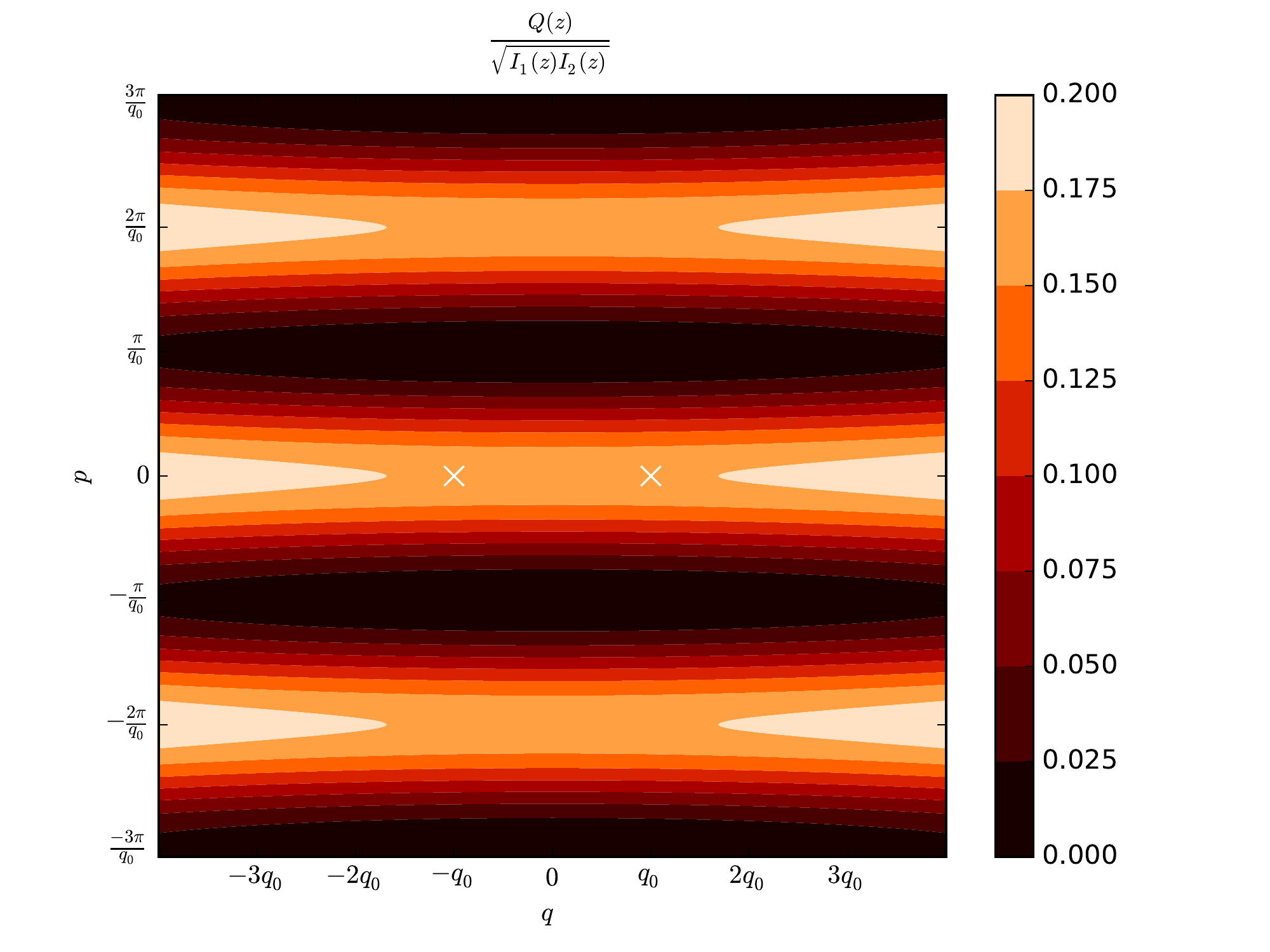}\label{fig:InPhase}}\quad
\subfigure[b][\scriptsize{ $\ket{\psi}=\frac{1}{N_2}(\ket{-q_0,0}-\ket{q_0,0})$\newline
Out of phase superposition of two coherent states along the x -axis. $q_0$=0.4
}]{ \includegraphics[width=0.48\textwidth]{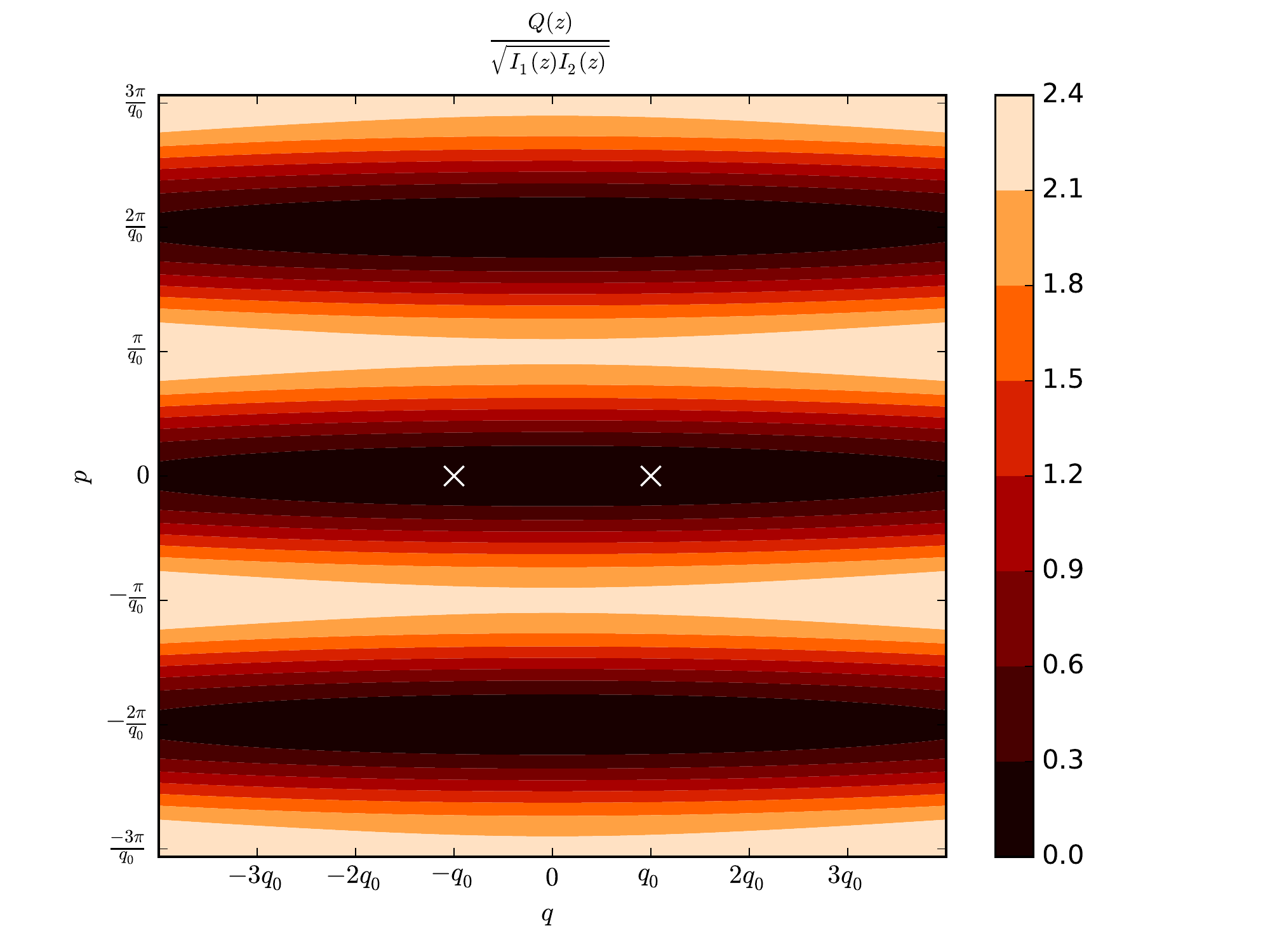}\label{fig:OutofPhase}}
\caption{Plot of $\frac{Q(z)}{\sqrt{I_1(z)I_2(z)}}$ in phase space for a superposition of two coherent states along the x-axis.
	We have divided the $Q$ distribution by $\sqrt{I_1(z)I_2(z)}$ to make the interference fringes more obvious.
	The positions of the coherent states are denoted by white crosses on the figure. Look at Eqn~\eqref{eqn:qfunction}
	for the relevant expressions. 
}
\end{figure}

Consider the following non-normalized superposition of two coherent states
\begin{equation}\label{Superposition}
 \ket{\psi} =\ket{z_1} + e^{i\theta}\ket{z_2},
\end{equation}
 The Pancharatnam phase between the two terms is
\begin{equation}
 \phi_{P}(\ket{z_1},e^{i\theta}\ket{z_2}) = \delta_{0}(\theta)=\theta + A(0,0;q_1,p_1;q_2,p_2),
\end{equation}
and it affects the norm of $\ket{\psi}$:
\begin{equation}
 \braket{\psi}{\psi} = 2(1+e^{-\frac{1}{2}{|z_1-z_2|}^2}\cos\delta_{0}(\theta)).
\end{equation}
This is a maximum for an in-phase superposition for which $\delta_{0}(\theta)$ vanishes:
\begin{subequations}
\begin{align}
 \theta =& - A(0,0;q_1,p_1;q_2,p_2)=-\phi_P(\ket{z_1},\ket{z_2}):\label{eqn:notesimilarity}\\
\ket{\psi}=&\ket{z_1} + e^{-i A(0,0;q_1,p_1;q_2,p_2)}\ket{z_2},\nonumber\\
\braket{\psi}{\psi}=& 2(1+e^{-\frac{1}{2}{|z_1-z_2|}^2}).
\end{align}
\end{subequations}
Equation \eqref{eqn:notesimilarity} is expected and is just \eqref{constructive_Interference} applied to coherent states.
Returning to the general case Eqn. (\ref{Superposition}) with variable $\theta$, it is instructive to calculate the 
corresponding $Q$ distribution. After simplifications we find:
\begin{align}
 Q_{\psi}(z) &= \frac{1}{{\pi||\psi||}^2}{|\braket{z}{\psi}|}^2\nonumber\\
&=\frac{1}{{\pi||\psi||}^2}\{I_1(z) + I_{2}(z) + 2\sqrt{I_{1}(z)I_{2}(z)}\cos\delta(z)\},\nonumber\\
I_{j}(z) &= \exp(-{|z-z_j|}^2),\quad j =1,2 \,\,~,\nonumber\\
\delta(z) &= \delta_{0}(\theta) + A(q_1,p_1;q,p;q_2,p_2).
\label{eqn:qfunction}
\end{align}
Equation~\eqref{eqn:qfunction} leads us to one of the key points in this paper. 
We overlook for the moment the comment in Section~\ref{Intro} that $Q(z)$ is not a physically 
meaningful probability distribution, and note the following points.
\begin{enumerate}[label=(\alph*)]
\item The $Q$ distribution can be interpreted as describing the interference pattern in the intensity distribution 
arising out of a superposition of coherent states which act as point sources. The intensity at a point $z$ due to the source 
at $z_i$ is given 
as $I_i(z)$. 
\item  The interference fringes are determined by the cosine of the angle $\delta(z)$, which is determined by {\em areas} in phase space, 
in contrast to the propagation distances which determine fringes in a Young's double slit experiment. 
Contours of equal $\delta(z)$ here run parallel to the line joing $z_2$ and $z_1$, in contrast to Young's interference 
fringes where they tend to be perpendicular to the line joing the point sources. The spacing between lines of constructive interference, corresponding to
$\delta(z)=2\pi n$, is proportional to ${|z_2-z_1|}^{-1}$. 

To make the fringes more clearly visible, we have divided the $Q$ function by the exponential factor
$\sqrt{I_1(z)\,I_2(z)}$ in  Figure~\ref{fig:InPhase} and ~\ref{fig:OutofPhase}.
For a superposition of two coherent states at $(-q_0,0)$ and $(q_0,0)$ the horizontal fringes are separated 
from each other by $\frac{2\pi}{q_0}$ as expected.

What is important is that phase space areas, rather than physical space distances, prove to be relevant here.
\item We note that since by Eqn.~\eqref{eqn:qfunction}, the intensity $I_i$ due to the point source $i$ decreases exponentially with 
distance from $z_i$, points with equal $\delta(z)$ are not points of equal intensity. 

However, when $\delta_0(\theta)=0$, the line joining $z_2$ and $z_1$ becomes a point of constructive interference, $\delta(z)=0$ on the line. 
This provides substance to the intuition that if we put a continuum of coherent states locally in phase with each other, 
the maximum of their $Q$ distribution will stay put along the line of their
superposition. (upto quantization caveat for closed orbits, illustrated in section~\ref{Hamiltonian})

We will see in the next section that this is indeed true for the $Q$ distribution for the position, 
momentum and Fock states expressed as in - phase superpositions of coherent states.
\end{enumerate}

The superposition in Eqn.~(\ref{Superposition}) is also physically significant in that it shows nonclassical features in a 
dramatic fashion, in the uncertainties in $\hat{q}$ and $\hat{p}$. 
Since the quadrature uncertainties, in fact the variance matrix as a whole, are invariant under phase space displacements and 
transform simply under phase space rotations, we can simplify the algebra, without loss of generality, by choosing special 
values for $z_1$ and $z_2$ in Eqn.~(\ref{Superposition}).
Thus we take a normalized superposition of two coherent states on the real co-ordinate axis placed symmetrically about the origin:
\begin{align}\label{Simple_Superposition}
\ket{\psi} &=\frac{1}{N}\left(\ket{-q_0,0}+e^{i\theta}\ket{q_0,0}\right), q_0\geq 0 :\nonumber\\
N^{2} &=2(1+c^4\cos\theta), c=e^{-\frac{1}{4}q_0^2}\leq 1.
\end{align}
In this state we find the following expectation values and spreads for $\hat{q}$ and $\hat{p}$:
\begin{align}\label{variances}
 <\hat{q}> &=0;\,\, {(\Delta q)}^2 =\frac{1}{2} + 2 \left({\frac{q_0}{N}}\right)^2;\nonumber\\
<\hat{p}> &= 2\frac{q_0}{N^2}c^4\sin\theta,\,\, <{\hat{p}}^2> = \frac{1}{2}-2{\left(\frac{q_0 c^2}{N}\right)}^2\cos\theta,
\nonumber\\
({\Delta p})^2 &= \frac{1}{2} - {\left(\frac{2 q_0 c^2}{N^2} \right)}^2 (c^4 +\cos\theta)\nonumber\\
\Delta(q,p) &=0.
\end{align}
Comparing with the vacuum values ${(\Delta q)}^2 = {(\Delta p)}^2 = \frac{1}{2}$, we see that for the particular two coherent 
state superposition Eqn.~(\ref{Simple_Superposition}) ``along the real axis", $\hat{q}$ never shows squeezing; however the `imaginary' 
quadrature $\hat{p}$ shows squeezing as long as $\cos\theta >-c^4$.
In fact this squeezing is maximum for $\theta =0$, which 
corresponds to an in phase superposition; and this gradually weakens and disappears at 
$\theta = \cos^{-1}(-c^4)\in(\frac{\pi}{2},\pi)$. We note that this form of squeezing is different from squeezing resulting from the action 
of the unitary squeezing operator on the coherent states, in that here $\Delta q\Delta p\neq \frac{1}{2}$.

That superposition of coherent states can lead to squeezing has been known for a while,\cite{Janzky1} and has been interpreted 
\cite{Schleich_1991,*Buzek1991,*Janszky_1993} using interference in phase space
 and the Wigner function of the superpositions. We will see later that this type of squeezing arising from in-phase superpositions
becomes more dramatic as more and more coherent states are included.
 At the risk of repetition, we note that the purpose of the above exercise was to help the reader develop the following intuition, 
 which will be borne out in the examples treated in the next Section.
\begin{itemize}
 \item In-phase superposition of coherent states along a curve, tends to lead to the maximum of the $Q$ function lying on 
 the curve.
 \item In-phase superposition along a curve leads to squeezing in the direction perpendicular to the curve.
 \item We saw in the above example that a shift in the relative phase between the two states in the superposition
 moves the fringes in a direction perpendicular to the line of superposition, without  destroying the squeezing.
 In the continuum case, it continues to be expected that a phase gradient along the curve will shift the maximum in a direction 
 perpendicular to the curve, maintaining the squeezing perpendicular to the curve. 
\end{itemize}

\section{Phase space integral representations for position, momentum, and Fock states}{\label{sec:h4generators}}
\begin{figure}
\centering
\subfigure[b][\,Orbit associated with $\hat{p}$]{
\includegraphics[width=0.30\textwidth]{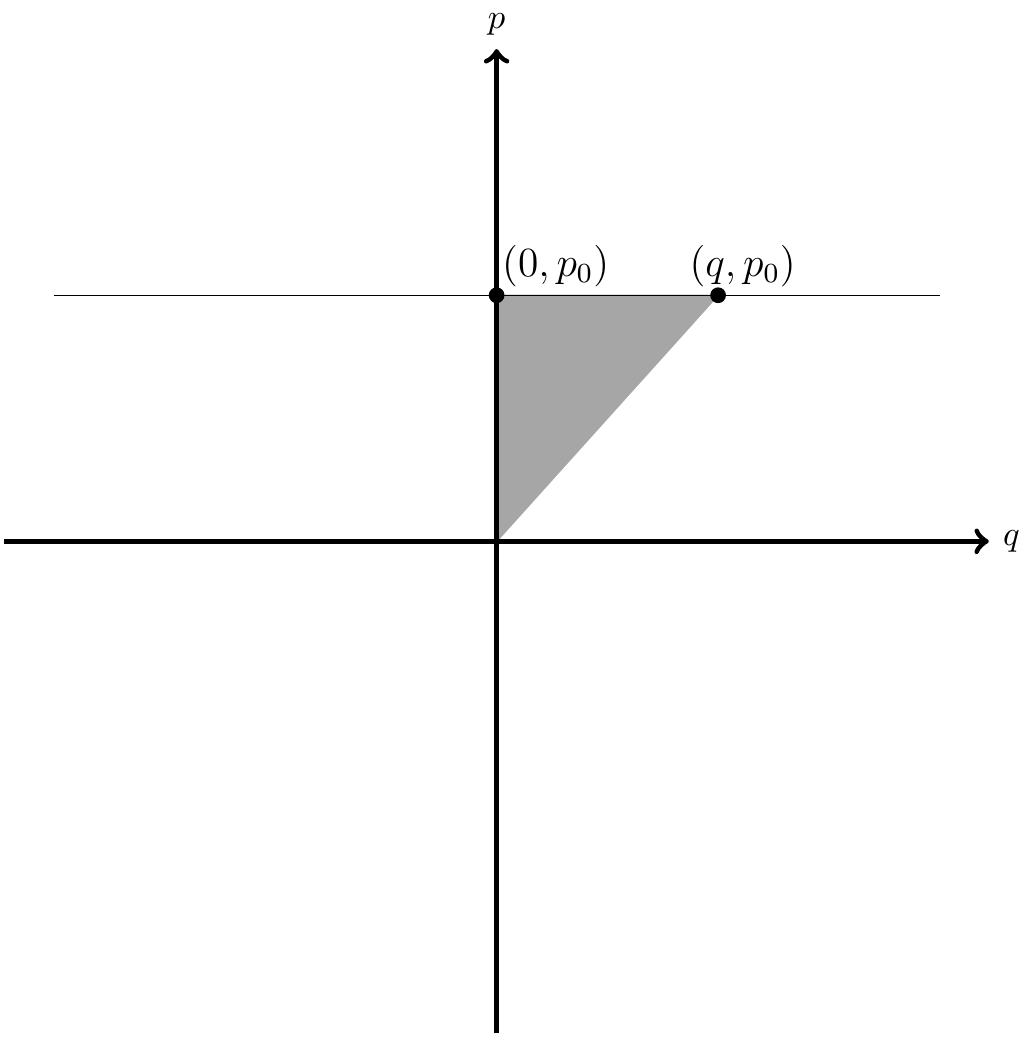}\label{fig:orbitp}}\qquad
\subfigure[b][\, Orbit associated with $\hat{q}$]{
\includegraphics[width=0.30\textwidth]{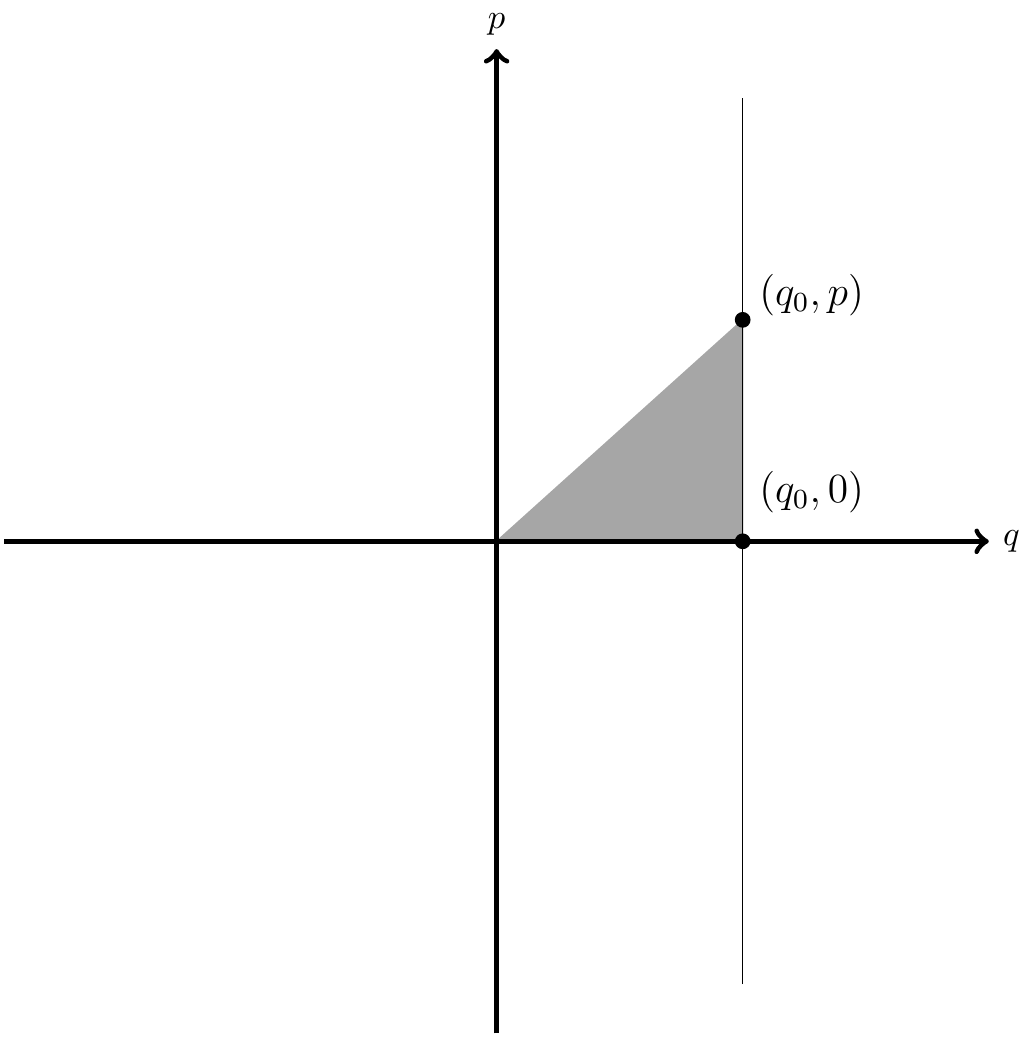}\label{fig:orbitq}}\\
\subfigure[\,Orbit associated with $\hat{n}$]{
\includegraphics[width=0.30\textwidth]{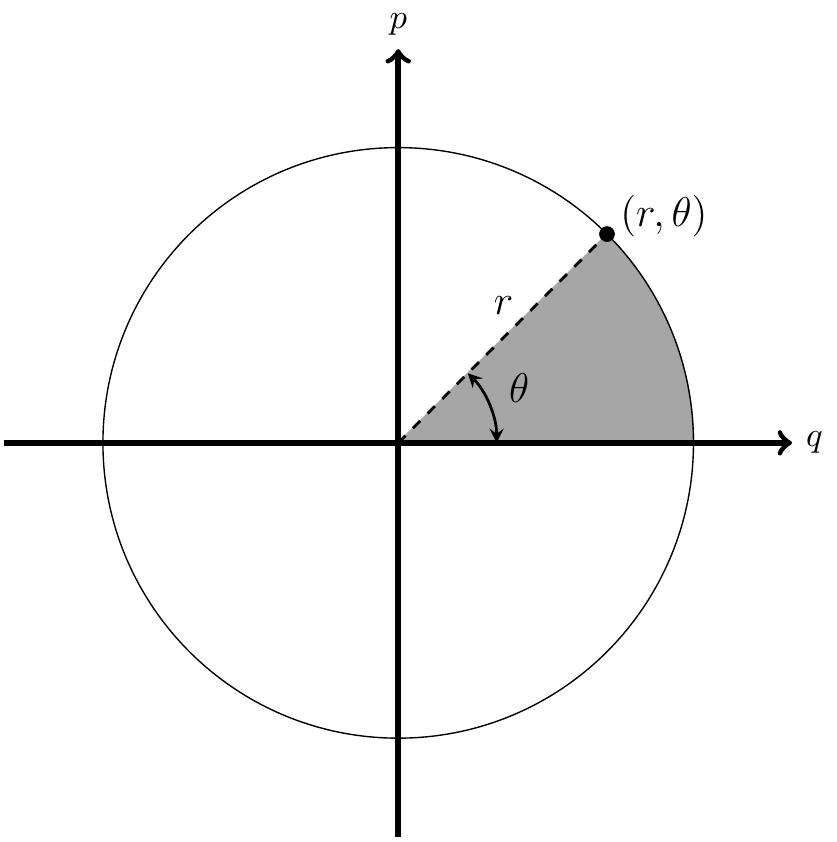}\label{fig:orbitn}}
\caption{Phase plane orbits associated with the position, momentum and rotation
generators in two dimensional phase space. The shaded areas (with appropriate
signs) correspond to the
phase factors which appear in the in-phase superpositions in Eqns. \eqref{In_phase},\eqref{In_phase_momentum},\eqref{Fock_in_phase}, as determined
by Eqn.~\eqref{eqn:coherentstatehorizontal}}
\label{fig:Orbitgenerators}
\end{figure}

The position and momentum operators generate translations in phase space, while the number operator generates rotations. 
We will look at them in the sequence $\hat{p}$ (translations in $q$), $\hat{q}$
(translations in $p$), and the number operator $\hat{n}=\hat{a}^{\dagger}\hat{a}$
(rotations about the origin).
 These actions are associated with the corresponding orbits in the phase plane, leading to preferred in-phase 
integral representations for their eigenvectors. 
The first two will turn out to have global in-phase expansions, 
while for $\hat{n}$, we will obtain a local in-phase superposition.

\subsection{Momentum eigenstates}
The momentum operator $\hat{p}$ shifts the co-ordinate variable, leaving momentum unchanged:
{\allowdisplaybreaks\begin{align}
e^{-i q' \hat{p}}&= D(q',0),\nonumber\\
D(q',0)\ket{q,p_0} &= e^{-\frac{i}{2}q'p_0}\ket{q'+q,p_0}.
\end{align}}
Thus, the orbits in the phase plane associated with $\hat{p}$ 
 action are, for each fixed $p_0$, the (horizontal) straight line $ p_0 \,=$ constant
parallel to the $q$ axis. (Fig. \ref{fig:orbitp})

Each such orbit, labelled by $p_0$, is a characteristic set. Therefore in principle, any $\ket{\psi}$ can be expanded in the form 
of a one-dimensional integral:
\begin{equation}\label{expansion}
 \ket{\psi}=\int_{-\infty}^{\infty}dq\,\,u(q)\ket{q,p_0},\,\,p_0\,\, \text{fixed}.
\end{equation}
Taking the overlaps with $\bra{q',\text{pos}}$ and $\bra{p',\text{mom}}$ leads to
\begin{align}\label{overlap}
 \psi(q') &= \frac{1}{{\pi}^{\frac{1}{4}}}\int_{-\infty}^{\infty} dq\,\,u(q) \exp[-\frac{1}{2}({q-q'})^2 +
 i p_0(q'-\frac{q}{2})],\nonumber\\
\tilde{\psi}(p') &= \frac{1}{{\pi}^\frac{1}{4}}\exp[-\frac{1}{2}{(p_0-p')}^2]\int_{-\infty}^{\infty}dq
\,\,u(q)\exp[-i q(p'-\frac{p_0}{2})].
\end{align}
The first relation is rather unwieldy, so we use the second. By Fourier inversion, we can express $u(q)$
in terms of $\tilde{\psi}(p')$:
\begin{align}\label{Fourier_momentum}
 u(q) = \frac{1}{2\pi^{3/4}}e^{-\frac{i}{2}q p_0}\int_{-\infty}^{\infty}dp'\,\,\tilde{\psi}(p')e^{\frac{1}{2}{(p'-p_0)}^2 + 
 i p'q}~~. 
\end{align}
In general, for a square integrable $\tilde{\psi}(p')$, $u(q)$ is a distribution.
For any fixed $p_{0}$, as $u(q)$ is unambiguously determined by $\ket{\psi}$, the one--dimensional family of coherent 
 states $\{ \ket{q,p_0}, -\infty<q<\infty\}$ is linearly independent, and the expansion in Eqn. (\ref{expansion}) is unique.

For the momentum eigenstate,
\begin{align}\label{momentum_eigenstate}
\ket{\psi} &= \ket{p'_{0},\text{mom}},\quad \tilde{\psi}(p') =\delta(p'-p'_0) :\nonumber\\
u(q) &= \frac{1}{2\pi^{3/4}}\,\exp[\frac{1}{2}{(p_0-p'_0)}^2 + i q (p'_{0}-\frac{p_0}{2})],\nonumber\\
\ket{p'_{0},\text{mom}}&= \frac{1}{2\pi^{3/4}}\,\exp[\frac{1}{2}{(p_0-p'_0)}^2]\int_{-\infty}^{\infty}dq 
\exp[i q (p'_{0}-\frac{p_0}{2})]\ket{q,p_0}.
\end{align}
If we examine the integrand here, we find that in general these vectors are out of phase :
\begin{align}
 \phi_{P}\left(\exp[i q'(p'_{0}-\frac{1}{2} p_{0})]\,\,\ket{q',p_0},\exp[i q(p'_{0}-\frac{1}{2} p_{0})]
 \,\,\ket{q,p_0}\right) =-(q'-q)(p'_{0}-p_0).
\end{align}
For given $\ket{p'_{0},\text{mom}}$, the preferred expansion via such integrals is obtained when $p_{0}=p'_{0}$, so that it 
becomes an in-phase superposition:
\begin{equation}\label{In_phase}
 \ket{p'_{0},\text{mom}} = \frac{1}{2\pi^{3/4}}\int_{-\infty}^{\infty}dq \exp[\frac{i}{2} q p'_{0}]\,\,\ket{q,p'_{0}}.
\end{equation}
Now that we have constructed the in-phase superposition, we note the following points
\begin{enumerate}[label=(\alph*)]
 \item In Eqn.~\eqref{momentum_eigenstate}, the momentum eigenstate $\ket{p'_0,\text{mom}}$ is concentrated along $p'_0$, while the 
 superposition which results in it is along the line $p_0$ in phase space. Splitting the phase factor 
 $\exp[i q (p'_{0}-\frac{p_0}{2})]$
 as $\exp[i q p_{0}/2]\exp[i q (p'_{0}-{p_0})]$, we see that in the absence of the phase gradient $\exp[i q (p'_{0}-{p_0})]$,
 the R.H.S. of Eqn.~\eqref{momentum_eigenstate}, would have been an in-phase superposition and the result would have stayed put 
 at the location $p_0$. The phase gradient drives the superposition precisely by the amount $p'_0-p_0$ to $p'_0$. 
 Since the amplitude of the coherent states on $p_0$ is small on $p'_0$ by the factor 
 $\exp\left[-\frac{1}{2}(p_0-p'_0)^2\right]$, we need the extra compensatory 
 factor of $\exp\left[\frac{1}{2}(p_0-p'_0)^2\right]$on the R.H.S. in Eqn.~\eqref{momentum_eigenstate}. Hence, the superposition of $\ket{p'_0,\text{mom}}$ along any other line 
 $p_0\ne p'_0$ is inefficient. 
 \item Since we have an integral along a straight line, any three points on it form a degenerate triangle with vanishing area, 
so this is {\em{a globally in-phase superposition}} of coherent states. We stress that while $p'_{0}$ and $p_0$ are independent 
in the valid expansion Eqn.~(\ref{momentum_eigenstate}), from the insights of geometric phase theory, we find the in-phase 
superposition Eqn.~(\ref{In_phase}) a preferred one.
\item Note that a globally in-phase superposition along the straight line $p=p_0^{\prime}$ necessitates the phase $\frac{1}{2} q p'_{0}$ in 
Eqn.~\eqref{In_phase}. This can be easily seen using Eqn.~\eqref{eqn:inphaseBargmanncoherent}.
\begin{align}
	\phi_P(\ket{0,p_0},\ket{q,p^\prime_0})&=A(0,0;0,p^\prime_0;q,p^\prime_0)=\frac{qp^\prime_0}{2}
\label{eqn:}
\end{align}
\item A momentum eigenstate is one in which the momentum is infinitely squeezed. We saw in Section~\ref{sec:coherentgeometricreview} that in-phase superposition 
of two coherent states $\ket{\pm q_0,0}$ on the $q$ axis leads to some squeezing in $\hat{p}$, as seen in Eqn.\,(\ref{variances}), with maximum 
squeezing when $\theta =0$. Standing in between these two cases, Eqn.\,(\ref{Simple_Superposition}) with $\theta =0$ and Eqn.~(\ref{In_phase})
is this example involving a globally in phase superposition, but with Gaussian weight. Apart from overall normalization,
\begin{align}
\ket{\psi} &= \int_{-\infty} ^{\infty}dq  \exp[-\frac{\sigma^2 q^2}{2} + \frac{i}{2} q p_{0}]\,\ket{q,p_0},\nonumber\\
{\tilde{\psi}(p)}&= \braket{p,\text{mom}}{\psi}= \sqrt{2}\frac{{\pi}^{\frac{1}{4}}}{\sigma}
\exp[-\frac{1}{2}{(p-p_0)}^2(1+\frac{1}{\sigma^2})].
\end{align}
This shows a variable degree of squeezing in momentum, gradually tending to a momentum eigenstate(infinite squeezing) as 
$\sigma \rightarrow 0$.
\item Also note that while a phase gradient shifts the superposition, 
it does not affect the squeezing, as anticipated in the last paragraph in subsection~\ref{subsec:preparatory}.
\end{enumerate}

We can in principle seek an expansion of the form of Eqn.~(\ref{expansion}) even for a position eigenstate; 
using Eqn.~(\ref{Fourier_momentum}),
\begin{align}
 \ket{\psi} &= \ket{q',\text{pos}}: \tilde{\psi}(p')=\sqrt{\frac{1}{2\pi}} \exp[-i q'p'],\nonumber\\
u(q) &= \frac{1}{2{\pi}^{3/4}}\frac{1}{\sqrt{2\pi}} \exp[-\frac{i}{2}q\,p_0]
\int_{-\infty} ^{\infty}dp'\exp[\frac{1}{2}{(p'-p_0)}^2+i p'(q-q')].
\end{align}
However, this is a quite badly behaved distribution, hence better avoided.
\subsection{Position eigenstates}
The position operator $\hat{q}$ shifts the momentum leaving position unchanged:
\begin{align}
  e^{i p' \hat{q}} = D(0,p'),\nonumber\\
D(0,p')\ket{q_{0},p} &= e^{\frac{i}{2}p'q_0}\ket{q_0,p'+p}.
\end{align}
 Now, the phase plane orbits are, for each fixed $q_0$, a vertical straight 
line parallel to the $p$ axis. (Fig. \ref{fig:orbitq})\,\,
Each such orbit is again a characteristic set, so in place of Eqns. (\ref{expansion}), (\ref{overlap}), we have:
\begin{align}\label{Expansion_momentum}
\ket{\psi} &= \int_{-\infty} ^{\infty}dp\,\,v(p)\ket{q_0,p},\nonumber\\
 \psi(q') &= \frac{1}{{\pi}^{\frac{1}{4}}}\exp[-\frac{1}{2}({q_0-q'})^2]\int_{-\infty}^{\infty} dp\,\,v(p) 
 \exp[i p(q'-\frac{q_0}{2})],\nonumber\\
\tilde{\psi}(p') &= \frac{1}{{\pi}^{\frac{1}{4}}}\int_{-\infty}^{\infty} dp\,\, 
v(p)\exp[-\frac{1}{2}({p-p'})^2-i q_{0}(p'-\frac{p}{2})].
\end{align}
This time the first relation is easier to handle and gives:
\begin{equation}
\label{3.12}
 v(p) = \frac{1}{2{\pi}^{3/4}}e^{\frac{i}{2}\,\, q_0 p}\int_{-\infty}^{\infty}dq'\,\,\psi(q')
 \exp[\frac{1}{2}(q'-q_0)^2 -i q'p].
 \end{equation}
We draw conclusions similar in spirit to those from Eqn.~\eqref{Fourier_momentum}:
For a general square integrable $\psi(q')$, $v(p)$ is a distribution.
Given any fixed $q_{0}$, as $v(p)$ is unambiguously given by $\psi(q')$, the one dimensional family of 
coherent states $\{ \ket{q_0,p}, -\infty<p<\infty\}$ is linearly independent and the 
expansion in Eqn.~(\ref{Expansion_momentum}) is unique.

For a general position eigenstate, we find :
\begin{align}\label{position_eigenstate}
\ket{\psi} &= \ket{q_{0}^{'},\text{pos}},\,\,{\psi}(q') =\delta(q'-q'_{0});\nonumber\\
v(p) &= \frac{1}{2\pi^{3/4}}\exp[\frac{1}{2}{(q_0-q'_0)}^2 - i p (q'_{0}-\frac{q_0}{2})],\nonumber\\
\ket{q'_{0},\text{pos}}&= \frac{1}{2\pi^{3/4}}\exp[\frac{1}{2}{(q_0-q'_0)}^2]\int_{-\infty}^{\infty}dp\,\, 
\exp[-i p (q'_{0}-\frac{q_0}{2})]\,\,\ket{q_0,p}.
\end{align}
If $q_{0}\neq q'_{0}$ this is not an in phase superposition. It becomes one (in fact globally so) if we choose $q_{0}=q'_{0}$, 
and then we get the preferred expansion
\begin{equation}\label{In_phase_momentum}
 \ket{q'_{0},\text{pos}}=\frac{1}{{2\pi}^{\frac{3}{4}}}\int_{-\infty}^{\infty}dp\,\,e^{-i \frac{q'_{0}p}{2}}\,\,\ket{q'_{0},p}.
\end{equation}
As in the case of momentum eigenstates, we see that the phase $\exp[-i p (q'_{0}-\frac{q_0}{2})]$ in Eqn.~\eqref{position_eigenstate}
can be split up into
\[\exp[-i p (q'_{0}-{q_0})]\exp[-ipq_0/2]\]
But for the factor of $\exp[-i p (q'_{0}-{q_0})]$, we would have obtained an in-phase superposition along $q_0$. This phase gradient 
along $q_0$ drives the superposition from $q_0$ to $q'_0$. Also, the inefficient superposition of $\ket{q'_0,\text{pos}}$
along $q_0$ necessitates the exponentially large compensatory factor $\exp\left[\frac{1}{2}(q_0-q'_0)^2\right]$, to make up
for the exponentially small overlap of coherent states at $q_0$ on $q_0'$.

Again, the phase in Eqn.~\eqref{In_phase_momentum} for global in-phase superposition along the straight line $q=q_0$ can be determined
by using Eqn.~\eqref{eqn:inphaseBargmanncoherent} 
\begin{align*}
	\phi_P(\ket{q_0,0},\ket{q_0,p})&=A(0,0;q_0,0;q_0,p)=-\frac{q_0p}{2}
\end{align*}
where the negative sign accounts for the fact that the vertices are traversed clockwise.
\subsection{Fock states}\label{Hamiltonian}
The operator 
$\hat{n}=\hat{a}^\dagger\hat{a}$ generates rotations in phase space 
\begin{align}\label{Generator of rotation}
 e^{i\theta \hat{n}}\ket{z}&=\ket{ze^{i\theta}}\nonumber\\
  e^{i\theta \hat{n}}\ket{q,p} &= \ket{q\cos\theta-p\sin\theta,p\cos\theta+q\sin\theta}.
\end{align}
The associated orbits are circles (Fig. \ref{fig:orbitn}) centred on $q=p=0$. In essence, this describes what is ``coherent'' about coherent states-- 
under time evolution by the Hamiltonian 
\begin{align}
 H&=\hat{a}^\dagger\hat{a}+\frac{1}{2}\nonumber\\
 e^{-i\hat{H}t}\ket{z}&= e^{-it/2}\ket{ze^{-i\theta}}
 \label{eqn:HamiltonianSHO}
\end{align}
the coherent state remains an eigenstate of $\hat{a}$ while being rotated around the classical constant energy circle $\frac{1}{2}({q}^2+{p}^2)$. 
This example is very different from the two previous ones since the orbit here is a closed curve.

Each such orbit is a characteristic set, so in principle any $\ket{\psi}$ can be expanded as an integral over a circle of 
coherent states for any fixed $r>0$. However, such a subset is not a linearly independent set
(in contrast to what we found in the cases of $\hat{p}$ and $\hat{q}$), as
\begin{equation}\label{linear_independence_circle}
 \int_{0}^{2\pi} d\theta~ e^{i m \theta}\ket{r\cos\theta,r\sin\theta}=0,\,\,m\,=\, 1,2,3\cdots
\end{equation}
These states are overcomplete. Therefore in expanding $\ket{\psi}$ as an integral over them we can limit the 
`expansion coefficients' suitably :
\begin{align}\label{eqn:limitcoefffock}
 \ket{\psi}= \int_{0}^{2\pi}d\theta\, v(\theta)\ket{r\cos\theta,r\sin\theta},\nonumber\\
v(\theta)= \sum_{m=0}^{\infty}v_{m}e^{-i m \theta}.
\end{align}
(Note that in Eqn.
\eqref{eqn:limitcoefffock} negative values of $m$ do not appear in the expansion for 
$v(\theta)$.)

Then $v(\theta),v_m$ are  unique for given $\ket{\psi}$:
\begin{align}
 \ket{\psi}&= \int_{0}^{2\pi}d\theta\sum_{m=0}^{\infty}v_{m}e^{-i m\theta}\sum_{n=0}^{\infty}\frac{e^{-\frac{r^2}{4}}}
 {\sqrt{n!}}{\left(\frac{r}{\sqrt{2}}\right)}^n\, e^{i n\theta}\ket{n}\nonumber\\
&=2\pi e^{-\frac{r^2}{4}}\sum_{n=0}^{\infty}\frac{v_n}{\sqrt{n!}}{\left(\frac{r}{\sqrt{2}}\right)}^n\ket{n};\nonumber\\
v_n &=\frac{{e^{\frac{r^2}{4}}}}{2\pi}\sqrt{n!}{\left(\frac{\sqrt{2}}{r}\right)}^n\, \psi_{n},\quad\psi_{n}=\braket{n}{\psi}.
\end{align}

The result is that while $\{\psi_{n}\}$ is always an $l^2$ sequence for normalisable $\ket{\psi}$, in general $\{v_n\}$ is not so. 
This makes $v(\theta)$ in general non-$L^{2}$, but a distribution. Formally, we have
\begin{equation}\label{expansion_co-efficient}
v(\theta)=\frac{1}{2\pi} e^{\frac{r^2}{4}}\sum_{n=0}^{\infty}\sqrt{n!}{\left(\frac{\sqrt{2}}{r}\right)}^n e^{-i n\theta}\psi_{n},
\end{equation}
and in all of  Eqns.~(\ref{linear_independence_circle}) to (\ref{expansion_co-efficient}), $r$ can be chosen and kept 
fixed at any value.

Now, we choose $\ket{\psi}$ to be a Fock state $\ket{n}$, an eigenstate of $\hat{n}$ with eigenvalue $n$. 
Then $v(\theta)$ is just a single term in the sum in Eqn.~(\ref{expansion_co-efficient})
and we have the integral representation (valid for any $r\,\,>$\,  0). This is well-known, see, for example \cite{gardiner1983handbook}.

\begin{equation}
\ket{n} = \frac{e^{\frac{r^2}{4}}}{2\pi}\sqrt{n!}{\left(\frac{\sqrt{2}}{r}\right)}^n\, \int_{0}^{2\pi}d\theta 
e^{-i n\theta}\ket{r\cos\theta,r\sin\theta}.
\label{eqn:Fockarbitrarysuperposition}
\end{equation}
In general this is not an in-phase superposition since
\begin{equation}
 \phi_{P}\left(e^{-i n\theta}\ket{r\cos\theta,r\sin\theta},e^{-i n(\theta+\delta\theta)}
 \ket{r\cos(\theta+\delta\theta),r\sin(\theta+\delta\theta)}\right) =\left(\frac{r^2}{2}-n\right)\delta\theta.
\end{equation}
We get a locally in phase superposition if given $n$, we choose $r^2=2n$; then we obtain the preferred integral representation
\begin{equation}
 \ket{n}=\frac{e^{n/2}}{2\pi}\sqrt{n!}n^{-n/2}\int_{0}^{2\pi}d\theta e^{-i n\theta}\ket{\sqrt{2n}\cos\theta,
 \sqrt{2n}\sin\theta}
\label{Fock_in_phase}
\end{equation}
for Fock states. Since the integration is over a circle (not a straight line as in Eqns. (\ref{In_phase}), 
(\ref{In_phase_momentum}), any three distinct points on it enclose a non-degenerate triangle with non-zero area. 
Thus, in contrast with the previous two examples, we have here only a locally in-phase superposition, not a global
one.

Also note that in Eqn.~\eqref{eqn:Fockarbitrarysuperposition}, we could have split the phase factor as in the momentum and 
position eigenstates as 
\[e^{-in\theta}=e^{-i\frac{r^2\theta}{2}}e^{i(\frac{r^2}{2}-n)\theta}\]
But for the phase gradient of $\left(\frac{r^2}{2}-n\right)\theta$ along the circle, the superposition would have been in phase and 
stayed put at $r$. The phase gradient drives the superposition away from the circle at radius  $r$ to radius $\sqrt{2n}$. 


We point out that the preferred in-phase superpositions for the Fock states satisfy the condition
\begin{align}
 \oint pdq=2\pi n
\label{eqn:BohrSommerfeldInPhase}
\end{align}
This would be the Bohr-Sommerfeld quantization condition of ``old'' quantum theory! (the area is $2\pi n$ instead of $2\pi (n+\frac{1}{2})$.)
Further, this is not a semiclassical result, it is exact.

We now return to the discussion around Eqn. \eqref{eqn:qfunction}, where we had made a case for
curves of in-phase superpositions being the maxima of $Q$ functions.
We now explicitly demonstrate  
that this is indeed the case for the three states considered.

The $Q$ distribution of the position, momentum and Fock states
$\ket{q_0,\text{pos}}$, $\ket{p_0,\text{mom}}$ and  $\ket{n}$ are
\begin{align}
	Q_{\ket{q_0,\text{pos}}}(q,p)&=\frac{1}{2\pi^{3/2}}\exp\left[-(q-q_0)^2\right]\nonumber\\	
	Q_{\ket{p_0,\text{mom}}}(q,p)&=\frac{1}{2\pi^{3/2}}\exp\left[-(p-p_0)^2\right]\nonumber\\	
	Q_{\ket{n}}(q,p)&=\frac{1}{2\pi n!}\left[\frac{1}{2}(p^2+q^2)\right]^n
	\exp\left[-\frac{1}{2}(p^2+q^2)\right]
\label{eqn:QdistributionH4}
\end{align}
which, as can easily be verified, are maximized along the preferred curves corresponding to in-phase superpositions, i.e. 
$q=q_0$, $p=p_0$ and $p^2+q^2=2n$ respectively. 

\begin{figure}
\centering
\subfigure[b][\scriptsize{(colour online) In-phase superposition along a circle of radius
	$\sqrt{2n}=\sqrt{10}$. Maximum of $Q$-function (black circles) located on the line of superposition. (white
circles).
}]{ \includegraphics[width=0.48\textwidth]{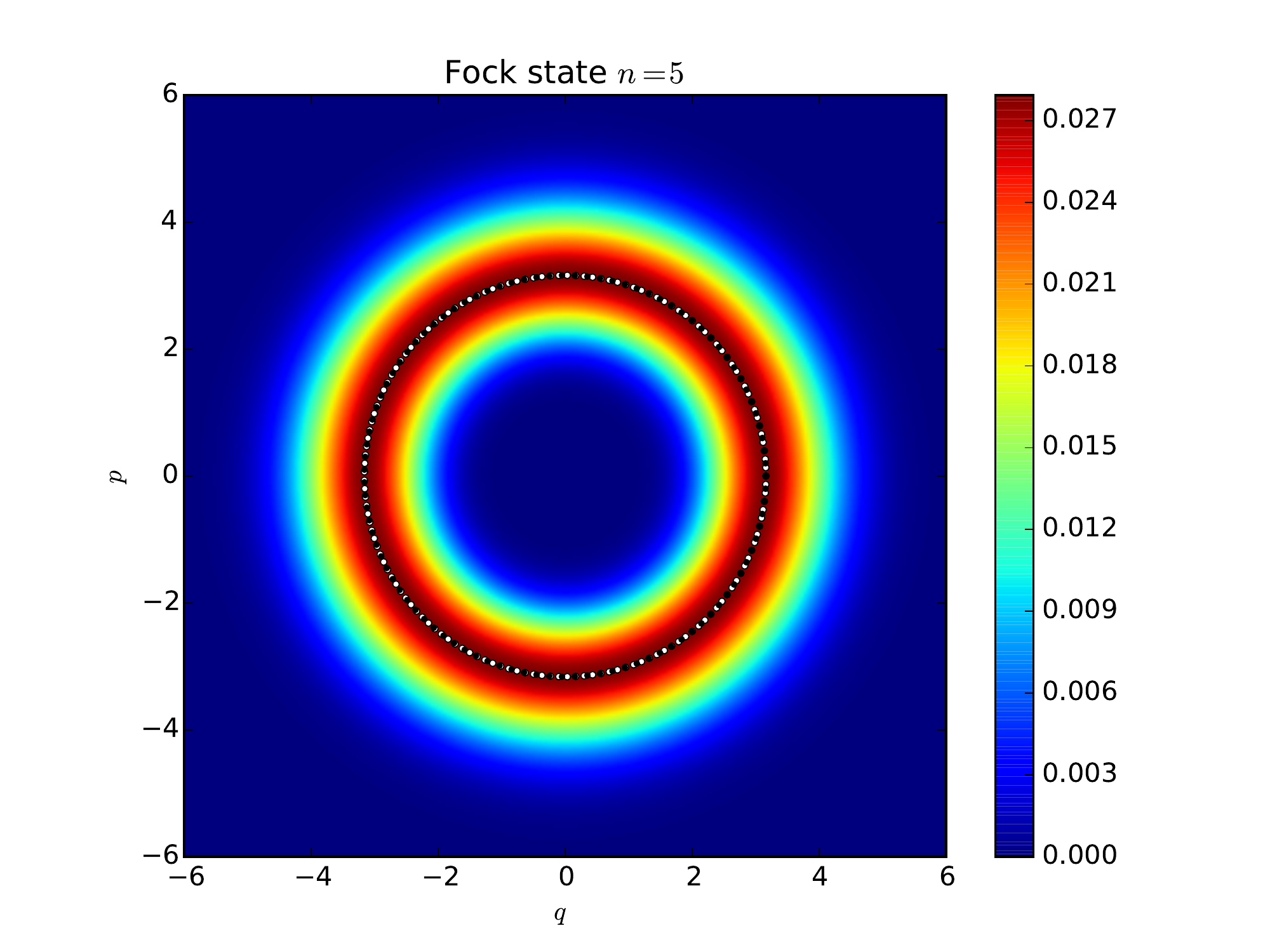}\label{fig:FockinPhase}}\quad
\subfigure[b][\scriptsize{(colour online) Superposition along a circle of radius
$5$ (white circles). Maximum of Q-function is at $\sqrt{10}$, denoted by black
circles
}]{ \includegraphics[width=0.48\textwidth]{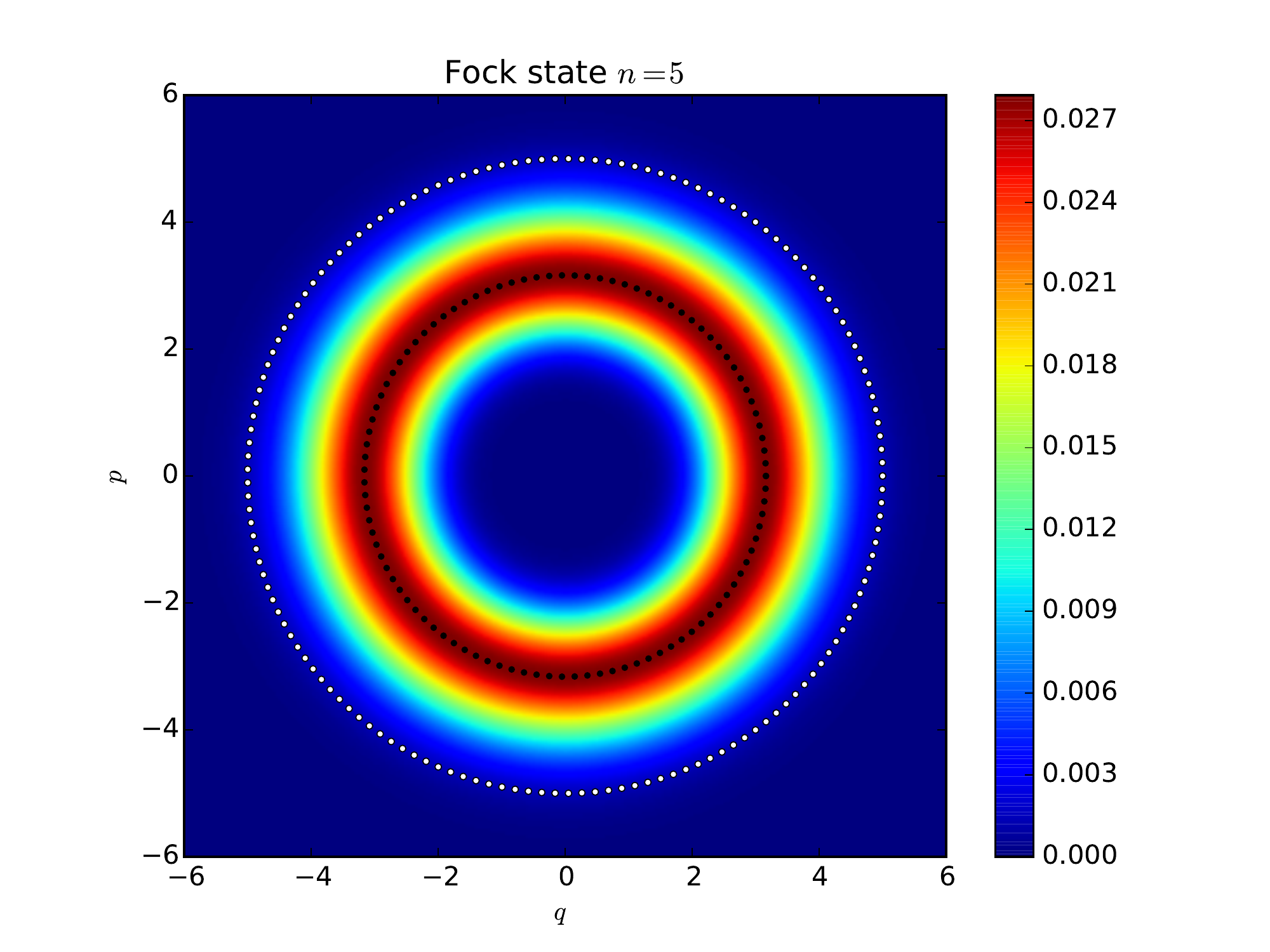}\label{fig:Fockoutofphase}}
\end{figure}

For the particular case of of the Fock state $n=5$ ,
we discretize the integral representation in Eqn.~\eqref{eqn:Fockarbitrarysuperposition} as 
\begin{equation*}
\ket{n} =
\frac{e^{\frac{r^2}{4}}}{2\pi}\sqrt{n!}{\left(\frac{\sqrt{2}}{r}\right)}^n\,
\sum\limits_{j=0}^{N} 
e^{-i \frac{2\pi nj}{N}}\ket{\,r\cos\frac{2\pi j}{N},r\sin\frac{2\pi
j}{N}\,}
\end{equation*}
and take $N=500$. In Figures~~\ref{fig:FockinPhase},~\ref{fig:Fockoutofphase} we display the 500 states appearing in the sum on 
the RHS for two values of $r$ together with the $Q$ function for the Fock state $n=5$. 

In Figure~\ref{fig:FockinPhase}, the states are put on a circle of radius $\sqrt{2n}=\sqrt{10}$ (denoted by white dots) corresponding to their 
in-phase superposition. Note that in this case, the $Q$ function has a maximum (black circles) along the in-phase superposition. In
contrast, as can be seen from Figure \ref{fig:Fockoutofphase}, this ceases to be so in the case when the states  are put on a 
circle of radius $r=5$. 

We conclude this section with a few clarifying comments. 
The overcompleteness of coherent states leads to enormous freedom 
in expanding a general vector $\ket{\psi}$ as an integral or sum over these states.
With each of the generators $\hat{p},\hat{q},\hat{n}$, we are led in the first instance to 
consider one dimensional integral expansions for a general $\ket{\psi}$ over one of their associated orbits. 
For momentum eigenstates  $\ket{p_0,\text{mom}}$, in looking for an in-phase superposition we are led to a 
particular $\hat{p}$ orbit determined by $p_0$; similarly for the states $\ket{q_0,\text{pos}}$ we are led to an 
in-phase integral 
over a corresponding $\hat{q}$ orbit determined by $q_0$; and lastly for Fock states $\ket{n}$ and corresponding 
$\hat{n}$ orbits, 
namely circles centred at the origin. The preferred eigenstate expansions in each case pick out a particular orbit of the 
concerned generator.

More generally, any locally in-phase sum or integral expansion, for any vector $\ket{\psi}$, of the form
\begin{align}
 \ket{\psi} &= \ket{\psi_1}+\ket{\psi_2}+\cdots \,\, +\ket{\psi_{j}}+\cdots, \quad \text{arg}\braket{\psi_j}{\psi_{j+1}}=0,\nonumber\\
\ket{\psi} &=\int ds \ket{\psi(s)},\quad\text{arg}\braket{\psi(s)}{\psi(s + \delta s)}=0,
\end{align}
retains these properties under any unitary transformation $U$, as inner products are preserved. As an example, if we start with 
the in-phase expansion in Eqn.~(\ref{In_phase_momentum}) for a position eigenstate
\begin{equation}
\label{Position_Eigenstate}
 \ket{q,\text{pos}} = \frac{1}{2\pi^{\frac{3}{4}}}\int_{-\infty}^{\infty}dp\,\,e^{-i \frac{qp}{2}}\ket{q,p},
\end{equation}
and apply the harmonic oscillator time evolution operator, we get using $\hat{H}= \hat{a}^\dagger\hat{a}+\frac{1}{2}$ 
along with Eqn. (\ref{Generator of rotation}):
\begin{equation}\label{SHO kernel action}
e^{-i \hat{H}t} \ket{q,\text{pos}} = \frac{1}{2\pi^{\frac{3}{4}}}\int_{-\infty}^{\infty}dp\,e^{-i \frac{qp}{2}-
i \frac{t}{2}}\ket{q\cos t+p\sin t,p\cos t-q\sin t},
\end{equation}
and it is guaranteed that the right hand side is an in-phase superposition of a line of coherent states obtained from the 
line in Eqn. (\ref{Position_Eigenstate}) by a clockwise rotation of amount $t$. Similarly, starting from the in-phase integral 
representation Eqn. (\ref{Fock_in_phase}) for Fock state $\ket{n}$ and applying the unitary displacement operator $D(q,p)$ 
(using Eqn (\ref{Displacement})\,\,), we get the guaranteed-to-be in-phase superposition:
\begin{equation}\label{displacement_fock}
 D(q,p)\ket{n}= \frac{e^{n/2}}{2\pi}\sqrt{n!}n^{-n/2}\int_{0}^{2\pi}d\theta\,e^{-i n\theta
 +i\sqrt{\frac{n}{2}}(p\cos\theta-q\sin\theta)}\ket{q+\sqrt{2n}\cos\theta,p+\sqrt{2n}\sin\theta}.
\end{equation}
So the integration is over a circle of radius $\sqrt{2n}$ with center $(q,p)$. We will exploit Eqns. (\ref{SHO kernel action}) 
and (\ref{displacement_fock}) in the next section.
\section{SOME APPLICATIONS OF THE FORMALISM}\label{sec:applications}
In this section, we present some sample calculations of overlaps of
chosen state vectors, and some nontrivial matrix elements of
interesting unitary operators of relevance to quantum optics, to convey the important features and
advantages of the coherent state and geometric phase based in-phase
superpositions. We illustrate how the one-dimensional integral in-phase representations of the position, 
momentum and number eigenstates
derived in the previous section can be used to understand 
oscillations in state overlaps and make contact with previous work on interference in phase space. 
\subsection{{Sample state vector overlaps and matrix elements}}\label{sec:matrixelements}
In this subsection, we consider four examples of increasing complexity, to see how the ideas
laid out in a general context in the preceding sections apply to specific state overlaps and matrix elements.

We have seen that the maximum of the $Q$ distribution lies along in-phase superpositions
of coherent states.
Guided by the intuition that state overlap is dominated
by regions where the maximum of the corresponding $Q$ functions intersect,
in each of the cases considered below, we calculate the critical point of the integrand. We see that it matches the \emph{geometric
intersection} of the in-phase integral representations, as we would expect without any calculation. 
We determine the overlap by expanding the integral about the critical point.

The first two examples, overlap of position-momentum eigenstates and the propagator for the harmonic oscillator, involve a single critical point and 
Gaussian integrals, and we recover well known exact results as expected.

The subsequent examples: Schr\"odinger wavefunction of a Fock state, and matrix elements of displacement operator in the Fock basis involve
two critical points.
Here, we get interference between the two contributions, resulting in oscillations.
We connect the oscillations to areas by obtaining asymptotic results from the exact integral representations in the semiclassical limit
of large quantum numbers. We demonstrate explicitly that the leading order contributions
to the oscillatory part of the overlap can be determined from geometry and the in-phase
superpositions alone, without actually evaluating  any integrals.

\begin{enumerate}[label=(\alph*),leftmargin=*]
\item We start with position-momentum overlap. It serves to illustrate some of
the basic ideas outlined above in a simple setting.

Use of the integral representations Eqns.~\eqref{In_phase}, \eqref{In_phase_momentum} followed by Eqn.~\eqref{eqn:coherentinnerproduct} leads to
\begin{align}
\langle q,\text{pos}|p,\text{mom}\rangle =&{1\over4\pi^{3/2}} \int_{-\infty}^\infty
\int_{-\infty}^\infty dq^\prime dp^\prime e^{{i\over2}(q^\prime p+p^\prime q)}
\langle q,p^\prime|q^\prime,p\rangle\nonumber\\
=&{1\over4\pi^{3/2}} e^{-{1\over4}(q^2+p^2)+{i\over2}qp} \int_{-\infty}^\infty
\int_{-\infty}^\infty dq^\prime dp^\prime e^{-\chi(q^\prime,p^\prime; q,p)}\,,\nonumber\\
\chi(q^\prime,p^\prime; q,p)=&{1\over4}(q^{\prime2}+p^{\prime2}) -
{1\over2}(q^\prime q+p^\prime p) + {i\over2}(q^\prime p^\prime - q^\prime p - p^\prime q)\,.
\label{eqn:posmomoverlap1}
\end{align}

 The
conditions
${\partial\chi\over\partial
q^\prime} =
{\partial\chi\over\partial p^\prime}
= 0$ solve to give one critical
point: $q^\prime=q$, $p^\prime=p$.
This is precisely the point at which the phase plane orbits involved
in the in-phase integral representations for $|p,\text{mom}\rangle$
and $|q,\text{pos}\rangle$ intersect, as was to be expected. 

Expanding about the point of intersection, we can rewrite the integral as
\begin{align*}
\langle q,\text{pos}|p,\text{mom}\rangle =& \frac{e^{iqp}
}{4\pi^{3/2}}\int_{-\infty}^\infty
\int_{-\infty}^\infty dq^\prime dp^\prime \exp(-\frac{1}{4}\left[(q^\prime-q)^2+(p^\prime-p)^2-2i(q^\prime-q)(p-p^\prime)\right])
\end{align*}
Shifting the origin to the critical point, 
\begin{align}
\langle q,\text{pos}|p,\text{mom}\rangle =& \frac{e^{iqp}
}{4\pi^{3/2}}\int_{-\infty}^\infty
\int_{-\infty}^\infty dq^\prime dp^\prime e^{-\frac{1}{4}\left[{q^\prime}^2+{p^\prime}^2+2iq^\prime p^\prime\right]}\nonumber\\
&=\frac{e^{iqp}}{4\pi^{3/2}}\int_{-\infty}^\infty
\int_{-\infty}^\infty dq^\prime dp^\prime \exp\left[-\frac{1}{2}
	\begin{pmatrix}q^\prime&p^\prime\end{pmatrix}\begin{pmatrix}\frac{1}{2}&\frac{i}{2}\\\frac{i}{2}&\frac{1}{2}\end{pmatrix}
	\begin{pmatrix}q^\prime\\p^\prime\end{pmatrix}
\right]\nonumber\\
&=\frac{e^{iqp}}{4\pi^{3/2}}\times\frac{2\pi}{\sqrt{1/2}}\qquad \det\begin{pmatrix}\frac{1}{2}&\frac{i}{2}\\\frac{i}{2}&\frac{1}{2}\end{pmatrix}=1/2
\nonumber\\
&=\frac{e^{iqp}}{\sqrt{2\pi}}
\label{eqn:posmomoverlap}
\end{align}
In this case, we get the exact result as we are dealing with a Gaussian integral. Also, note that the phase of the overlap is determined 
entirely by the critical point.
\item Next, we consider the propagator of the
simple harmonic oscillator  in the position basis.
\begin{figure}[b]
 \includegraphics[width=0.3\textwidth]{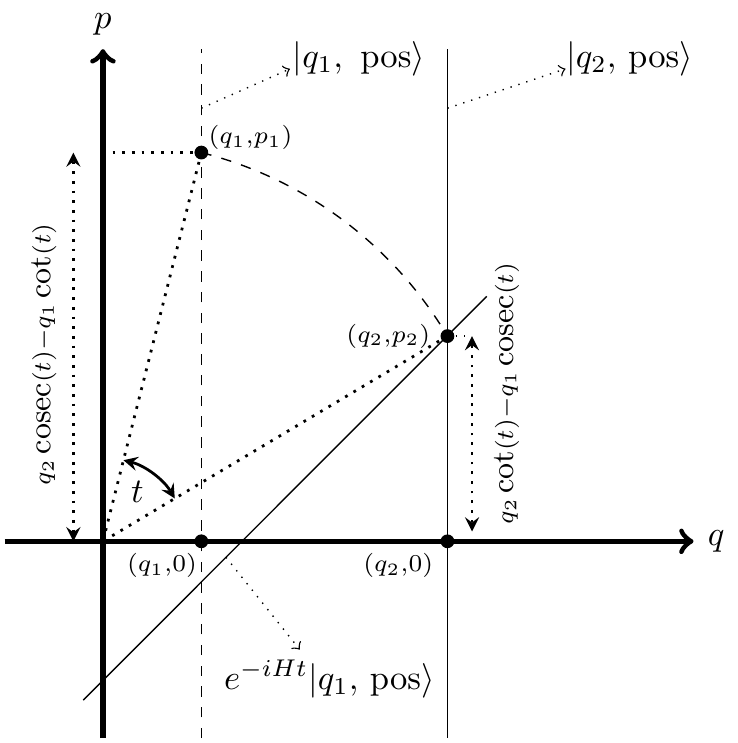}
 \caption{Geometrical interpretation of the inner product $\langle q_2,\,\mbox{pos}\vert e^{-iHt}\vert q_1,\,\mbox{pos}\rangle$. 
 The initial state $\vert q_1\rangle$ is given by an in phase superposition along the vertical line $q=q_1$ by Eqn. 
 \eqref{In_phase_momentum}. $e^{-iHt}\vert q_1,\,\mbox{pos}\rangle$ is determined by 
 the rotated line. The stationary point of the exponent in the integrand, Eqn.~\eqref{eqn:intersectiontilted} has a 
 geometrical interpretation as the 
 point of intersection of the in-phase superposition corresponding to $\vert q_2,\,\mbox{pos}\rangle$ and the rotated 
 line corresponding to $e^{-iHt}\vert q_1,\,\mbox{pos}\rangle$.
}
 \label{fig:rotatedintersection}
\end{figure}
The effect of oscillator time evolution on a position
eigenvector is given in Eqn.~\eqref{SHO kernel action}. Combining this with Eqn.~\eqref{In_phase_momentum} and
\eqref{eqn:coherentinnerproduct}, we have the following configuration space kernel
\begin{align}
\langle q_2,\text{pos}|e^{-i\hat{H}t}|q_1,\text{pos}\rangle = &
{e^{-it/2}\over4\pi^{3/2}} \int_{-\infty}^\infty dp_1
\int_{-\infty}^\infty dp_2 ~e^{{i\over2}(q_2 p_2-q_1p_1)}
\langle q_2,p_2|q_1C+p_1S, p_1C-q_1S\rangle\nonumber\\
= & {e^{-it/2}\over4\pi^{3/2}}
\int_{-\infty}^\infty dp_1 \int_{-\infty}^\infty dp_2
\,\,e^{-Z}\,,\nonumber\\
Z =  {1\over4}(p_1^2+p_2^{2})-{1\over2} p_1p_2 e^{-it} &+
{i\over2}\left[(q_1-q_2
e^{-it})p_1-(q_2-q_1e^{-it})p_2\right]+\frac{1}{4}(q_1^2+q_2^2)-\frac{1}{2}q_1q_2e^{-it}\,,\nonumber\\
C = & \cos t~,~~S=\sin t\,.
\label{eqn:expandSHOKernel}
\end{align}
Alternatively, remembering how $\hat{n}$ rotates the coherent state $\ket{q,p}$, Eqn.~\eqref{Generator of rotation}
\[e^{i\theta \hat{n}}\ket{q,p} = \ket{q\cos\theta-p\sin\theta,p\cos\theta+q\sin\theta}.\]
This leads to Fig.~\ref{fig:rotatedintersection}.
Some geometry shows that the point of intersection has the co-ordinates
\begin{align}
 p_1^0=q_2\mbox{ cosec}(t)-q_1\mbox{cot}(t);~~~ 
 p_2^0=q_2\mbox{cot}(t)-q_1\mbox{ cosec}(t)
 \label{eqn:intersectiontilted}
\end{align}
where $p_1^0$ is measured along the unrotated $\ket{q_1,\text{ pos}}$ superposition. (Fig.~\ref{fig:rotatedintersection}) 

Indeed, we see that the stationary point of $Z(p_1,p_2)$ which would contribute
maximally to the inner product leads to the same critical point, thus validating the geometrical interpretation.
\begin{align}
	\frac{\partial Z}{\partial p_1}\bigg\vert_{(p_1^0,p_2^0)}=\frac{p_1^0}{2}-\frac{p_2^0}{2}e^{-it}+\frac{i}{2}(q_1-e^{-it}q_2)=0;
 ~~\frac{\partial Z}{\partial p_2}=\left(\frac{\partial Z}{\partial p_1}\right)^{\ast}=0
 \label{eqn:statHamilevol}
\end{align}
Solving the real and imaginary parts of Eqn.~\eqref{eqn:statHamilevol},
we again get back the stationary point with coordinates given by Eqn.~\eqref{eqn:intersectiontilted}. 

Expanding $Z$ about the critical point we get
\begin{align}
	Z(p_1,p_2)-Z(p_1^0,p_2^0)&=\frac{1}{4}\left[(p_1-p_1^0)^2+(p_2-p_2^0)^2-2\,(p_1-p_1^0)(p_2-p_2^0)e^{-it}\right]\nonumber\\
	Z(p_1^0,p_2^0)&=-i\left[\frac{\left(q_1^2+q_2^2\right)\cos t-2q_1q_2}{2\sin t}\right] 
\end{align}
Thus, we get
\begin{align*}
 \langle q_2,\text{pos}|e^{-i\hat{H}t}|q_1,\text{pos}\rangle = &
 {e^{-it/2}\over4\pi^{3/2}}\exp\left[-Z(p_1^0,p_2^0)\right]\int_{-\infty}^{\infty}dp_1\int_{-\infty}^{\infty}dp_2\nonumber\\
 &\exp\left(-\frac{1}{4}\left[(p_1-p_1^0)^2+(p_2-p_2^0)^2-2\,(p_1-p_1^0)(p_2-p_2^0)e^{-it}\right]\right)\nonumber\\
\end{align*}
Shifting the origin to the critical point,
\begin{align}
	\langle q_2,\text{pos}|e^{-i\hat{H}t}|q_1,\text{pos}\rangle &=\exp\left[i\frac{(q_1^2+q_2^{2}) \cos t
 - 2q_1q_2}{2 \sin t}\right]{e^{-it/2}\over4\pi^{3/2}}\times\nonumber\\
 \int_{-\infty}^{\infty}dp_1\int_{-\infty}^{\infty}dp_2&\,\,\exp\left(-\frac{1}{4}\left[p_1^2+p_2^2-2p_1p_2e^{-it}\right]\right)\nonumber\\
&=(2\pi i \sin
 t)^{-1/2}\exp\left[i\frac{(q_1^2+q_2^{2}) \cos t
 - 2q_1q_2}{2 \sin t}\right] 						      
\label{eqn:propagatorSHO}
\end{align}

The special advantages of using  in-phase integral representations
for position eigenvectors when expanded in terms of coherent states
should be evident.
\item In this third case, we encounter the first example where we have more than one critical point, and we get interference between their contributions.
For the Schr\"odinger wave function of a Fock state, use of Eqns.~\eqref{In_phase_momentum}, \eqref{Fock_in_phase}  followed by
\eqref{eqn:coherentinnerproduct} gives
\begin{align}
\langle q_0,\text{pos}|n\rangle = & {\mathcal{N}_n\over4\pi^{7/4}}
\int_{-\infty}^\infty dp_1~ e^{{i\over2}q_0p_1} \int_0^{2\pi} d\theta~ e^{-in\theta}
\langle q_0,p_1|\sqrt{2n}\,C,\,\sqrt{2n}\,S\rangle\nonumber\\
=&{\mathcal{N}_n\over4\pi^{7/4}} \int_0^{2\pi} d\theta~ e^{-in\theta-
{1\over4}(q_0-\sqrt{2n}C)^2+{i\over2}q_0\sqrt{2n}S}\nonumber\\
&\times \int_{-\infty}^\infty dp_1~ e^{-{1\over4}(p_1-\sqrt{2n}S)^2 +
{i\over2}(q_0-\sqrt{2n}C)p_1}\,,\nonumber\\
C = & \cos\theta\,, S=\sin\theta\,,\,\mathcal{N}_n=e^{n/2}n^{-n/2}\sqrt{n!}
\end{align}
Shifting $p_1$, carrying out the Gaussian integral, remembering $n\ge
0$ and simplification by scaling leads to
\begin{subequations}
\begin{align}
\langle q_0,\text{pos}|n\rangle = & {\mathcal{N}_n\over2\pi^{5/4}}
\int_0^{2\pi} d\theta~ e^{-in\theta}
e^{-{1\over2}(q_0-\sqrt{2n}C)^2+iq_0\sqrt{2n}S-inCS}\nonumber\\
 = & {\mathcal{N}_n\over2\pi^{5/4}}e^{-{n\over2}-{q_0^2\over2}}
\int_0^{2\pi} d\theta ~e^{-in\theta}
e^{q_0\sqrt{2n} e^{i\theta}-{n\over2}e^{2i\theta}}\label{eqn:intermediateforhermite}\\
=& {\mathcal{N}_n\over\pi^{1/4}}e^{-{n\over2}-{q_0^2\over2}}\left[\frac{1}{2\pi i}\oint d\xi \frac{e^{q_0\sqrt{2n}\xi-\frac{n}{2}\xi^2}}{\xi^{n+1}}\right];\quad \xi=e^{i\theta}\label{eqn:intermediateforhermite2}\\
 = & {\mathcal{N}_n\over\pi^{1/4}}
{e^{{-n\over2}-{q_0^2\over2}}\over n!}\left({\partial^n\over\partial\xi^n}
e^{q_0\sqrt{2n}\xi-{n\over2}\xi^2}\right)_{\xi=0} \nonumber\\
 = & {\mathcal{N}_n\over\pi^{1/4}}
{e^{{-n\over2}-{q_0^2\over2}}\over n!} \left({n\over 2}\right)^{n/2} 
\left({\partial^n\over\partial\xi^n}
e^{2q_0\xi-\xi^2}\right)_{\xi=0}\nonumber\\
= &{\mathcal{N}_n\over\pi^{1/4}}
{e^{{-n\over2}-{q_0^2\over2}}\over n!} \left({n\over 2}\right)^{n/2} 
H_n(q_0)\left(\mbox{using  } \quad e^{2xt-t^2}=\sum_{n=0}^{\infty}H_n(x)\frac{t^n}{n!}\right)\nonumber\\
= & {1\over\pi^{1/4}2^{n/2}\sqrt{n!}}~e^{-{q_0^2\over2}} H_n(q_0)
\end{align}
\end{subequations}
where $H_n(q_0)$ are the Hermite polynomials. Again this example is
meant only to illustrate the present formalism, yet it appears to be
a shade more substantial than the previous one.

Now, we use the in-phase integral representations of the Fock and the position eigenstates to give a geometric interpretation to 
the inner product $\langle q_0,\text{pos}|n\rangle$ and simultaneously derive an asymptotic formula for the Hermite polynomials valid in the limit 
of large $n$.

We start with Eqn.~\eqref{eqn:intermediateforhermite2}. Analyticity of the integrand allows us to use the steepest descent formula. 
(accessible introductions can be found in \cite{bleistein1986asymptotic,*krzywicki1996mathematics}) We approximate
\begin{align}
	I&=\frac{1}{i}\oint d\xi \frac{e^{q_0\sqrt{2n}\xi-\frac{n}{2}\xi^2}}{\xi^{n+1}}\nonumber\\
	 &=\frac{1}{i}\oint d\xi \frac{e^{nf(\xi)}}{\xi};\qquad f(\xi)=2C\xi-\frac{\xi^2}{2}-\log(\xi);\,q_0=\sqrt{2n}C\nonumber
\end{align}
In the limit of large $n$, the integral is dominated by contributions from the saddle point
\begin{align*}
f^\prime(\xi_0)=0;&\quad\implies
\xi_0=C\pm\sqrt{C^2-1}
\end{align*}
We are interested in the regime where the orbits of $\ket{n}$ and $\ket{q_0,\text{pos}}$ intersect (at the two saddle points), 
and we get oscillatory behaviour due to interference from their contributions.
Hence, we take $\vert C\vert <1$ and parametrize it as $C=\cos\theta_0$.

Thus, the two saddle points are 
\begin{align}
\xi_0&=e^{\pm i\theta_0}
\end{align}
Both these points satisfy $q_0=\sqrt{2n}\cos(\theta_0)$, as expected from the intersection of the orbits in Fig.~~\ref{fig:hermitestat}.
$\pm\theta_0$ correspond to the points $P_1$ and $P_2$, respectively.
\begin{figure}[htbp]
 \includegraphics[width=0.30\textwidth]{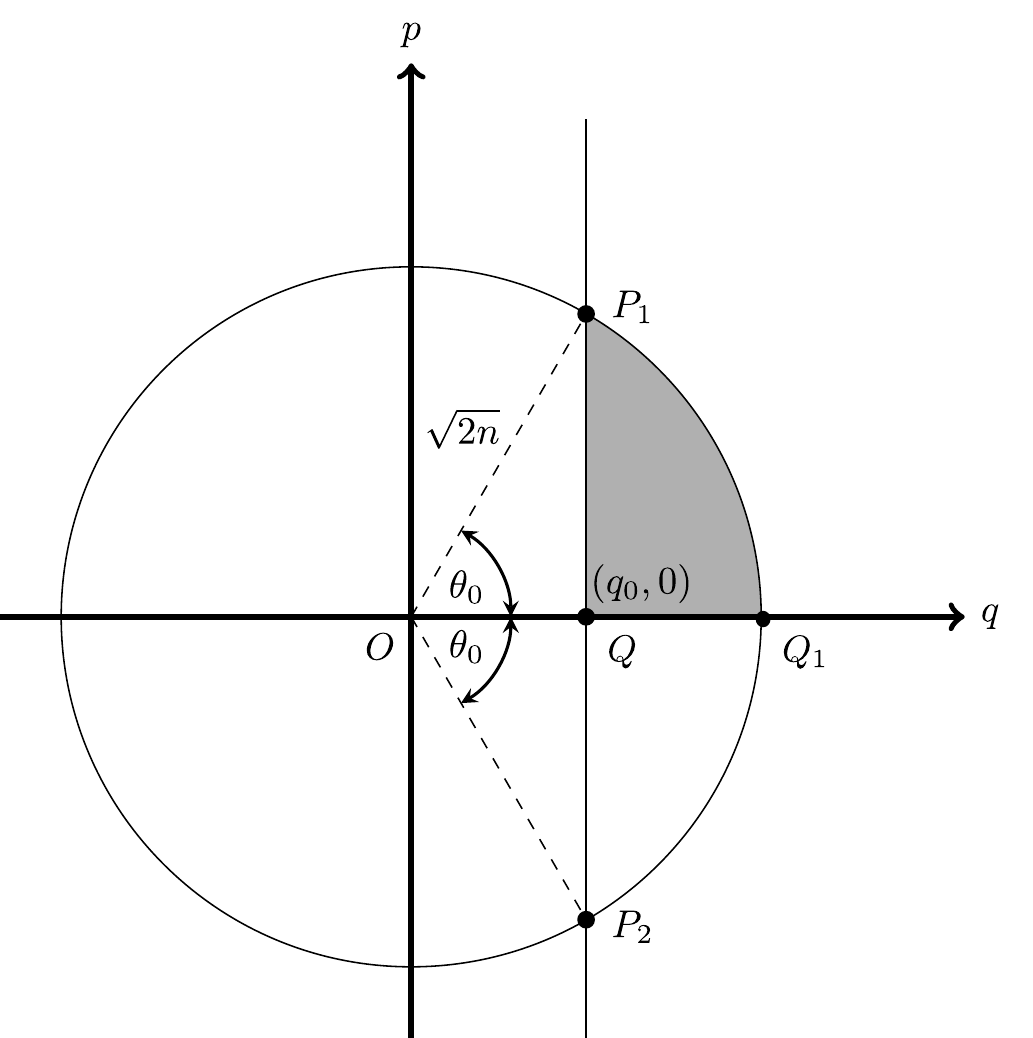}
\caption{The inner product $\langle q_0,\text{pos}|n\rangle$ is determined in the steepest descent approximation by the two points $P_1$ and $P_2$ corresponding to the points of intersection of the in-phase superpositions of the Fock state $\vert n\rangle $ and the position eigenstate 
$\vert q_0,\,\mbox{pos}\rangle$.}
\label{fig:hermitestat}
\end{figure}

In the neighbourhood of the critical points $\pm\theta_0$, $f^{\prime\prime}(e^{\pm i\theta_0})=\sin\theta_0\exp\left(\mp i(\theta_0+\frac{\pi}{2})\right)$.
\begin{align}
	f(\xi)=f(\xi_0)+r^2\sin(\theta_0)e^{i\left[2\phi\mp(\frac{\pi}{2}+\theta)\right]};\quad \text{where }\,
	\xi-\xi_0=re^{i\phi};\quad \xi_0=e^{\pm i\theta_0}
	 \label{eqn:diffhermite}	 
\end{align}
The direction of steepest descent $\phi$ in Eqn.~\eqref{eqn:diffhermite} at the points $P_1$ and $P_2$ compatible with an anticlockwise contour 
turn out to be 
\begin{equation*}
	\phi{(\theta_0)}=\frac{3\pi}{4}+\frac{\theta_0}{2};\qquad\phi{(-\theta_0)}=\frac{\pi}{4}-\frac{\theta_0}{2}
\end{equation*}
These are the directions along which the imaginary part stays constant and the real part decays. This allows us to extract the phase contribution
from the critical point before integrating the Gaussian fluctuations.
Adding up the contributions,
\begin{align}
	I&\approx\frac{e^{i(\frac{3\pi}{4}+\frac{\theta_0}{2})}}{ie^{i\theta_0}}\left[\int^{\infty}_{-\infty}dr e^{-nr^2\sin\theta_0}\right]e^{nf(e^{i\theta_0})}+
\frac{e^{i(\frac{\pi}{4}-\frac{\theta_0}{2})}}{ie^{-i\theta_0}}\left[\int^{\infty}_{-\infty}dr e^{-nr^2\sin\theta_0}\right]e^{nf(e^{-i\theta_0})}\nonumber\\
e^{n f(e^{\pm i\theta_0})}&=\exp\left[n\left(1+\frac{\cos2\theta_0}{2}\right)\pm ni\left(\frac{\sin2\theta_0}{2}-\theta_0\right)\right]\nonumber
\end{align}
After simplification,
\begin{align*}
	I&\approx2\sqrt{\frac{\pi}{n\sin\theta_0}}e^{n(1+\frac{\cos2\theta_0}{2})}\cos\left[n\left(\frac{\sin 2\theta_0}{2}-
	\theta_0\right)+\frac{\pi}{4} -\frac{\theta_0}{2}\right]
\end{align*}
 and hence
 \begin{align}
 \langle q_0,\text{pos}|n\rangle \approx& \frac{n^{-\frac{n}{2}
}\sqrt{n!}}{\pi^{3/4}}\frac{ e^{\frac{n}{2}}}{\sqrt{n\sin \theta_0}}\cos\left[\frac{\pi}{4}-\frac{\theta_0}{2}+
n\left(\frac{\sin 2\theta_0}{2}-\theta_0\right)\right]\nonumber\\
&\approx \left(\frac{2}{\pi^2 n}\right)^{1/4}\frac{\cos\left[\frac{\pi}{4}-\frac{\theta_0}{2}+n\left(\frac{\sin 2\theta_0}{2}-\theta_0\right)\right]}{\sqrt{\sin \theta_0}}\label{eqn:Hermiteasymp2}
\end{align}
where in the last line we have used the Stirling approximation for $n!$.

Now, note that the term proportional to $n$ in the argument of the cosine in Eqn.~\eqref{eqn:Hermiteasymp2}. 
is precisely half the area caught between the curves of in-phase superposition and is represented by the shaded portion
in Fig.~~\ref{fig:hermitestat}. It is
\begin{align*}
	\frac{1}{2}(\sqrt{2n})^2\theta_0-\frac{1}{2}(\sqrt{2n}\cos\theta_0)(\sqrt{2n}\sin\theta_0)=n\left(\theta_0-\frac{1}{2}\sin2\theta_0\right)
\end{align*}
This is anticipated, considering the phase information in terms of areas was built into our integral representation 
through Eqn.~\eqref{eqn:coherentstatehorizontal}. Phase associated with the Fock state at the point $P_1$ is
$-n\theta_0$ = -Area ($P_1OQ_1$) subtended by the arc $P_1Q_1$ at the origin. Similarly, phase at $P_1$ for the position state 
is $-\frac{1}{2}q_0p(q_0)=-\frac{n}{2}\sin(2\theta_0)$. (the area of the triangle $P_1OQ$) 

So, the phase contribution to the inner product $\langle q_0,\text{pos}|n\rangle$
is 
\[-\text{Area } (P_1OQ_1)+ \text{Area }(P_1OQ)=-\text{ Area }(P_1QQ_1) \]
which is precisely the negative of the shaded area. 

The extra term of $\frac{\pi}{4}$ in the argument of the cosine can be attributed to the turning point. The term $\frac{\theta_0}{2}$
comes with the factor of $\frac{\pi}{4}$ to enforce boundary conditions \newline $H_n(q_0=0)\propto \cos\frac{n\pi}{2}$ 
(for $\theta_0=\frac{\pi}{2},\,q_0=0$.)

Using the exact relation $\langle q_0,\text{pos}|n\rangle={1\over\pi^{1/4}2^{n/2}\sqrt{n!}}~e^{-{q_0^2\over2}} H_n(q_0)$, we are led to an asymptotic expression of 
the Hermite polynomial 
\begin{align}
 e^{-\frac{q_0^2}{2}}H_{n}(q_0)&\approx \left(2e\over n\right)^\frac{n}{2}\frac{n!}{\sqrt{\pi n\sin\theta_0}}\cos\left[\frac{\pi}{4}-\frac{\theta_0}{2}+n\left(\frac{\sin 2\theta_0}{2}-\theta_0\right)\right];\quad q_0=\sqrt{2n}\cos(\theta_0)\label{eqn:Hermiteapprox1}
 \end{align}
 Using Stirling's approximation for $n!$ we can simplify it to
 \begin{align}
 H_n(\sqrt{2n\cos \theta_0})&=\sqrt\frac{2}{\sin\theta_0}\left(2n\right)^\frac{n}{2}e^{\frac{n}{2}\cos 2\theta_0}\cos\left[\frac{\pi}{4}-\frac{\theta_0}{2}+n\left(\frac{\sin 2\theta_0}{2}-\theta_0\right)\right]
 \label{eqn:Hermiteapprox2}
 \end{align}
Note that this expression is exactly the same as that obtained recently in Ref.~\cite{dominici2007asymptotic} 
through different techniques and was used for the estimation of zeros.
\begin{figure}[htbp]
 \includegraphics[width=0.8\textwidth]{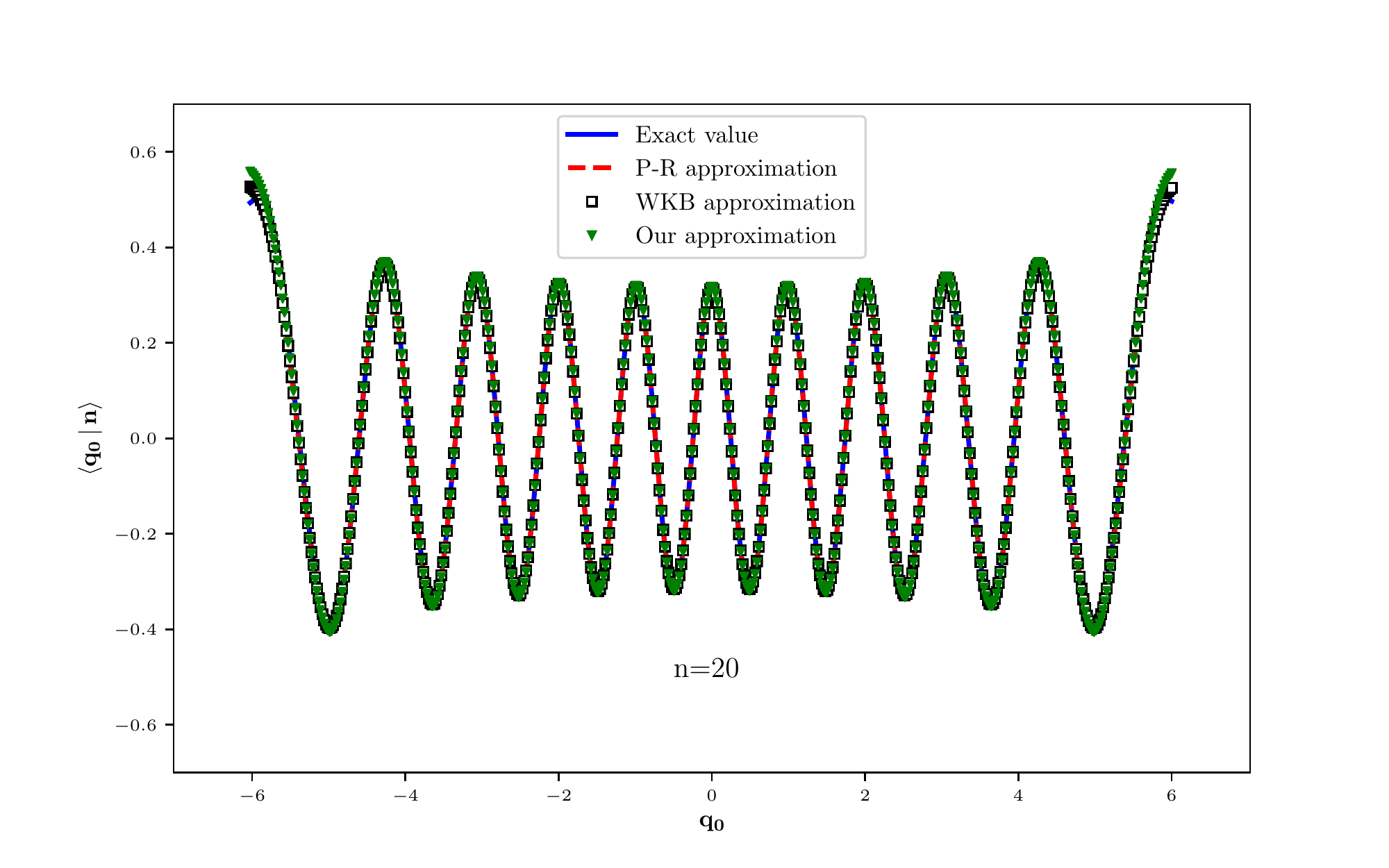}
 \caption{Plot of $\langle q_0\vert n\rangle$ for n=20 using the exact value of the Hermite polynomial(solid blue line), 
 Plancherel-Rotach approximation Eqn. \eqref{eqn:szegoapproxhermite} (dashed red line), WKB approximation Eqn.~\eqref{eqn:WKBHermite}
 (black squares) and our approximation Eqn.~\eqref{eqn:Hermiteasymp2} (inverted green triangles). We interpolated between 
 $q_0\in(-\sqrt{40}+0.3,\sqrt{40}-0.3) $ using 512 uniformly spaced points.}
\label{fig:Plothermiapprox}
\end{figure}

We compare our results with the standard asymptotic expression due to Plancherel and Rotach \cite{Plancherel1929} ( referred to hereafter as the 
P-R approximation) available in the literature for $H_n(q_0)$ as $n\rightarrow\infty$. 
\begin{align}
 e^{-\frac{q_0^2}{2}}H_{n}(q_0)&\approx \frac{2^{n/2+1/4}\sqrt{n!}\left(\pi n\right)^{-\frac{1}{4}}}{\sqrt{\sin\theta}}
\left\{\cos\left[\frac{\pi}{4}+(n+\frac{1}{2})\left(\frac{\sin 2\theta}{2}-\theta\right)\right]+O(n^{-1})\right\}
\label{eqn:szegoapproxhermite}
\\
q_0&=\sqrt{2n+1}\cos\theta\nonumber
\end{align}

We also compare these with the WKB approximation, which yields
\begin{align}
 \bra{q_0,\text{ pos}}n\rangle&= \sqrt{\frac{2}{\pi\,p_n(q_0)}}\cos\left(S_n(q_0)-\frac{\pi}{4}\right) 
\label{eqn:WKBHermite}\\
 p_n(q_0)&= \sqrt{2n+1-q_0^2}\nonumber\\
 S_n(q_0)&= (n+\frac{1}{2})\tan^{-1}\left(\frac{p_n(q_0)}{q_0}\right)-\frac{1}{2}q_0p_n(q_0)\nonumber 
\end{align}
In Fig.~~\ref{fig:Plothermiapprox},
we display the plot for the exact $\langle q_0,\text{pos}|n\rangle$ along with those corresponding to our asymptotic results, the P-R and WKB 
approximations.
To the bare eye, all the approximations seem to be in good agreement with the exact result.
\begin{table}[htbp]
 \begin{center}
\begin{tabular}{ |c|c|c|c| } 
 \hline
 $\,\,n\,\,$ &RMSE (Our result) & RMSE(P-R) & RMSE (WKB)\\ 
	     &Eqn.~\eqref{eqn:Hermiteasymp2}&Eqn.~\eqref{eqn:szegoapproxhermite}&Eqn.~\eqref{eqn:WKBHermite}\\
 \hline
 20& 0.0089  & 0.0053 & 0.0048  \\ 
 \hline
 30& 0.0065 &  0.0041 & 0.0038\\ 
 \hline
 40& 0.0052 & 0.0035 & 0.0033\\ 
 \hline
 50 & 0.0044 & 0.0030 & 0.0029\\ 
 \hline
\end{tabular}
\end{center}
\caption{Root mean squared error for our, P-R and WKB approximation for ${1\over\pi^{1/4}2^{n/2}\sqrt{n!}}~e^{-{q_0^2\over2}}H_n(q_0)$ using $512$ equally 
spaced points for  $q_0\in(-\sqrt{2n}+0.3,\sqrt{2n}-0.3)$ when compared with the exact result}
\label{tab:rmseHermite}
\end{table}
 To probe the comparison between the three approximations further, we do a naive root mean squared error estimation
from the exact value for a range of $n$ using $512$ equally spaced points between 
$(-\sqrt{2n}+0.3,\sqrt{2n}-0.3)$. These are summarised  in Table \ref{tab:rmseHermite}.

We see that the WKB approximation is better than the rest. Schleich~\cite{Schleich_2001} has shown
that the WKB approximation is equivalent to steepest descent such that the saddle points lie at a radial
distance of $\sqrt{2n+1}$. Thus, it seems that for determining the values of Hermite polynomials, doing steepest descent over
a contour of radius $\sqrt{2n+1}$ might be advantageous over $\sqrt{2n}$. As remarked before, Ref.~\cite{dominici2007asymptotic} obtains
asymptotic expansions identical to ours for the purpose of estimating zeros of Hermite polynomials. 
Numerical approximations are not the subject of this paper, and we drop further investigations.

We note here that all numerical results and plots presented in this section were generated with 
the mpmath library \cite{mpmath} with default precision, which translates to 15 decimal places.

\item
{\allowdisplaybreaks	
The last case before moving on to squeezed states
is the calculation of the Fock state matrix elements of
the unitary displacement operator. In the now familiar fashion,
using Eqns. \eqref{Fock_in_phase}, \eqref{displacement_fock}, \eqref{eqn:coherentinnerproduct} and after simplification we have:
\begin{align}
\langle m|D(q,p)|n\rangle = & {\mathcal{N}_m\mathcal{N}_n\over
4\pi^2} \int_0^{2\pi} d\theta~ e^{im\theta} \int_0^{2\pi}
d\theta^\prime~ e^{-in\theta^\prime +
i\sqrt{{n\over2}}(pC^\prime-qS^\prime)}\nonumber\\
&\times \langle\sqrt{2m}C, \sqrt{2m} S|q+\sqrt{2n}C^\prime,
p+\sqrt{2n}S^\prime\rangle\nonumber\\
=& {\mathcal{N}_m\mathcal{N}_n\over4\pi^2}
e^{-{1\over2}(m+n)-{1\over4}(q^2+p^2)}\int_0^{2\pi}d\theta
~e^{im\theta} \int_0^{2\pi}d\theta^\prime~ e^{-in\theta^\prime}
e^Y\,,\nonumber\\
Y = & \sqrt{m}ze^{-i\theta} - \sqrt{n}z^*e^{i\theta^\prime} +
\sqrt{mn} e^{i(\theta^\prime-\theta)},\,\nonumber\\
C=&\cos\theta,~S=\sin\theta,~C^\prime=\cos\theta^\prime,~
S^\prime=\sin\theta^\prime,~z={1\over\sqrt{2}}(q+ip)\,.
\label{a}
\end{align}
Since $m,n\ge0$ and $Y$ is a polynomial in $e^{-i\theta}$ and $e^{i\theta^\prime}$, putting $\xi=e^{-i\theta}$ and $\eta=e^{i\theta'}$
\begin{subequations}
\begin{align}
	\langle m|D(q,p)|n\rangle ={\mathcal{N}_m\mathcal{N}_n\over4\pi^2}e^{-{1\over2}(m+n)-{1\over4}(q^2+p^2)}& 
\oint\limits_{\text{\tiny{clockwise}}}\frac{d\xi}{\xi^{m+1}}\oint\limits_{\text{\tiny{anticlockwise}}}\frac{d\eta}{\eta^{n+1}} e^{\sqrt{mn}\xi\eta +\sqrt{m}z\xi-\sqrt{n}z^*\eta}\label{eqn:LaguerreContourIntegral1}\\
={\mathcal{N}_m\mathcal{N}_n\over m!n!} e^{-{1\over2}(m+n)-{1\over4}(q^2+p^2)}& 
{\partial^m\over\partial\xi^m} {\partial^n\over\partial\eta^n} e^{\sqrt{mn}\xi\eta +\sqrt{m}z\xi
-\sqrt{n}z^*\eta} |_{\xi=\eta=0}\nonumber\\
(\sqrt{m}\xi\rightarrow\xi,\,\sqrt{n}\eta\rightarrow\eta)\qquad &=  {e^{{1\over4}(q^2+p^2)}\over\sqrt{m!n!}} {\partial^m\over\partial\xi^m} {\partial^n\over\partial\eta^n}
e^{(\eta+z)(\xi-z^*)}|_{\xi=\eta=0}\label{eqn:intermediate_laguerre}\\
&=  {e^{{1\over4}(q^2+p^2)}\over\sqrt{m!n!}} {\partial^m\over\partial\xi^m} (\xi-z^*)^n 
e^{z(\xi-z^*)}|_{\xi=0}\nonumber\\
\left(\chi=z^\ast-\xi\right)\qquad&=  {e^{{1\over4}(q^2+p^2)}\over\sqrt{m!n!}} (-1)^{n-m} {\partial^m\over\partial \chi^{m}} (\chi^{n}e^{-z\chi})\vert_{\chi=z^\ast}\nonumber\\
z\chi\rightarrow\chi\qquad &=  {(-z)^{m-n}\over\sqrt{m!n!}}  e^{{1\over2}z^*z}\left({d^m\over d\chi^m}(e^{-\chi}\chi^n)\right)_{\chi=z^*z}\,\nonumber\\
&=  {\sqrt{m!}\over\sqrt{n!}} (-z^*)^{n-m} e^{-{1\over2}z^*z} L_m^{n-m}(z^*z)\,.
\end{align}
\end{subequations}}
Alternatively, taking the $\xi$ derivative first in Eqn.~\eqref{eqn:intermediate_laguerre} yields
\begin{align}
	\langle m|D(p,q)|n\rangle\,&=\,\sqrt{\frac{n!}{m!}}z^{m-n}e^{-\frac{1}{2}z^\ast z}L_{n}^{m-n}\left(z^\ast z\right)
\end{align}
The matrix element of the displacement operator in the Fock basis 
matches the form calculated in~\cite{GlauberLaguerre}.

We next use the in-phase integral representation in Eqn.~$(\ref{a})$ for $\langle m|D(q,p)|n\rangle$  to obtain an asymptotic formula 
for $\langle m|D(q,p)|n\rangle$ valid in the limit of large $m,n$. This matrix element has been computed using phase space methods 
cited earlier \cite{Dowling_1991},\cite{Mundarain1}. 

Since $\langle m|D(d\cos\phi ,d\sin\phi)|n\rangle= e^{i\phi(m-n)}\langle m|D(d,0)|n\rangle$
it suffices to focus our attention on $\langle m|D(d,0)|n\rangle$, for which Eqn.~\eqref{eqn:LaguerreContourIntegral1} yields: 
\begin{align}
	\lefteqn{\langle m|D(d,0)|n\rangle =}\nonumber\\
	&&\frac{{\cal N}_m {\cal N}_n}{4\pi^2}
e^{-{1\over2}(m+n)-{1\over4}d^2}\oint\limits_{\text{\tiny{clockwise}}}\frac{d\xi}{\xi^{m+1}}\oint\limits_{\text{\tiny{anticlockwise}}}\frac{d\eta}{\eta^{n+1}} e^{\sqrt{mn}\xi\eta +\sqrt{m}z\xi-\sqrt{n}z^*\eta}\nonumber
\end{align}
Rescaling, $\xi\rightarrow\sqrt{m}\xi$ and $\eta\rightarrow\sqrt{n}\eta$, 
\begin{align}
	\langle m|D(d,0)|n\rangle &=C_1
	\oint\limits_{\text{\tiny{clockwise}}}d\xi \oint\limits_{\text{\tiny{anticlockwise}}}d\eta\frac{e^{f(\xi,\eta)}}{\xi\eta}
\label{eqn:contourLaguerre1}
\end{align}
$C_1=\frac{{\cal N}_m {\cal N}_n}{4\pi^2} m^{-\frac{m}{2}}n^{-\frac{n}{2}}e^{-{1\over2}(m+n)-{1\over4}d^2}$,
$f(\xi,\eta)=mn\xi\eta+m\xi\frac{d}{\sqrt{2}}-n\eta\frac{d}{\sqrt{2}}-m\log(\xi)-n\log(\eta)$

Henceforth, we will drop the clockwise and anticlockwise beneath the integrals. Now, the contours are defined as
\begin{align}
	\xi=\frac{e^{-i\theta}}{\sqrt{m}};\quad& \eta=\frac{e^{i\theta^{\prime}}}{\sqrt{n}}
\label{eqn:xietaval}
\end{align}
We take advantage of the analyticity of the integrand in ($\xi,\eta$) to calculate the matrix element in the semiclassical
limit of large $m,n$ using the method of steepest descent.
The saddle point of $f(\xi,\eta)$ yields
\begin{align}
\frac{\partial f(\xi,\eta)}{\partial\xi}\bigg\rvert_{\xi_0}=0.&\qquad\mbox{  Solving, }\xi_0=\frac{1}{n\eta+\frac{d}{\sqrt{2}}},\qquad 
\frac{\partial^2f(\xi,\eta)}{\partial\xi^2}\bigg\rvert_{\xi_0}=\frac{m}{\xi_0^2}
\label{eqn:saddle1laguerre}
\end{align}
To integrate with respect to $\xi$, we expand $f(\xi,\eta)$ around the saddle point $\xi_0$ (which is now a function of $\eta$) 
and determine the direction of steepest descent
\begin{align*}
f(\xi,\eta)-f(\xi_0,\eta)&=\frac{m}{2\xi_0^2}r^2e^{2i\phi}\qquad(\xi-\xi_0=re^{i\phi})
\end{align*}
Choice of $\phi$ compatible with a clockwise contour is $\phi=\mbox{arg}(\xi_0)-\frac{\pi}{2}$. 
We calculate
\begin{align}
	\bra{m}D(d,0)\ket{n}&=C_1\oint \frac{d\eta}{\xi_0\eta}e^{f(\xi_0(\eta),\eta)}\left(\int_{-\infty}^{\infty}dr
	e^{-\frac{mr^2}{2\vert\xi_0\vert^2}}\right)(-i)e^{i\left[\text{{\small arg} }\xi_0\right]}\nonumber\\
	&=(-i)C_1\oint\frac{d\eta}{\xi_0\eta}e^{f(\xi_0(\eta),\eta)}\sqrt{\frac{2\pi}{m}}\vert\xi_0\vert e^{i\text{\small arg }(\xi_0)}\nonumber\\
  &= (-i)C_1\sqrt{\frac{2\pi}{m}}\oint\frac{d\eta}{\eta}e^{f(\xi_0(\eta),\eta)}
\label{eqn:integralovereta}
\end{align}
\begin{figure}[t]
\centering
\includegraphics[width=0.40\textwidth]{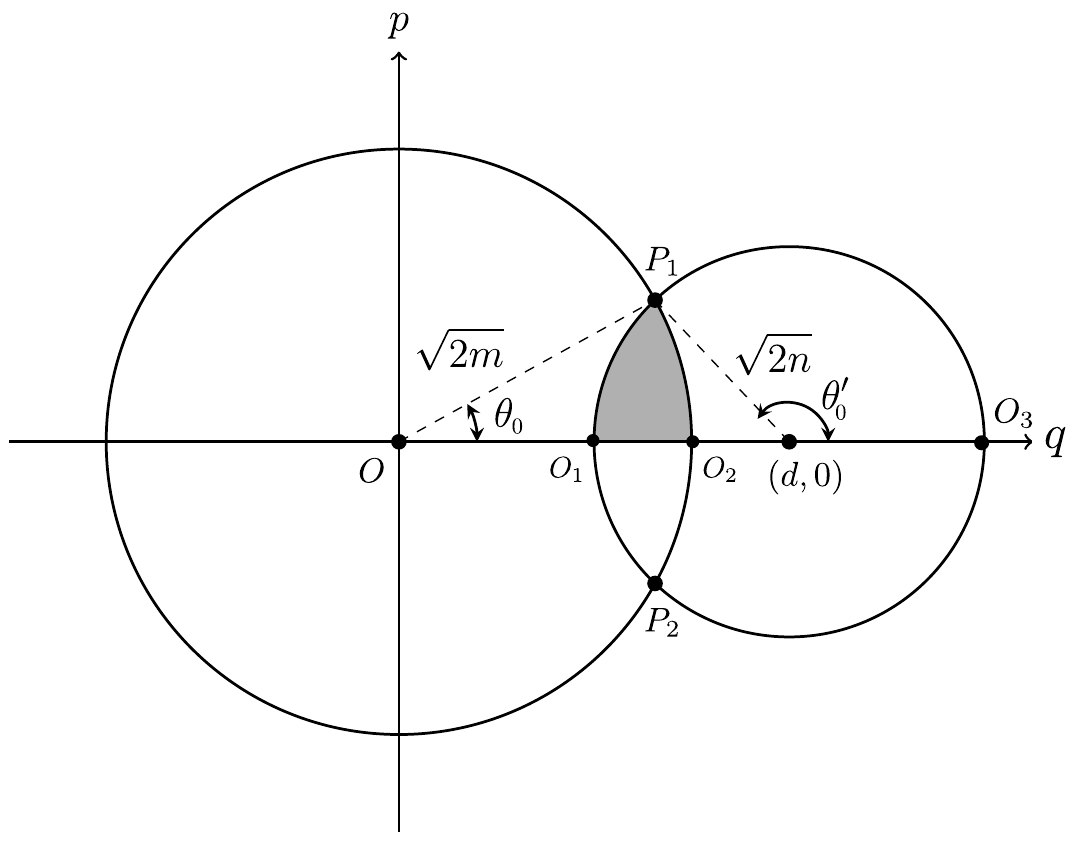}
\caption{Intersecting orbits of Fock state $m$ with another displaced Fock state $D(d,0)\vert n\rangle$}
\label{fig:Fockdisplaced}
\end{figure}
Using Eqns.~\eqref{eqn:saddle1laguerre} and ~\eqref{eqn:contourLaguerre1}, we get
\begin{align}
	f(\xi_0(\eta),\eta)&=m-n\eta\frac{d}{\sqrt{2}}+m\log\left(n\eta+\frac{d}{\sqrt{2}}\right)-n\log\eta
	\label{eqn:fxi0eta}
\end{align}
Calculating the saddle point of $f(\xi_0(\eta),\eta)$ to compute the integral in $\eta$, we get
\begin{align}
\frac{\partial f(\xi_0,\eta)}{\partial\eta}\bigg\rvert_{\eta_0}=0\,\,\implies\,\,\frac{1}{\eta_0}+\frac{d}{\sqrt{2}}=m\xi_0
\label{eqn:saddlelaguerre}
\end{align}
Putting the values from $\xi,\eta$ from Eqn.~\eqref{eqn:xietaval} in ~\eqref{eqn:saddlelaguerre} we get
\begin{align}
	\sqrt{2m}e^{-i\theta_0}-\sqrt{2n}e^{-i\theta_0^{\prime}}=d&	\Leftrightarrow\nonumber\\
	d=\sqrt{2m}\cos\theta_0-\sqrt{2n}\cos\theta_0^\prime~~;& ~~\sqrt{2m}\sin\theta_0 =\sqrt{2n}\sin\theta_0^\prime.
\label{eqn:saddleLaguerreAngle}
\end{align}
Now, let us take a moment to appreciate this result in the light of Fig.~\ref{fig:Fockdisplaced}. This is precisely the 
figure we would have drawn to depict the states $\ket{m}$ and $D(d,0)\ket{n}$, and would have been led to the same result as
Eqn.~\eqref{eqn:saddleLaguerreAngle}. In other words, we could have anticipated the saddle point from geometry, guided by
in-phase representations alone.

As $\theta_0,\theta^\prime_0$ are varied independently from $0$ to $2\pi$
\begin{align*}
 \text{ max. of}~&|\sqrt{2m} e^{-i\theta_0}-\sqrt{2n}e^{-i\theta_0^\prime}| = \sqrt{2m}+\sqrt{2n},\\
 \text{min. of}~&|\sqrt{2m} e^{-i\theta_0}-\sqrt{2n}e^{-i\theta_0^\prime}| = |\sqrt{2m}-\sqrt{2n}|.
\end{align*}
Therefore, the necessary and sufficient conditions for existence of real solutions to the above conditions are 
\begin{equation}
 |\sqrt{2m} -\sqrt{2n}|~\leq d~\leq \sqrt{2m} +\sqrt{2n}
 \label{eqn:conditionsforexistence}
\end{equation}
This is the regime we are going to be interested in.
From Fig.~~\ref{fig:Fockdisplaced}, we can calculate the angles
\begin{align}
	\theta_0&=\cos^{-1}\left(\frac{d^2+2m-2n}{2\sqrt{2m}d}\right)&\theta_0^\prime=\pi-\cos^{-1}\left(\frac{d^2+2n-2m}{2\sqrt{2n}d}\right)
	\label{eqn:theta0values}
\end{align}
Using Eqns.~\eqref{eqn:saddle1laguerre},~\eqref{eqn:saddlelaguerre} in ~\eqref{eqn:fxi0eta} we get
\begin{align}
f(\xi_0,\eta_0)&=m+n-mn\xi_0\eta_0-m\log\xi_0-n\log\eta_0\nonumber\\
	       &=m+n-\sqrt{mn}e^{\pm i(\theta_0^{\prime}-\theta_0)}\pm i(m\theta_0-n\theta_0^\prime)+\log(m^{\frac{m}{2}}n^{\frac{n}{2}})\text { at }\pm(\theta_0,\theta_0^{\prime})
\label{eqn:fxi0eta0}
\end{align}
\begin{align}
\frac{\partial^2f(\xi_0,\eta)}{\partial\eta^2}\bigg\rvert_{\eta_0}&=
\frac{n}{\eta_0^2}-\frac{mn^2}{\left(n\eta_0+\frac{d}{\sqrt{2}}\right)^2}=\frac{n}{\eta_0^2}-\frac{mn^2}{\xi_0^2}\nonumber
\end{align}
Evaluating the value at the saddle points $\pm(\theta_0,\theta_0^{\prime})$
\begin{align}
\frac{\partial^2f(\xi_0,\eta)}{\partial\eta^2}\bigg\rvert_{\pm(\theta_0,\theta_0^{\prime})}&=2n^2\sin(\theta_0^{\prime}-\theta_0)
e^{\mp i\left(\theta_0+\theta_0^\prime+\frac{\pi}{2}\right)}
\end{align}
We return to the integral
\begin{align*}
	\bra{m}D(d,0)\ket{n}&= (-i)C_1\sqrt{\frac{2\pi}{m}}\oint\frac{d\eta}{\eta}e^{f(\xi_0,\eta)}\\
	\text{Let } I&=\oint\frac{d\eta}{\eta}e^{f(\xi_0,\eta)}
\end{align*}
In the neighbourhood of the critical points $\pm(\theta_0,\theta_0^\prime)$ 
\begin{align}
f(\xi_0,\eta)-f(\xi_0,\eta_0)=r^2n^2\sin(\theta_0^\prime-\theta_0)e^{i\left[2\phi\mp(\frac{\pi}{2}+\theta_0+\theta_0^\prime)\right]};\quad \text{where }\,
	\xi-\xi_0=re^{i\phi};
	\label{eqn:difflaguerre}
\end{align}
The directions of steepest descent at the points
$\pm(\theta_0,\theta_0^\prime)$ compatible with a counterclockwise 
contour are $\frac{3\pi}{4}+\frac{\theta_0+\theta_0^\prime}{2}$ and
$\frac{\pi}{4}-\frac{\theta_0+\theta_0^\prime}{2}$. 
Calculating the integral in $\eta$
\begin{align}
I&=\int_{-\infty}^{\infty}dr e^{-n^2r^2\sin(\theta_0^\prime-\theta_0)}\left(
\frac{e^{f(\xi_0,\eta_0)}}{\eta_0}\bigg\vert_{(\theta_0,\theta_0^\prime)}
e^{i\left[\frac{3\pi}{4}+\frac{\theta_0+\theta_0^\prime}{2}\right]}\nonumber
+\frac{e^{f(\xi_0,\eta_0)}}{\eta_0}\bigg\vert_{(-\theta_0,-\theta_0^\prime)}
e^{i\left[\frac{\pi}{4}-\frac{\theta_0+\theta_0^\prime}{2}\right]}\right)\\
&=\sqrt{\frac{\pi}{n\sin(\theta_0^\prime-\theta_0)}}\left(
e^{f(\xi_0,\eta_0)}\bigg\vert_{(\theta_0,\theta_0^\prime)}
e^{i\left[\frac{3\pi}{4}+\frac{\theta_0-\theta_0^\prime}{2}\right]}\nonumber
-e^{f(\xi_0,\eta_0)}\bigg\vert_{(-\theta_0,-\theta_0^\prime)}
e^{-i\left[\frac{3\pi}{4}+\frac{\theta_0-\theta_0^\prime}{2}\right]}\right)\nonumber
\end{align}
Using 
\begin{align*}
	f(\xi_0,\eta_0)\bigg\vert_{(\theta_0,\theta_0^\prime)}=f(\xi_0,\eta_0)\bigg\vert_{(-\theta_0,\theta_0^\prime)}^\ast
\end{align*}
\begin{align}
I&=2i\sqrt{\frac{\pi}{n\sin(\theta_0^\prime-\theta_0)}}\text{ Im}
\left[e^{i\left(\frac{3\pi}{4}+\frac{\theta_0-\theta_0^\prime}{2}\right)}\times \exp{f(\xi_0,\eta_0)}\bigg\vert_{(\theta_0,\theta_0^\prime)}
\right]\nonumber\\
&=2i\,m^{\frac{m}{2}}n^{\frac{n}{2}}\sqrt{\frac{\pi}{n\sin(\theta_0^\prime-\theta_0)}}\exp\left[m+n-\sqrt{mn}
\cos(\theta_0^\prime-\theta_0)\right]\nonumber\\
\times &\sin\left[(m+\frac{1}{2})\theta_0-(n+\frac{1}{2})\theta_0^\prime+\frac{3\pi}{4}-\sqrt{mn}\sin(\theta^\prime_0-\theta_0)\right]
\label{eqn:I}
\end{align}
Using $\bra{m}D(d,0)\ket{n}= (-i)C_1\sqrt{\frac{2\pi}{m}}I$,
and simplifying, we get
\begin{align}
\langle m|D(d,0)|n\rangle & \simeq \frac{n^{-\frac{n}{2}
}m^{-\frac{m}{2}
}}{\pi} \sqrt{\frac{m!n!}{2mn}}\,e^{n+m-\left[\frac{d^2}{4}+\sqrt{mn}\cos(\theta_0^\prime-\theta_0)\right]}\nonumber\\
&\times \dfrac{\cos[(m+{1\over2})\theta_0-(n+{1\over2})\theta_0^{\prime}+{\pi\over4}
 -\sqrt{mn}\sin(\theta_0^\prime-\theta_0)]}{\sqrt{\sin(\theta_0^\prime-\theta_0)}}
\label{eqn:displacementapprox1}
\end{align}
Use of Stirling's formula ${\cal N}_n/{2\pi}\simeq
(n/(2\pi)^3)^{1/4}$ for large $n$ and likewise for ${\cal N}_m/{2\pi}$ for large $m$, we obtain:
\begin{align}
 \langle m|D(d,0)|n\rangle & \simeq \left(1\over{\pi^2 mn}\right)^{1/4}e^{{1\over2}(\sqrt{m}-\sqrt{n})^2-
 {d^2\over4}}\times e^{\sqrt{mn}(1-\cos(\theta_0^\prime-\theta_0^))}\nonumber\\
 &\times \dfrac{\cos[(m+{1\over2})\theta_0-(n+{1\over2})\theta_0^\prime+{\pi\over4}
 -\sqrt{mn}\sin(\theta_0^\prime-\theta_0)]}{\sqrt{\sin(\theta_0^\prime-\theta_0)}}.
 \label{eqn:displacementapprox2}
\end{align}
First, note that the leading order (in $m,n$) contributions to the oscillatory
part of the overlap could have been obtained using the in-phase superpositions in Eqns.~\eqref{Fock_in_phase} and
~\eqref{displacement_fock} of
$\ket{m}$ and $D(d,0)\ket{n}$ respectively, applied to the saddle point. 
Geometry determines the saddle point and hence leads to Eqn.~\eqref{eqn:saddleLaguerreAngle}.
Thus, the contribution to the oscillatory phase due to the saddle points would be
\begin{align*}
	\text{Phase }D(d,0)\ket{n} -\text{ Phase }\ket{m}\,\bigg\vert_{P_1}&=m\theta_0-n\theta_0^\prime-\sqrt{n}\sin\theta_0^\prime\frac{d}{\sqrt{2}}\\
	\text{Using Eqn.~\eqref{eqn:saddleLaguerreAngle} }\qquad&=m\theta_0-n\theta_0^\prime-\sqrt{mn}\sin(\theta_0^\prime-\theta_0)
\end{align*}
which matches the leading order terms in the cosine, as expected. The phase at $P_2$ would just be negative of the term at $P_1$.
The extra factors of $\frac{\pi}{4}$ can again be attributed to the turning point.

Alternatively, we calculate the shaded area in Fig.~~\ref{fig:Fockdisplaced}
\begin{align}
	&n\pi+m\theta_0-n\theta_0^\prime +\frac{1}{2}(n\sin 2\theta_0^\prime-m\sin 2\theta_0)\nonumber\\
	=&n\pi+m\theta_0-n\theta_0^\prime-\sqrt{mn}\sin(\theta_0^\prime-\theta_0)
	\label{eqn:LaguerreArea}
\end{align}
The extra factor of $n\pi$ when we calculate areas, can be attributed to the fact that we determine the phase at the point $P_1$
for in-phase superpositions using the area swept out by the radius starting from $O_3$ to $P_1$, (with extra phase factors due to the 
action of $D(d,0)$) and not from $O_1$ to $P_1$. In fact, the in-phase superposition associates $0$ phase to $O_3$ and $-n\pi$ to $O_1$.

For numerical comparison of the asymptotic expression for $\langle m|D(d,0)|n \rangle$ with the exact expression
\begin{equation}
\langle m|D(d,0)|n\rangle 
=  {\sqrt{m!}\over\sqrt{n!}} \left(-{d\over{\sqrt{2}}}\right)^{n-m} e^{-{1\over4}d^2} L_m^{n-m}\left({d^2\over{2}}\right)\,,
\label{4.36}
\end{equation}
we consider the cases $m=n$ and $m\neq n$ separately 
\begin{description}[leftmargin=*]
\item[\bf Case I)  $m=n$ ]\hfill\\
In this case, the exact expression for $\langle m|D(d,0)|n\rangle$ becomes
\begin{equation}
\langle m|D(d,0)|m\rangle 
=   e^{-{1\over4}d^2} L_m\left({d^2\over{2}}\right)\,.
\end{equation}
Further, Eqn.~\eqref{eqn:theta0values} for $\theta_0$ and $\theta_0^\prime$ in terms of $d$ and $m$ reduces to
\begin{equation}
\theta_0= \cos^{-1}(d/2\sqrt{2m}),\quad\theta_0^\prime=\pi-\theta_0
\end{equation}
and the asymptotic expression for $\langle m|D(d,0)|m\rangle$ reads
\begin{align}
 \langle m|D(d,0)|m\rangle & \simeq \dfrac{m!}{\sqrt{2}\pi m^{m+1}}e^{2m+m\cos(2\theta_0)-\frac{d^2}{4}}\nonumber\\
 &\times \dfrac{\cos[(2m+1)\theta_0-(m+{1\over4})\pi -m\sin(2\theta_0)]}{\sqrt{\sin(2\theta_0)}}.
\end{align}
Using Stirling's approximation and simplifying:
\begin{align}
 \langle m|D(d,0)|m\rangle & \simeq \sqrt{\dfrac{1}{\pi m}}e^{-\frac{d^2}{4}+m(1+\cos(2\theta_0))}\nonumber\\
 &\times \dfrac{\cos[(2m+1)\theta_0-(m+{1\over4})\pi -m\sin(2\theta_0)]}{\sqrt{\sin(2\theta_0)}}.
 \label{eqn:laguerrestirling}
\end{align}
\item[\bf Case II)  $ m \neq n$\,\,]\hfill\\
	In this case, we rewrite the Eqns.~\eqref{eqn:theta0values} and ~\eqref{eqn:displacementapprox2} for convenience
\begin{align*}
	\theta_0&=\cos^{-1}\left(\frac{d^2+2m-2n}{2\sqrt{2m}d}\right)&\theta_0^\prime=\pi-\cos^{-1}\left(\frac{d^2+2n-2m}{2\sqrt{2n}d}\right)
\end{align*}
\begin{align*}
 \langle m|D(d,0)|n\rangle & \simeq \left(1\over{\pi^2 mn}\right)^{1/4}e^{{1\over2}(\sqrt{m}-\sqrt{n})^2-
 {d^2\over4}}\times e^{\sqrt{mn}\left[1-\cos(\theta_0^\prime-\theta_0)\right]}\nonumber\\
 &\times \dfrac{\cos[(m+{1\over2})\theta_0-(n+{1\over2})\theta_0^\prime+{\pi\over4}
 -\sqrt{mn}\sin(\theta_0^\prime-\theta_0)]}{\sqrt{\sin(\theta_0^\prime-\theta_0)}}.
\end{align*}

It is instructive to compare our results for $\langle m|D(d,0)|n\rangle $ with those obtained  using the asymptotic formulae 
for Laguerre polynomials due to Tricomi \cite{Tricomi}
\begin{align}
 &e^{-\frac{1}{2}x}L_n^{\alpha}(x) = 2(-1)^n(2\cos \theta)^{-\alpha}(\pi \nu \sin 2\theta)^{-1/2}\times \nonumber\\
 &\left(\sin \Theta+\frac{1}{12}\left(\frac{\nu \sin 2\theta}{4}\right)^{-1}\left[\frac{5}{4\sin^2\theta}-
 (1-3\alpha^2)\sin^2\theta-1\right]\sin(\Theta+3\pi/2)
 +O(n^{-2})\right)\nonumber\\ 
 &~~~~~~~~~~~~x=\nu \cos^2\theta;~~~\nu=(4n+2\alpha+2)\qquad \Theta=\frac{\pi}{4}+\frac{\nu(2\theta-\sin(2\theta))}{4}
 \label{eqn:Tricomi}
 \end{align}
valid in the `oscillatory region' $0<x<\nu$ for large $n$. Using this formula in $(\ref{4.36})$ one obtains 
\begin{align}
& \langle m|D(d,0)|n\rangle \simeq (-1)^n \sqrt{\frac{m!}{n!}}\left(\frac{2}{\pi^2d^2}\right)^{1/4}\left(\frac{n+m+1}{2}\right)^{\frac{2n-2m-1}{4}}
   (\sin\theta)^{-1/2}\nonumber\\ &\times 
   \left(\sin \Theta+\frac{1}{12}\left(\frac{\nu \sin 2\theta}{4}\right)^{-1}\left[\frac{5}{4\sin^2\theta}-(1-3\alpha^2)\sin^2\theta-1\right]\sin(\Theta+3\pi/2)
 \right)\nonumber\\
 &\nu=2(n+m+1);\alpha=n-m;\,\Theta=\frac{(n+m+1)}{2}(2\theta-\sin 2\theta)+\frac{\pi}{4};\,\,\theta=\cos^{-1}\left[\frac{d}{2\sqrt{n+m+1}}\right]
\label{eqn:Tricomidisplaced}
\end{align}
As in the Hermite case, we also consider the WKB approximation (Dowling and Schleich \cite{Dowling_1991}), 
who interpret their results in terms of interfering areas in phase space. 
\begin{align}\label{eqn:SchleichLaguerre}
	\bra{m}D(d,0)\ket{n}&=2\sqrt{A_{m,n}}\cos(\phi_{m,n})\\
	A_{m,n}=\frac{1}{2\pi p_{m}(x_c)};&\quad x_c=\frac{m-n}{d}+\frac{d}{2}\nonumber\\
	p_m(x)=\sqrt{2m+1-x^2};&\quad\phi_{m,n}=S_{m,n}(x_c)+\frac{\pi}{4}\nonumber\\
	S_{m,n}=-\left(m+\frac{1}{2}\right)\sin^{-1}\left[\frac{x_c}{\sqrt{2m+1}}\right]+&\left(n+\frac{1}{2}\right)\sin^{-1}
	\left[\frac{x_c-d}{\sqrt{2n+1}}\right]-\frac{d}{2}\,\,p_m(x_c)-(n-m)\frac{\pi}{2}\nonumber
\end{align}
\begin{figure}[htbp]
  \includegraphics[width=0.8\textwidth]{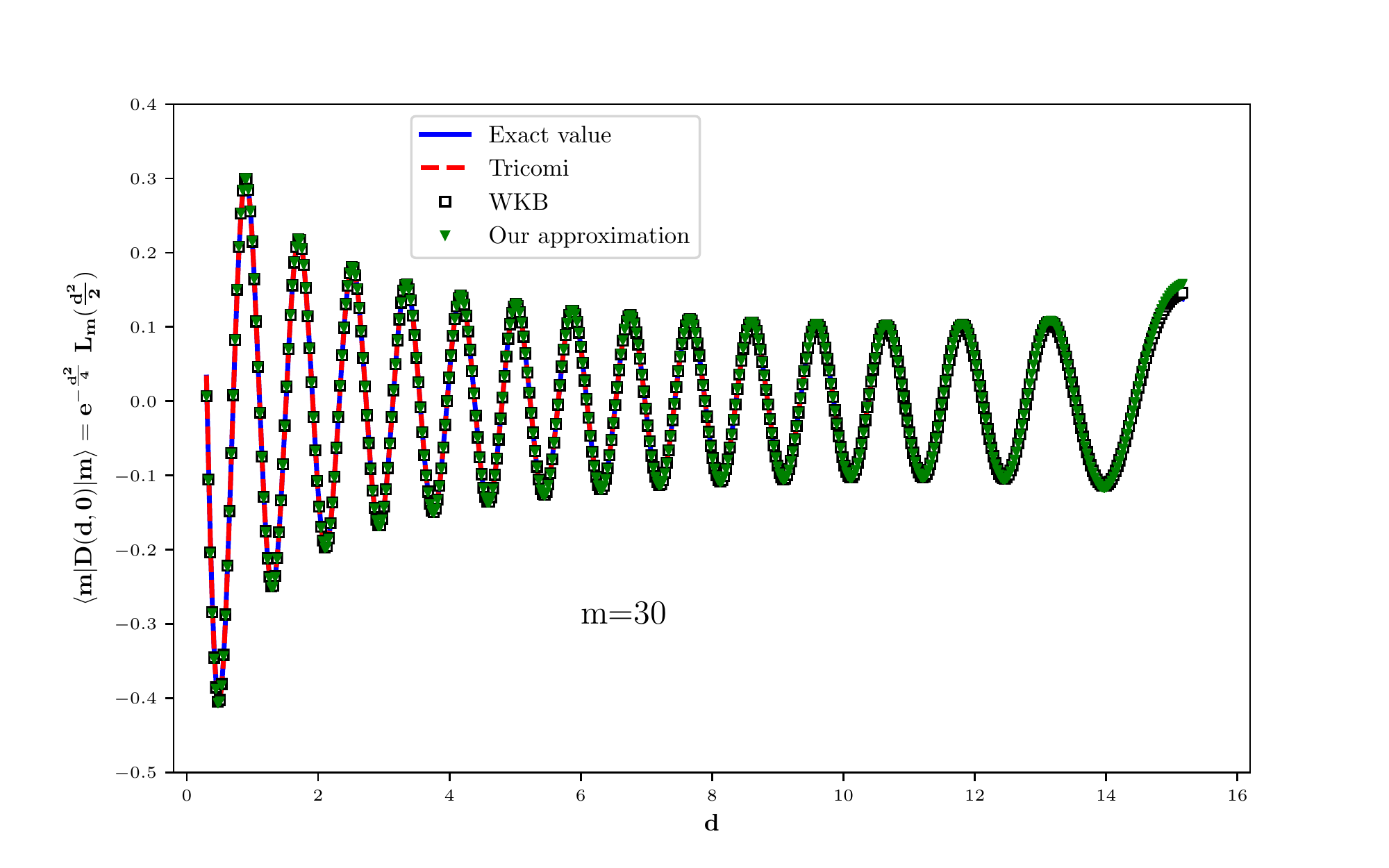}
  \caption{Plot of $\bra{m}D(d,0)\ket{m}$: Comparison of our approximation \eqref{eqn:laguerrestirling}, Tricomi \eqref{eqn:Tricomidisplaced},
  and WKB~\eqref{eqn:SchleichLaguerre} with the exact value.}
\label{fig:LaguerrequalWKB}
 \end{figure}
 
For $m=n=30$, Fig.~\ref{fig:LaguerrequalWKB} compares the exact results with the three approximations noted above.
They match each other very well.

For $m>n$ with $m=30$ and $n=20$, we compare our asymptotics with Tricomi in Fig.~\ref{fig:Laguerreunequal}.
Our approximation is visibly better than Tricomi's.
The performance of Tricomi's approximation is particularly bad near the lower end at $d= \sqrt{2m}-\sqrt{2n}$.
However, note that our results are similar to the WKB approximation in Fig.~\ref{fig:Laguerreunequalschleich}. 
  \begin{figure}[htbp]
  \includegraphics[width=0.8\textwidth]{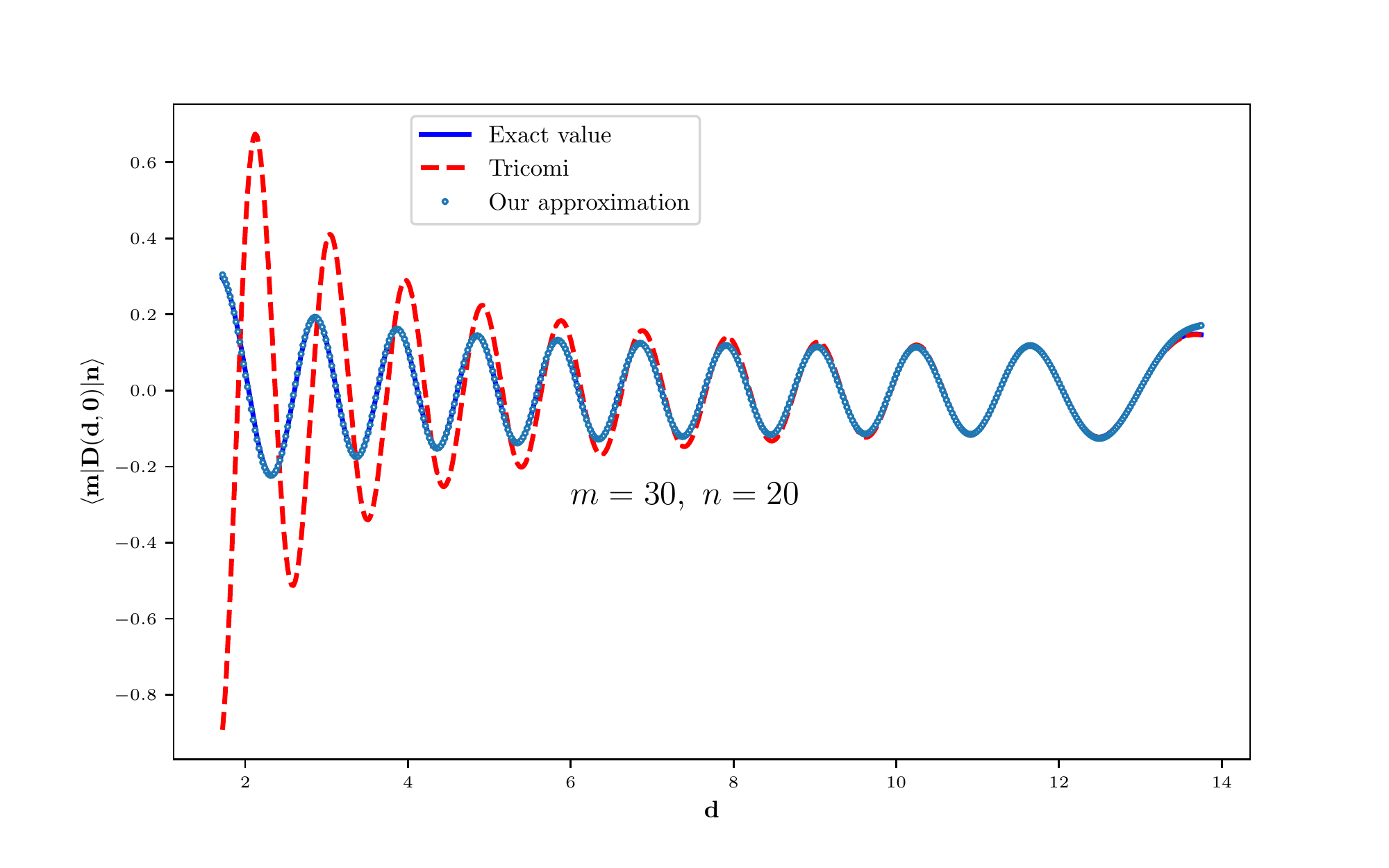}
  \caption{Plot of $\bra{m}D(d,0)\ket{n}$: Comparison of our approximation \eqref{eqn:displacementapprox2} and Tricomi \eqref{eqn:Tricomidisplaced}with the exact value.}
 \label{fig:Laguerreunequal}
 \end{figure}  
 \begin{figure}[H]
  \includegraphics[width=0.8\textwidth]{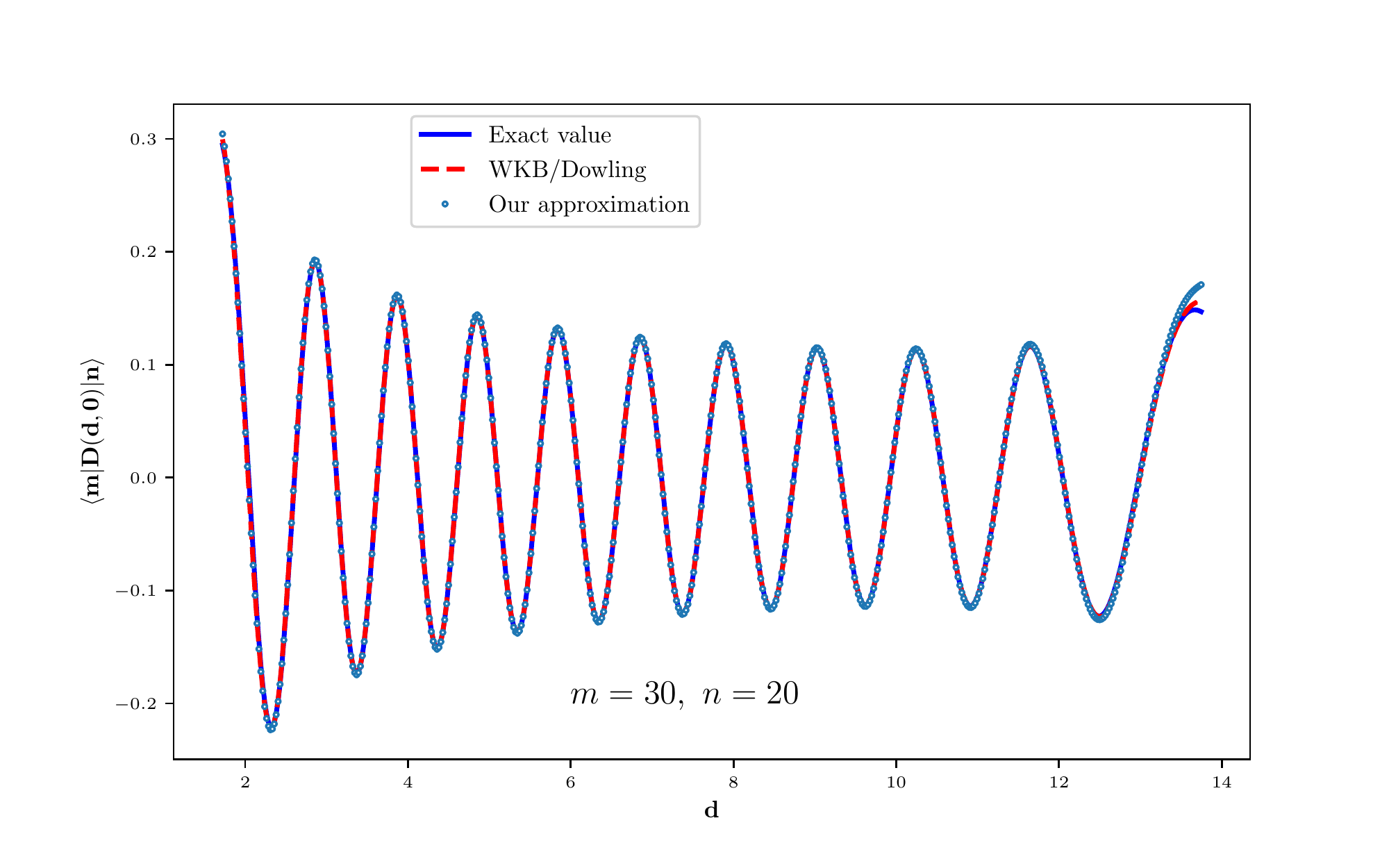}
  \caption{Plot of $\bra{m}D(d,0)\ket{n}$: Comparison of our approximation
  \eqref{eqn:displacementapprox2} and Dowling's \eqref{eqn:SchleichLaguerre}with the exact value.}
  \label{fig:Laguerreunequalschleich}
 \end{figure}

 To compare the closeness of the three approximations to the exact values  we perform a simple root mean squared error 
 analysis for $512$ equally spaced points in the 
 range (a) $(\epsilon,2\sqrt{2m}+\epsilon)$ for the case when $m=n$ and (b)  $(\sqrt{2m}-\sqrt{2n}+5,\sqrt{2m}+\sqrt{2n}-\epsilon)$ for the case when 
 $m>n$ with $\epsilon=0.3$ as before. The results are summarised in the Tables~\ref{tab:RMSELaguerreequal} and~\ref{tab:rmselag2}. 
 \begin{table}
 \begin{center}
\begin{tabular}{ |c|c|c|c| } 
 \hline
 $\,\,m\,\,$ &RMSE (Our result) & RMSE(Tricomi) &RMSE(WKB) \\
	     &Eqn.~\eqref{eqn:laguerrestirling}&Eqn.~\eqref{eqn:Tricomidisplaced}&Eqn.~\eqref{eqn:SchleichLaguerre}\\
 \hline
 20&0.0048   & 0.0009 &0.0036\\ 
 \hline
 30& 0.0032 & 0.0006 &0.0025 \\ 
 \hline
 40&  0.0023& 0.0005 &0.0018\\ 
 \hline
 50 & 0.0018 & 0.0004 &0.0013\\ 
 \hline
\end{tabular}
\end{center}
\caption{Root mean squared error for our, Tricomi's and the WKB approximation for \newline$\bra{m}D(d,0)\ket{m}$ using $512$ equally spaced points for  $d\in(\epsilon,2\sqrt{2m}-\epsilon)$ when compared with the \newline exact result. $\epsilon=0.3$ in the figure.}
\label{tab:RMSELaguerreequal}
\end{table}

 \begin{table}
 \begin{center}
\begin{tabular}{ |c|c|c|c|c|} 
 \hline
 $\,\,m\,\,$&$\,\,n\,\,$ &RMSE (Our result) & RMSE(Tricomi)&RMSE (WKB) \\
	    &&Eqn.~\eqref{eqn:displacementapprox2}&Eqn.~\eqref{eqn:Tricomidisplaced}&Eqn.~\eqref{eqn:SchleichLaguerre}\\
 \hline
 30&20& 0.0036 & 0.0143&0.0015\\ 
 \hline
 40&30& 0.0024 & 0.0099&0.0011 \\ 
 \hline
 50&40&0.0017  &0.0076&0.0008 \\ 
 \hline
 30&10 &0.0068  & 0.1198&0.0024\\ 
 \hline
\end{tabular}
\end{center}
\caption{Root mean squared error for our, Tricomi's and the WKB approximation for $\bra{m}D(d,0)\ket{n}$ using $512$ equally spaced points for  $d\in(\sqrt{2m}-\sqrt{2n}+\delta,\sqrt{2m}+\sqrt{2n}-\epsilon)$ when compared with the exact result. $\epsilon=0.3,\delta=5$ in the figure.We shift the lower limit
 asymmetrically so that Tricomi's approximation does not produce large errors because of its poorer performance near the lower limit.
}
\label{tab:rmselag2}
\end{table}

This analysis shows that for $m=n$ Tricomi's result works the best, for $m\neq n$,
the WKB approximation fares better than the others. However, in each case, the approximations derived here
have errors of the same order as the WKB method.

 As before, all the numerical results and plots in this section were generated with the mpmath library
\cite{mpmath} with default precision. (15 decimal places)

 \end{description}

\end{enumerate}

\subsection{{Selected squeezing operator matrix elements}}\label{sec:squeezing}
In this subsection, we work with squeezed states. They are obtained by the action of the squeezing operator
$S_0(\mu)$ on the vacuum state. We had remarked in the introduction that by the construction itself
the generators of the $h_4$ algebra $\{\hat{q},\hat{p},\hat{a}^\dagger\hat{a}\}$ have simple actions on 
coherent states. In contrast, the squeezing operator has non-trivial action on coherent states.
As a first step, we develop one dimensional in-phase integral representations in  phase space for the squeezed vacuum. 
For any real $\mu$ let us 
define :
\begin{align}
S_0(\mu)&=e^{\frac{\mu}{4}\left[{(\hat{a}^\dagger)}^2-\hat{a}^2\right]}\\ 
|\psi_\mu\rangle &= S_0(\mu)|0\rangle\,.
\end{align}
The squeezing operator acts on the position and momentum operators as reciprocal scaling
\begin{align}
S_0(\mu)(\hat{q},\hat{p})S_0(\mu)^{-1} = &( e^{-\mu/2} \hat{q},~e^{\mu/2}\hat{p})\,,\nonumber\\
S_0(\mu)D(q,p)S_0(\mu)^{-1} = & D(e^{\mu/2}q ,~e^{-\mu/2}p)\,,
\label{eqn:squeezedoperatorondisplacement}
\end{align}
From this the effects on position and momentum eigenstates follow:
\begin{align}
S_0(\mu)|q,\text{pos}\rangle = & e^{\mu/4}|e^{\mu/2}q,\text{pos}\rangle\,,\nonumber\\
S_0(\mu)|p,\text{mom}\rangle = & e^{-\mu/4}|e^{-\mu/2}p,\text{mom}\rangle\,.
\label{eqn:squeezingposmomstate}
\end{align}
As a result, for example, the Schr\"odinger wavefunction of a
squeezed coherent state is, from Eqn. \eqref{eqn:wavefncoherent}, read off immediately:
\begin{align}
\langle q^\prime, \text{pos}|S_0(\mu)|q,p\rangle  = {e^{-\mu/4}\over\pi^{1/4}}
\exp\left\{-{1\over2}(e^{-\mu/2}q^\prime-q)^2+ip(e^{-\mu/2}q^\prime-q/2)\right\}\,.
\label{4.6}
\end{align}

From Eqn.~\eqref{4.6}, the position and momentum space wavefunctions of the squeezed vacuum are:
\begin{align}
\psi_\mu(q) = {e^{-\mu/4}\over\pi^{1/4}} e^{-{1\over2}e^{-\mu}q^2}\,,\nonumber\\
\tilde{\psi}_\mu(p) = {e^{\mu/4}\over\pi^{1/4}} e^{-{1\over2} e^\mu p^2}\,.
\end{align}
Thus, consistent with Eqn.~\eqref{eqn:squeezingposmomstate}, for $\mu>0$ (respectively $\mu<0$), $|\psi_\mu\rangle$ is squeezed in momentum (respectively position). 
Therefore, we expect to obtain an integral representation along a horizontal line in the phase plane, as in Eqn.~\eqref{expansion}, if $\mu$ is positive; 
and along a vertical line as in Eqn.~\eqref{Expansion_momentum} if $\mu$ is negative. Indeed, following the steps given in Section~\ref{sec:h4generators} and 
using Eqns.~\eqref{Fourier_momentum},~\eqref{3.12} in the two cases, we find the integral representations:
\begin{subequations}
\begin{align}
\mu>0,k=&e^{\mu/2}:\nonumber\\
|\psi_\mu\rangle =&\sqrt{{k\over2\pi(k^2-1)}} \,e^{k^2p^2_0/2(k^2-1)} \int_{-\infty}^{\infty}dq~ e^{-q(q+i(k^2+1)p_0)
/2(k^2-1)} |q,p_0\rangle\,,\text{any}~p_0\,;\\
\mu<0,k=&e^{-\mu/2}:\nonumber\\
|\psi_\mu\rangle =&\sqrt{{k\over2\pi(k^2-1)}} \,e^{k^2q^2_0/2(k^2-1)} \int_{-\infty}^{\infty}dp~ e^{-p(p-i(k^2+1)q_0)
/2(k^2-1)} |q_0,p\rangle\,,\text{any}~q_0\,;
\end{align}
\end{subequations} 
For general choices of $q_0,p_0$ these are not in-phase integrals. For instance, for $\mu>0$, calculating the Pancharatnam phase between adjoining elements in the 
superposition  one obtains
\begin{align}
 \phi_{P}\left[ e^{-i(k^2+1)p_0 q\over 2(k^2-1)}\vert q,p_0\rangle, e^{-i(k^2+1)p_0 q'\over 2(k^2-1)}\vert q',p_0\rangle\right]=p_0(q-q')\frac{k^2}{k^2-1}
\end{align}
which vanishes only when $p_0=0$. Similarly for $\mu <0$ this happens when $q_0=0$.

So we have the simpler preferred in-phase integral representations:
\begin{subequations}
\begin{align}
\mu>0:~~
|\psi_\mu\rangle =&\sqrt{{k\over2\pi(k^2-1)}} \, \int_{-\infty}^{\infty}dq~e^{-q^2/2(k^2-1)}
|q,0\rangle\,,\label{eqn:squeezedmom}\\
\Delta p = & {1\over k\sqrt{2}} ~< ~{1\over\sqrt{2}}\,;\\
\mu<0:~~
|\psi_\mu\rangle =&\sqrt{{k\over2\pi(k^2-1)}} \, \int_{-\infty}^{\infty}dp~ e^{-p^2/2(k^2-1)}
|0,p\rangle\,,\label{eqn:squeezedpos}\\
\Delta q = & {1\over k\sqrt{2}} ~<~{1\over\sqrt{2}}\,;
\end{align}
\end{subequations}
for the squeezed vacuum, with $k=e^{|\mu|\over2}>1$. We note that these representations were first derived by Agarwal and Simon in \cite{Agarwal_1992}, 
though without any particular reference to the in-phase nature of the superpositions involved.

In the final two examples, we  study the overlap between Fock states and the squeezed coherent states and the  
squeezed number states. These have been extensively studied in the literature 
\cite{Yuen},\cite{Satyanarayana,*kral1990displaced} and we compare our results for the matrix elements with the ones reported there. 
A detailed  history of earlier works on these overlaps and comprehensive references may be found in \cite{nieto1997displaced}.
In particular, we note that both these matrix elements have also been calculated in \cite{Janszkysqueezeddispacednumberstate,*Janskysqueezedcoherent}
using a one dimensional representation of a squeezed displaced number state on a circle in conjunction with a one dimensional representation of 
a squeezed coherent state as a gaussian-weighted continuous superposition along the real axis.
\vskip3mm

\noindent a)
As noted earlier,  oscillations in the photon number distributions in the squeezed coherent states 
was one of the problems in which the notion of `interference in phase space' first arose
\cite{Schleich_1987,*Schleich_1988foundations},~\cite{Schleich_1988},~\cite{Schleich_2001},\cite{Mundarain1}. 
As a first application of the above integral representations, we compute the overlap between a Fock state and a squeezed coherent state 
 which is a ``mixed matrix element" of $S_0(\mu)$:
\begin{align}
\langle n|S_0(\mu)|q,p\rangle = \langle n|S_0(\mu)D(q,p)|0\rangle\,.
\end{align}
Using Eqn.~\eqref{eqn:squeezedoperatorondisplacement}, we can move the displacement operator to the left of the squeezing operator with a change in the parameters:
\begin{align}
S_0(\mu)D(q,p) = D(e^{\mu/2}q,~e^{-\mu/2} p)S_0(\mu)\,,
\end{align}
therefore we may equally well deal with the expression involving $|\psi_\mu\rangle$:
\begin{align}
\langle n|D(q,p)S_0(\mu)|0\rangle = \langle n|D(q,p)|\psi_\mu\rangle\,.
\end{align}
Let us assume $\mu>0$. Then using (the conjugate of) Eqn.~\eqref{displacement_fock}, followed by Eqns.~\eqref{eqn:squeezedmom} and 
~\eqref{eqn:coherentinnerproduct} we get:
\begin{align}
\langle n|D(q,p)|\psi_\mu\rangle = & {\mathcal{N}_n\over2\pi} \sqrt{{k\over2\pi(k^2-1)}} \int_{-\infty}^{\infty} 
dq^\prime \int_0^{2\pi}d\theta ~e^{in\theta+i\sqrt{n\over 2}(pC-qS)-{q^{\prime2}\over2(k^2-1)}}\nonumber\\
&\times \langle \sqrt{2n}C-q,  \sqrt{2n}S-p|q^\prime,0\rangle\,\nonumber\\
=&{\mathcal{N}_n\over2\pi}\sqrt{{k\over2\pi(k^2-1)}} e^{-{n\over2}-{1\over4}(q^2+p^2)} \int_{-\infty}^{\infty} 
dq^\prime \int_0^{2\pi} d\theta~ e^{in\theta} e^W\,,\nonumber\\
W = & \sqrt{{n\over2}}(q+ip+q^\prime)e^{-i\theta}-{q^\prime\over2}(q-ip) - {q^{\prime2}\over4}{k^2+1\over k^2-1}\,\nonumber\\
C=&\cos\theta,~S=\sin\theta\,.
\end{align}
As $n\ge 0$, the angular integration is immediate and leads to
\begin{align}
\langle n|D(q,p)|\psi_\mu\rangle = & {1\over\sqrt{2^n n!}} \sqrt{{k\over 2\pi(k^2-1)}} e^{-{1\over 4}(q^2+p^2)}\nonumber\\
&\times \int_{-\infty}^{\infty}dq^\prime(q^\prime+q+ip)^n e^{-{q^{\prime 2}\over4}{k^2+1\over k^2-1} - {q^\prime\over2}
(q-ip)}\,.
\end{align}
After a (complex) translation and then scaling the integration variable, we obtain a standard integral whose value involves a Hermite polynomial, and after simplification the final result is:
\begin{align}
\langle n|D(q,p)|\psi_\mu\rangle = & {i^n\over\sqrt{2^nn!}} \sqrt{{2k\over k^2+1}} \left({k^2-1\over k^2+1}\right)^{n/2}\nonumber\\
&\times \exp(-(q-ip)(q+ik^2p)/2(k^2+1)) H_n\left({k^2p-iq\over\sqrt{k^4-1}}\right)\,.\nonumber\\
k = & e^{\mu/2}~,~~\mu >0\,.
\label{4.52}
\end{align}
This result matches up with the one found in \cite{Yuen},\cite{Satyanarayana,*kral1990displaced}.
The case $\mu<0$ can be treated in a similar manner, but we forego the details.\\
b) ~As a last example of the application of our phase space methods, we consider thematrix elements of the squeezing operator 
$S_0(\mu)$ in the Fock basis. This has relevance to the oscillations in the photon number distribution in squeezed Fock states. 
Ref~\cite{KnightSqueezednumberstates} analyzes it using the phase space approach, and
finds four interfering regions instead of two as in the case of the photon distributions of 
squeezed coherent states.

Assume again $\mu >0$. The result Eqn.~\eqref{4.52} just obtained comes in handy. Using Eqn.~\eqref{Fock_in_phase} for the ket in the matrix element of interest,
we have with $k=e^{\mu/2}$, $C=\cos\theta$ and $S=\sin\theta$ as usual:
\begin{align}
\langle n|S_0(\mu)|m\rangle = &{\mathcal{N}_m\over2\pi} \int_0^{2\pi}d\theta~ e^{-im\theta}
\langle n|S_0(\mu)|\sqrt{2m}C,~\sqrt{2m}S\rangle\nonumber\\
 = &{\mathcal{N}_m\over2\pi} \int_0^{2\pi}d\theta ~e^{-im\theta}
\langle n|S_0(\mu)D(\sqrt{2m}C,\sqrt{2m}S)|0\rangle\nonumber\\
 =& {\mathcal{N}_m\over2\pi} \int_0^{2\pi}d\theta~ e^{-im\theta}
\langle n|D(\sqrt{2m} k C, {\sqrt{2m}\over k}S)|\psi_\mu\rangle .
\end{align}
Here, we used Eqn.~\eqref{eqn:squeezedoperatorondisplacement} in moving $S_0(\mu)$ to the right of the displacement operator. The matrix element inside the integrand is just what 
we computed in Eqn.~\eqref{4.52}, with the choices $q=\sqrt{2m}kC$, $p=\sqrt{2m}S/k$. After simplification of this expression to expose the $\theta$ 
dependencies, we have:
\begin{align}
\langle n|S_0(\mu)|m\rangle  = & {\mathcal{N}_m\over2\pi} {i^n\over\sqrt{2^nn!}} 
\sqrt{{}2k\over k^2+1} \left({k^2-1\over k^2+1}\right)^{n/2} e^{-m/2}\nonumber\\
&\times \int_0^{2\pi}d\theta~ e^{-im\theta} e^{-{m\over 2}{k^2-1\over k^2+1} e^{2i\theta}} 
H_n\left( {-ik\sqrt{2m}\over\sqrt{k^4-1}} e^{i\theta}\right)\,.
\end{align}
Note that $\langle n|S_0(\mu)|m\rangle$ is non-zero only when $m,n$ are
both even(odd).
The fact that this must be so follows from the fact that $S_0(\mu)$ is invariant under parity, and the eigenstates of the simple harmonic oscillator 
$\ket{m}$ and $\ket{n}$ have even(odd) parity depending upon whether $m,n$ is even(odd).

Carrying out the $\theta$-integration and after simplification we have the result for $\mu>0$:
\begin{align}\label{eqn:SqueezingpartialFockbasis}
\langle n|S_0(\mu)|m\rangle = & \sqrt{{2k\over k^2+1}} {i^n\over\sqrt{2^nn!}} \left({k^2-1\over k^2+1}\right)^{n/2}
{1\over\sqrt{m^mm!}} \nonumber\\
&\times {d^m\over d\xi^m} \left(e^{-{m\over2}{k^2-1\over k^2+1}\xi^2}
H_n\left({-ik\sqrt{2m}\over \sqrt{k^4-1}}\xi\right)\right)_{\xi=0}\,\nonumber\\
&=\frac{i^{n-m}}{\sqrt{2^n2^m}
\sqrt{m!n!}}\sqrt{\frac{2k}{k^2+1}}{\left(\frac{k^2-1}{k^2+1}\right)}^{\frac{n+m}{2}}
\frac{\partial^m}{\partial\xi^m}\left[e^{\xi^2}H_n\left(\frac{2k}{k^2-1}\xi\right)\right]_{\xi=0}
\end{align}
Using 
$H_n(x)={\left(-1\right)}^ne^{x^2}\frac{d^n}{dx^n}e^{-x^2}$ we get
\begin{equation}
	\left[\frac{d^{m-p}}{d\xi^{m-p}}\,\,e^{\xi^2}\right]_{\xi=0}=-H_{m-p}(0)
\end{equation}
Also,
\begin{equation}
	\frac{d^p}{d\xi^p}\left[H_n\left(\frac{2k}{k^2-1}\xi\right)\right]_{\xi=0}=\left(\frac{4k}{k^2-1}\right)^p\frac{n!}{(n-p
	)!}H_{n-p}(0)
\end{equation}
Using Leibnitz rule, we get
\begin{align}
\langle n|S_0(\mu)|m\rangle &=-\frac{i^{n-m}}{\sqrt{2^n2^m}
\sqrt{m!n!}}\sqrt{\frac{2k}{k^2+1}}{\left(\frac{k^2-1}{k^2+1}\right)}^{\frac{n+m}{2}}\times\\&\sum_{p=0}^{\text{min}(m,n)}
{m\choose p} \left(\frac{4k}{k^2-1}\right)^p\frac{n!}{(n-p)!}H_{n-p}(0)\,\,H_{m-p}(0)
\end{align}
Since,
\begin{equation}
H_{2n}(0)={\left(-1\right)}^n\frac{(2n)!}{n!}; \,\,H_{2n+1}(0)=0;	
\end{equation}
we see that $\langle n|S_0(\mu)|m\rangle\,\,\neq 0$ only when $m-p$ and $n-p$
are simultaneously even which is only possible when $m,n$ have the same parity
as previously noted.
Thus, if $m-n$ is even 
\begin{align}
\langle n|S_0(\mu)|m\rangle =-i^{n-m}&\frac{\sqrt{m!n!}}{\sqrt{2^n2^m}
}\sqrt{\frac{2k}{k^2+1}}{\left(\frac{k^2-1}{k^2+1}\right)}^{\frac{n+m}{2}}
\,\,\,\sum_{p=0}^{\text{min}(m,n)}\left(\frac{4k}{k^2-1}\right)^p\frac{\left(-1\right)^{\frac{m+n}{2}-p}}
{p!\left(\frac{n-p}{2}\right)!\left(\frac{m-p}{2}\right)!}
\label{eqn:SqueezingFockBasis}
\end{align}
where $p$ has the same parity as $m$ (or $n$). If $m-n$ is odd,
$\langle n|S_0(\mu)|m\rangle=0$. Our result matches with the formula given in Ref.~\cite{Satyanarayana,*kral1990displaced}.

\section{Concluding remarks}\label{sec:conclusions}
It is well known that owing to the overcompleteness of coherent states, a general vector in the Hilbert space can be expressed as a sum or 
integral over these states in many different ways. The question that we pose and answer in this work is : Can one find a useful general 
strategy that permits one to select a `preferred' expansion from this multitude of possibilities ? In this work, guided by ideas and insights gained from 
the geometric phase theory, we propose one such scheme, which consists in demanding that the expansion be locally in-phase in the sense explained
in the text. A particularly appealing feature of this strategy is that the property of an expansion being locally in phase is preserved 
under unitary transformations. The usefulness of the scheme proposed here is demonstrated by deriving exact and asymptotic expressions for 
a selection of overlaps and matrix elements of physical interest. Area considerations emerge naturally in this setting, and we discuss 
consequences of such superpositions on the behaviour of $Q$ functions. In-phase superpositions are seen to have bearing on squeezing too.
Although the  scheme developed in this work is in the context of the `harmonic oscillator' or the Schr\"odinger coherent states, we have no 
doubt that the notion of `in-phase' expansions based on the geometric phase theory will find useful applications in the context of other 
coherent state systems as well.
\section*{Acknowledgements}
Mayukh N. Khan was supported by NSF grant DMR 1351895-CAR while at the University of Illinois, Physics Department, ICMT, 
where he did some of the calculations. He is also grateful to the Institute of Mathematical Sciences, Chennai for hosting him as a visitor. 
N. Mukunda thanks the Indian National Science Academy for enabling this work through the INSA
Distinguished Professorship. 
The last author, R. Simon, acknowledges support from the
Science and Engineering Research Board (SERB), Government of India,
for support through a Distinguished Fellowship.
%
\end{document}